\documentclass[11pt]{article}

\usepackage[margin=1in]{geometry}
\usepackage{graphicx}
\usepackage{booktabs}
\usepackage{amsmath}
\usepackage[version=4]{mhchem}
\usepackage{lineno}
\usepackage[numbers,sort&compress]{natbib}
\usepackage[hidelinks]{hyperref}

\graphicspath{{figures/main/}{figures/extended_data/}}

\newcommand{\CIDERC}{CIDER26C}
\newcommand{\CIDERCD}{CIDER26C-D4}
\newcommand{\CIDERSS}{CIDER26SS}

\title{Machine-learned exchange-correlation functionals for molecules, solids, and reactive surfaces}

\author{
Mohamed S. Abdallah\textsuperscript{1,*},
Zhuotao Jin\textsuperscript{1,2},
Boris Kozinsky\textsuperscript{1,3},
and Kyle Bystrom\textsuperscript{4,1,\dag}
}
\date{}

\begin{document}
\maketitle

\begin{center}
\small
\textsuperscript{1}Harvard University, Cambridge, Massachusetts 02138, USA\\
\textsuperscript{2}Massachusetts Institute of Technology, Cambridge, Massachusetts 02139, USA\\
\textsuperscript{3}Robert Bosch LLC Research and Technology Center, Cambridge, Massachusetts 02139, USA\\
\textsuperscript{4}Initiative for Computational Catalysis, Flatiron Institute, New York, New York 10010, USA\\
\textsuperscript{*}e-mail: mabdallah@g.harvard.edu\\
\textsuperscript{\dag}e-mail: kbystrom@flatironinstitute.org
\end{center}

\begin{abstract}
    The application of density functional theory to heterogeneous catalysis is hindered by the shortcomings of conventional density functional approximations. We combine machine learning with explicitly non-local physically informed descriptors and introduce an exchange-correlation functional (CIDER26SS) framework regularized for wide transferability. CIDER26SS is size-extensive, highly efficient, provides a balanced and accurate description of both molecular and solid-state systems, and is specifically well-optimized for transition metal surface chemistry.  Surpassing existing conventional functionals, CIDER26SS resolves the CO/Pt puzzle, identifying the correct binding site for CO adsorption on the Pt(111) surface, along with an accurate adsorption energy, Pt lattice constant, and surface energy.  Predictions agree well with the experimental values, even when all bulk and surface data for Pt are excluded from the training set. Remarkably, CIDER26SS exceeds the accuracy of semilocal approximations even for systems far outside the training domain.
\end{abstract}

\begin{figure}[!t]
    \centering
    \includegraphics[width=\linewidth]{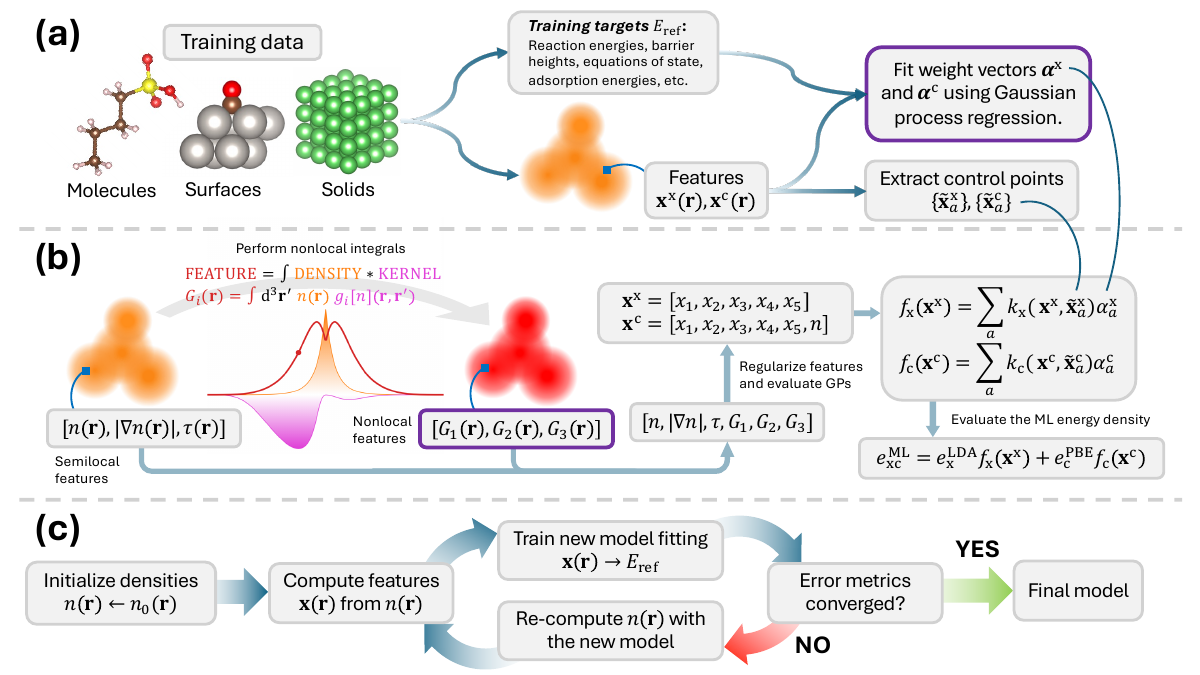}
    \caption{\textbf{CIDER learns the full exchange-correlation energy from molecular and
    periodic reference data in one nonlocal framework.}
    \textbf{a}, Workflow for a single training iteration. Training data provide target reaction
    energies, orbitals, and densities, from which hand-designed features are extracted. Gaussian
    process regression fits the reference energies as a function of the features, yielding learned
    weight vectors and a set of control points used during inference.
    \textbf{b}, Evaluation of the CIDER feature vector and prediction of the XC energy density.
    Semilocal ingredients are augmented with nonlocal integral features of the density, and feature vectors
    are regularized to enforce exact constraints and fed into the learned Gaussian process model.
    \textbf{c}, Self-consistent training loop. Starting from initial fixed densities, the model is
    re-trained repeatedly, using the densities of the current model as inputs, until the error
    metrics converge.}
    \label{fig:framework}
\end{figure}

Many modern technologies rely on a detailed understanding of heterogeneous chemical systems: small molecules interacting with transition metals and oxides for heterogeneous catalysis~\cite{thorarinsdottirSelfhealingOxygenEvolution2022,skubicReviewMultiscaleModelling2024}; interfaces between the electrodes and electrolyte in batteries~\cite{dingCoupledReactionDiffusion2025}; two-dimensional materials bound to substrates~\cite{dahalGrapheneNickelInterfaces2014} or decorated with adatoms~\cite{coulterEngineeringIdealHelical2024}; and so on. Computational modeling of these systems requires large simulation cells, placing them beyond the reach of accurate but computationally costly correlated wave function methods~\cite{bartlettCoupledclusterTheoryQuantum2007,kent2020qmcpack}. Simulations therefore tend to use density functional theory (DFT)~\cite{hohenberg1964inhomogeneous,kohn1965self}, the computationally efficient workhorse of electronic structure calculations. However, while DFT possesses the efficiency to simulate systems with thousands of electrons, the need to approximate the exchange-correlation (XC) contribution to the ground-state energy hinders its accuracy. For example, DFT predictions could in principle enable the computational screening and design of transition metal catalysts for ammonia synthesis and other industrial applications~\cite{norskov2009towards,wellendorff2015benchmark}, but deviations between different XC approximations can leave ammonia synthesis rates uncertain by up to two orders of magnitude~\cite{medford2014assessing}. Better XC functionals are needed to improve the precision of these predictions.

XC functionals can be categorized along two axes: the complexity (and computational cost) of the ``ingredients'' used to predict the XC energy~\cite{perdewJacobsLadderDensity2001}, and the level of empiricism used to design the functional~\cite{Kaplan2023TheTheory}. Semilocal functionals predict the XC energy density at a point in space using only the density, density gradient, and/or kinetic energy density at that point. This computationally efficient structure is convenient but limits the accuracy of semilocal DFT~\cite{mardirossian_thirty_2017-1}.
Hybrid functionals, on the other hand, explicitly incorporate nonlocal interactions between orbitals via the exact exchange operator to improve the description of the XC energy. In general, hybrid functionals are far more accurate than semilocal functionals for molecular systems~\cite{mardirossian_thirty_2017-1}, but the evaluation of the exact exchange operator is costly~\cite{neeseEfficientApproximateParallel2009a,linAdaptivelyCompressedExchange2016}.
The computational expense of hybrid DFT is most severe in plane-wave calculations~\cite{kresse1993ab}, which are the most widely used approach for condensed-phase systems. Critically, the exact exchange contribution to hybrid functionals must be appropriately screened by correlation effects in metallic systems~\cite{paier2006screened,paier2006screenederratum,paier2007b3lyp}, and this effect can only be captured by even more computationally expensive virtual orbital-based correlation~\cite{schimka2010accurate,schmidt2018benchmark}, not semilocal models. Therefore, most hybrid functionals are impractical for metals. In heterogeneous systems, such as a molecule on a metal surface, this leads to a difficult trade-off when choosing a simulation method. The metal would be better described by a semilocal model that implicitly screens out any exact exchange, while the molecule would be better described by hybrid DFT. Further, the molecule-surface interaction is intrinsically nonlocal, so an accurate treatment requires nonlocal density functionals~\cite{dionVanWaalsDensity2004,vydrov2010vv10}. In addition, semilocal functionals typically underestimate surface energies and overestimate chemisorption energies; modifying the functional often improves one property at the expense of the other~\cite{schimka2010accurate}.

A second conundrum arises when considering the level of empiricism used to design an XC model. Nonempirical functionals~\cite{perdew1996generalized,taoperdew2003tpss,sun2015strongly,furness2020accurate} are designed by interpolating between XC energies known from exact physical constraints and simple model systems like the uniform electron gas. Empirical functionals~\cite{becke1988exchange,lee1988development,becke1993density,becke1997systematic,mardirossian2014wb97xv,mardirossian2016wB97MV,grimme2006semiempirical} fit parametric functional forms to chemical reference databases consisting of experimental and/or high-accuracy computational data. Empirical functionals, especially empirical hybrids, tend to perform better than nonempirical models for molecular systems~\cite{mardirossian_thirty_2017-1}, for which the required training data is available in carefully curated benchmark databases~\cite{goerigk2017look,liang2025gscdb137}. Comparable benchmark data for extended systems are harder to generate, so most empirical functionals are trained only on molecules, with a few exceptions mostly focused on surface chemistry~\cite{wellendorffDensityFunctionalsSurface2012,Wellendorff2014MBEEF:Functional,trepte2022vcml}. In addition, the chemical composition of materials is arguably much more diverse than that of the molecular systems for which empirical functionals thrive, making it a difficult domain to fit empirically. As a result, nonempirical functionals built on robust and universal physical principles tend to be more reliable than empirical functionals for solids~\cite{Kaplan2023TheTheory}. Again, this causes problems in the case of heterogeneous systems: What is the best model to use when solids (especially metals) interact with molecules?

The above considerations suggest a recipe for XC functionals capable of simultaneously describing molecules, solids, and their interactions. First, exact exchange should be avoided if possible, due to its computational cost and problematic behavior for metals. Second, nonlocal exchange and correlation effects must be accounted for to mitigate self-interaction error and capture long-range dispersion interactions, respectively. Third, the functional should incorporate exact physical constraints to improve transferability across a broad range of systems while also leveraging empirical fitting to improve accuracy for molecular systems. Finally, accurate training data must be employed for solids and heterogeneous systems in addition to molecules, since the chemistry at play in these different cases can be quite distinct. In this work, we execute this recipe to design a machine-learned (ML) XC functional that predicts molecular properties with the accuracy of hybrid DFT while also simulating transition metal surface chemistry with better accuracy than semilocal DFT.

ML XC functionals~\cite{tozer1996exchange,snyder2012finding,snyder2013orbital,snyder2015nonlinear,li2016understanding,li2016pure,hollingsworth2018can,ma2022evolving,pokharel2022exact,cuierrier2021constructing} are a logical extension of empirical functional fitting. By substituting flexible ML architectures for the simpler regression models used by conventional empirical functionals, the problem of explicitly designing analytical forms for the functional can be bypassed, and the accuracy of the model is only limited by its training data and input descriptors. However, ML functionals have not yet achieved widespread use by DFT practitioners. Most of the existing ML functionals either use semilocal functional forms~\cite{dick2021highly,kasim2021learning,li2021kohn,nagai2022machine}, sharply limiting their accuracy, or use atomic geometry-dependent descriptors and restrict their focus to a narrow class of chemical systems~\cite{chen2021deepks,margraf2021pure,gao2024learning}. A few studies have introduced geometry-independent nonlocal density descriptors~\cite{lei2019design,nagai2020completing,riemelmoser2023machine,zhou2019toward,ryabov2020neural}, but these models have also not achieved widespread use beyond their initial development. The DM21 functional~\cite{kirkpatrick2021pushing} achieved high accuracy on molecular benchmarks by employing a local range-separated hybrid functional form, but its practical application was hindered by transferability issues, numerical stability problems, and high computational cost~\cite{palos2022density,zhao2024dm21,kulaev2025dm21}. Other machine-learned hybrids and double hybrids have been developed~\cite{liu2017improving,vargas2020bayesian,kauppDataDrivenLearningOptimal2025,kovacs2026doubly}, but they were trained for molecules and still exhibit the higher computational cost of hybrid DFT.

Recently, Microsoft Research released the Skala functional~\cite{luise2025accurate}, which attained state-of-the-art accuracy for molecular systems at similar cost to semilocal DFT by leveraging a training set of hundreds of thousands of molecular data. Skala incorporates an atomic structure-dependent nonlocal interaction layer, which improves accuracy at the cost of abandoning exact constraints and the universal, geometry-independent nature of the exact functional~\cite{hohenberg1964inhomogeneous}. Skala also does not yet support DFT calculations of periodic systems. As discussed above, training data is much harder to obtain for solids than for molecules, and nonempirical functionals tend to perform better for solids. This makes geometry-independent, physically-informed models an intellectually and pragmatically important complement to deep learning approaches like Skala.

In this work, we adopt the compressed scale-invariant density representation (CIDER) developed for learning exchange functionals~\cite{bystrom2022cider,bystrom2024nonlocal,bystrom2024training}, and we extend it to fit the full exchange-correlation energy. Figure~\ref{fig:framework} gives an overview of the structure and training procedure for a CIDER XC functional. The key distinguishing features of CIDER models are the Gaussian process training framework and the nonlocal descriptors. Gaussian process (GP) regression~\cite{Rasmussen2005GaussianLearning} is a Bayesian learning technique that places a prior over predictive functions and yields a posterior predictive distribution that quantifies the uncertainty of model predictions. GP covariance and noise hyperparameters control the smoothness of the learned function and the weighting of noisy training data, which helps avoid the numerical stability issues encountered in some ML functionals. The other distinctive feature of CIDER functionals is nonlocal descriptors of the density distribution. In addition to providing necessary information about medium-range exchange and correlation effects, these descriptors use a local characteristic length-scale to physically constrain how the XC energy changes under scaling (i.e., compression or expansion) of the density~\cite{bystrom2024nonlocal}. In addition to these scaling rules, we fit all of the functionals in this work to the uniform electron gas~\cite{ceperley1980ground,perdewAccurateSimpleAnalytic1992} across a wide range of densities, using a minuscule noise hyperparameter so that this system is effectively fit exactly.

The original CIDER formalism was only built for exchange functionals, so a few significant modifications are introduced in this work. First, separate exchange and correlation enhancement factors are trained simultaneously, with spin-separable exchange and spin-coupled correlation treated by the corresponding parts of the functional. Second, the model is trained self-consistently, which has been shown to significantly improve model fidelity~\cite{dick2020machine,luise2025accurate}.  Self-consistent training is required because the choice of functional influences the density through the variational minimization (self-consistent field) procedure, so the trained model will produce variational densities different from the fixed densities used in training. We therefore iteratively retrain the model using densities obtained with the previous model until the validation error converges, as shown in Figure~\ref{fig:framework}(c). Finally, we develop a high-quality theoretical database of bulk transition metal and surface science data using the random phase approximation (RPA)~\cite{ren2012random}, which is considered a high-fidelity reference method for surface science~\cite{schimka2010accurate,schmidt2018benchmark}. The dataset includes properties relevant to heterogeneous catalysis, such as adsorption energies, surface energies, and equations of state.

We demonstrate the utility of the CIDER XC framework in two stages. First, we train two functionals, one without (CIDER26C) and one with (CIDER26C-D4) empirical D4 dispersion correction~\cite{caldeweyher2019d4}, on existing molecular data only. We show that in spite of the lack of exact exchange incorporated into the CIDER models, both new functionals exceed the accuracy of many commonly used hybrid functionals on unseen test data. Both exhibit similar performance, suggesting that the nonlocal CIDER features can learn a significant portion of the dispersion energy. Next, we train a third functional (CIDER26SS) on both the molecular data and the RPA surface science dataset. With only a marginal decrease in accuracy for molecular systems compared to CIDER26C, CIDER26SS accurately fits bulk and surface properties of metals as well as the adsorption energies of small molecules on metal surfaces.

To test the transferability of CIDER26SS and demonstrate its promise for practical applications, we remove all Pt bulk and surface data from the training set and compute the adsorption energy of CO on the Pt(111) surface. This is the famous CO/Pt puzzle~\cite{feibelman2001co}: Standard semilocal functionals predict that CO prefers high-coordination hollow sites on Pt(111), whereas experiment finds low-coverage adsorption at atop sites~\cite{feibelman2001co}. Nonlocal van der Waals functionals can shift the site preference toward the atop site~\cite{lazic2010density}, but the original vdW-DF family is known to overestimate lattice constants for solids~\cite{klimes2011van}, so obtaining the adsorption energy, site preference, and Pt lattice constant simultaneously remains a demanding test of a production-level functional. Even without Pt training data, CIDER26SS predicts the correct atop site at low coverage, gives an accurate adsorption energy, and predicts a Pt lattice constant close to the EXX+RPA and experimental values. To our knowledge, this is the only scalable, non-geometry-dependent density functional to obtain this combination of properties for CO/Pt(111) without the higher computational cost of EXX+RPA or other quantum chemistry methods.\footnote{DFT+$U$ has been shown capable of identifying the correct binding site~\cite{kresseSignificanceSingleelectronEnergies2003,maxson2026dftu}, but this approach is geometry-dependent and requires system-specific tuning of the $U$ parameters. Also, dynamic effects can affect the adsorption energy, even correcting the site preference of some functionals~\cite{guoFirstPrinciplesDeterminationCO2018,li2021thermal,weiResolvingCOPuzzle2025}. However, adsorption energy predictions with conventional functionals are often inaccurate even when accounting for dynamic effects, and calculations with random phase approximation correlation suggest that the top site should be preferred by CO on Pt(111) even when dynamic effects are not considered~\cite{schimka2010accurate}. The CO/Pt puzzle therefore serves as an important benchmark for electronic structure methods despite these subtleties.}

CIDER26SS also transfers to chemical and materials-science problems far outside its training domain: the adsorption of graphene on Ni(111), the interlayer binding of graphite, the bulk and electronic-structure properties of transition metal oxides, and the protonation energetics of iron--sulfur clusters. Across these systems, spanning dispersion-bound layered materials, strongly correlated oxides, and open-shell transition-metal chemistry, CIDER26SS consistently improves on semilocal DFT, in several cases reaching the accuracy of far more expensive hybrid or beyond-DFT methods.

\section*{Results and Discussion}

\subsection*{Overview of Datasets and Metrics}

Our molecular benchmark data draws from two existing databases, GMTKN55~\cite{goerigk2017look} and the recently published GSCDB137~\cite{liang2025gscdb137}. All of GMTKN55, along with the transition-metal chemistry and total-atomic-energy data from GSCDB137, was used to train the CIDER models. The remainder of GSCDB137 was split between validation and testing. Duplicate data points between GSCDB137 and GMTKN55 were carefully removed from the validation/test pool to ensure that these sets were not contaminated with training data, as described in the Methods section. Both databases draw primarily on reference values of coupled-cluster quality. Errors in this work are reported with three metrics. WTMAD-2 is a weighted total mean absolute deviation (MAD) introduced for GMTKN55, which combines the 55 subset MADs with the benchmark's standard weights~\cite{goerigk2017look}. NER is the unitless normalized error ratio introduced with GSCDB137, in which each subset error is divided by the official GSCDB137 standard error and the ratios are averaged, so that a value near one corresponds to a strong hybrid-functional baseline~\cite{liang2025gscdb137}. MoM-MAE is the unweighted mean of the per-subset mean absolute errors (MAEs), in units of kcal~mol$^{-1}$. We additionally use root mean square error (RMSE) metrics, NER-RMSE and MoM-RMSE, to characterize error spread. See the Methods section for further details on these error metrics.

In Figure~\ref{fig:benchmarks}(c,d), we plot both the test-only and combined validation+test error metrics for GSCDB137, with functionals sorted by validation+test error. The validation data was exclusively used to monitor self-consistent training, not to fit hyperparameters, and the self-consistent training shows no risk of overfitting (Extended Data Fig.~1). Therefore, the validation data can be considered nearly as reliable as test data for model assessment. Test-only errors are also displayed for comparison.

\subsection*{Self-consistent training}

Starting from initial ``seed'' models fit on fixed PBE and PBE0 densities (see Methods), we apply the self-consistent training loop of Figure~\ref{fig:framework}(c) to the molecular training problem, tracking iterations 0--4 for the dispersion-free and D4-corrected variants. The seed models' fixed-density WTMAD-2 values on GMTKN55 are already good, 3.20 and 3.18~kcal~mol$^{-1}$, but running the same models self-consistently increases the errors to 4.48 and 4.41~kcal~mol$^{-1}$ (Extended Data Fig.~1). This reflects the density-distribution mismatch that the self-consistent procedure targets. The model is accurate on the densities used to fit it, but orbital relaxation drives each self-consistent calculation to a different density distribution corresponding to the model's own variational minimum.

One training update on regenerated densities closes the gap, and the increased accuracy carries over to the validation set (Extended Data Fig.~1). The GSCDB137 validation NERs of the seed models fall from 2.169 and 2.023 to 1.459 and 1.470, and the later iterations remain in the narrow 1.46--1.47 range. The isolated-atom and molecular transition-metal densities were held fixed during this loop to limit cost, so we carried out two further updates in which they were also regenerated. These final iterations move the validation error just below the earlier plateau, and we select the resulting models, denoted CIDER26C (C standing for chemistry) and CIDER26C-D4, at validation NERs of 1.427 and 1.446.

We then turn to the combined model, which augments the molecular data with our new surface-science dataset computed with exact exchange plus RPA correlation (EXX+RPA). Because the first one or two iterations eliminate the fixed-density/self-consistent gap in the molecular case, we stop this loop at iteration 2 and denote the model CIDER26SS (SS standing for surface science). Its GMTKN55 WTMAD-2 is 4.10~kcal~mol$^{-1}$ at the fixed training densities and 4.12~kcal~mol$^{-1}$ self-consistently, and its MAEs for every reaction class of the RPA training data agree between the two evaluations to within a few hundredths of an eV, so the density mismatch is resolved in the combined training.

\subsection*{Benchmarks on molecular systems}

\begin{figure}[!t]
    \centering
    \includegraphics[width=0.78\linewidth]{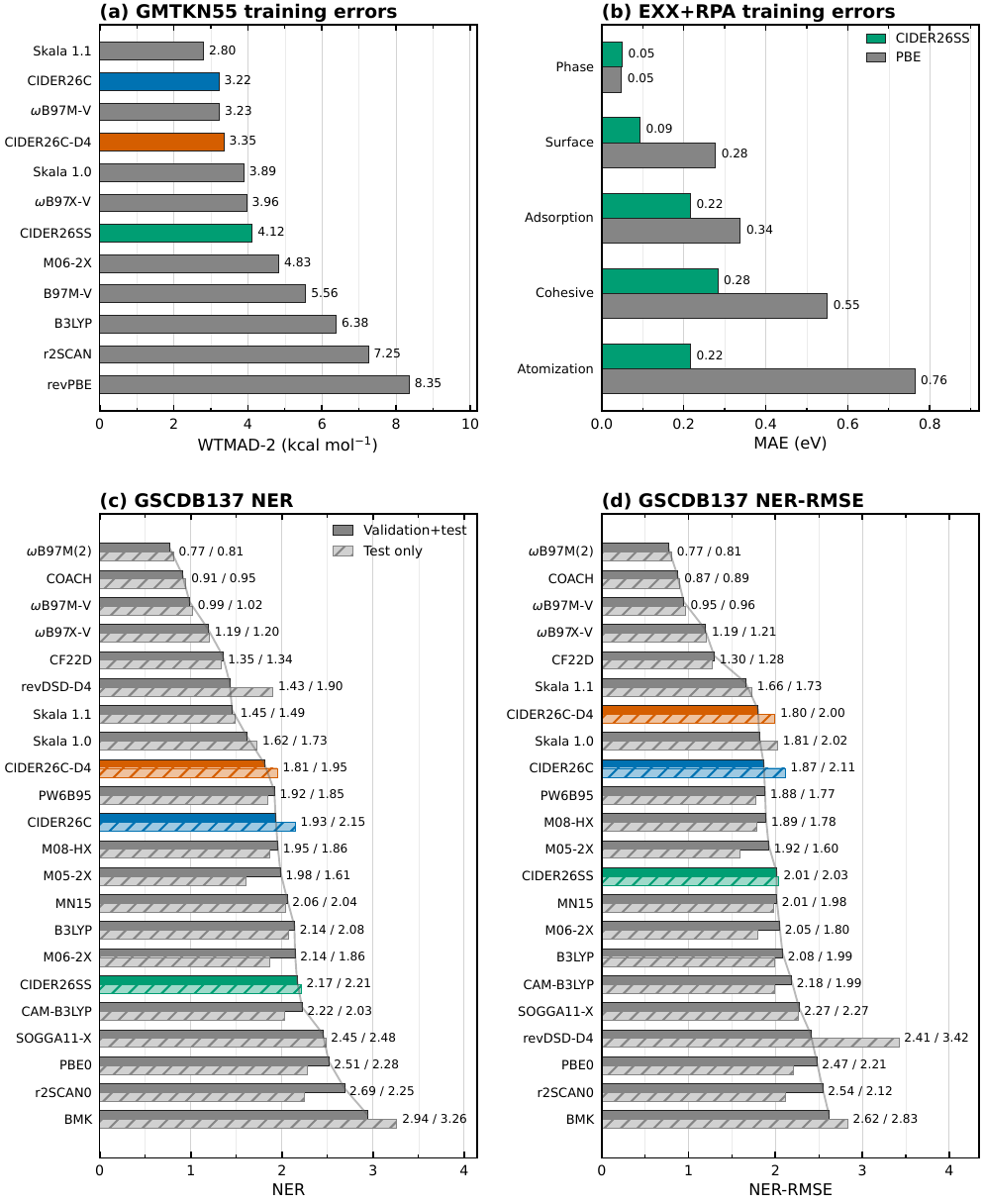}
    \caption{\textbf{One functional spans molecular and surface benchmarks at hybrid-level
    accuracy.} \textbf{a}, GMTKN55 WTMAD-2; GMTKN55 is training data for the CIDER
    models. \textbf{b}, Self-consistent \CIDERSS{} MAEs against the EXX+RPA training data for the five
    non-EOS reaction classes, with the PBE baseline evaluated on the same reactions for
    comparison. \textbf{c,d}, GSCDB137 NER and NER-RMSE; solid bars give the combined value over
    the full held-out panel (validation+test, 68 subdatasets) and hatched bars the value on the
    test split alone (3044 reactions), with labels giving combined/test values. Rows are ordered
    by the combined value, which the grey line traces as a guide to the eye. Skala~1.1 covers
    3041 of the test reactions; the other displayed CIDER and Skala models cover both complete
    splits. Lower is better in all panels. The validation and test splits are shown separately in
    Extended Data Fig.~3, and MoM-MAE and MoM-RMSE counterparts in Extended Data Fig.~2.
    Reference values are from
Refs.~\cite{luise2025accurate,luise2025accuratev3,liang2025gscdb137,liang2026coach}.}
    \label{fig:benchmarks}
\end{figure}

Figure~\ref{fig:benchmarks} compares the final models with other XC functionals on GMTKN55 and on the GSCDB137 validation and test sets, with reference values from Refs.~\cite{luise2025accurate,liang2025gscdb137,liang2026coach}. On the GMTKN55 training data, CIDER26C gives a WTMAD-2 of 3.22~kcal~mol$^{-1}$, matching the state-of-the-art hybrid functional $\omega$B97M-V~\cite{mardirossian2016wB97MV} (3.23~kcal~mol$^{-1}$), and CIDER26C-D4 remains close at 3.35~kcal~mol$^{-1}$ (Figure~\ref{fig:benchmarks}(a)). Skala~1.1 gives the lowest WTMAD-2 among the external models shown, at 2.80~kcal~mol$^{-1}$. On the blind GSCDB137 test set, both CIDER models sit in the accuracy range of established hybrid functionals. Their test NERs of 2.15 and 1.95 coincide with the range spanned by CAM-B3LYP~\cite{yanai2004new}, MN15~\cite{yu_mn15_2016}, and B3LYP~\cite{becke1993density,lee1988development,stephens1994abinitio} (2.03--2.08) (hatched bars in Figure~\ref{fig:benchmarks}(c)), and their MoM-MAEs are comparable to or lower than those of the same hybrids (Extended Data Fig.~2). On the combined validation+test panel shown as the solid bars in Figure~\ref{fig:benchmarks}(c), both models rank ahead of all three hybrids (1.93 and 1.81 versus 2.06--2.22).

The comparison between CIDER26C and CIDER26C-D4 is most informative in the noncovalent categories of GMTKN55, which assess the van der Waals (vdW) interactions that the D4 correction~\cite{caldeweyher2019d4} treats explicitly. The D4 correction does not improve CIDER on these subsets: The intermolecular noncovalent-interaction MAD is 0.29~kcal~mol$^{-1}$ for CIDER26C compared to 0.31~kcal~mol$^{-1}$ for CIDER26C-D4, and the intramolecular values are 0.28 and 0.30~kcal~mol$^{-1}$, respectively (Extended Data Table~1). The nonlocal density features of CIDER therefore learn a substantial part of the medium-range nonlocal binding from the molecular training data, though long-range dispersion models may still matter in the asymptotic regime, which these benchmarks do not thoroughly test.

The results presented so far probe reaction energies, but a functional can perform well in benchmark energy metrics while generating inaccurate ground-state densities, since density errors can cancel in energy differences~\cite{kim2013understanding}. This motivates the use of dipole moments as an assessment of the accuracy of the density distributions because the dipole is a direct moment of the density~\cite{hait2018dipoles}. On the Dip146 subset of GSCDB137, CIDER26C, CIDER26C-D4, and CIDER26SS give NERs of 1.30, 1.30, and 1.32, within the span of established hybrid functionals and far ahead of semilocal functionals such as r$^2$SCAN~\cite{sun2015strongly,furness2020accurate} (1.79), B97M-V~\cite{mardirossian2015b97mv} (2.07), and revPBE~\cite{zhang_comment_1998} (2.44)~\cite{liang2025gscdb137} (Extended Data Table~2), attesting to the quality of the ground-state densities generated by CIDER26C and CIDER26C-D4.

Skala~\cite{luise2025accurate}, another ML XC model trained for molecular chemistry, provides a useful point of comparison with the CIDER models. However, the GSCDB137 data partition was selected independently for CIDER, and some of their reactions overlap data listed in the Skala~1.1 model card~\cite{skala_hf_model_card}. Therefore, our test set data was unseen by CIDER during training but not necessarily unseen by Skala. With this caveat in mind, we note that Skala~1.1 gives lower aggregate errors on the GSCDB137 test set (1.49, compared to 1.95 for CIDER26C-D4 and 2.15 for CIDER26C). The overall NER is weighted toward noncovalent interactions because there are more noncovalent subsets compared to other data types, so we also report NERs for each category of reaction in Extended Data Fig.~4. On the combined validation+test panel, Skala~1.1 achieves lower NERs on barrier heights, isomerization energies and thermochemistry, while the CIDER functionals are more accurate for vibrational frequencies.

The possibility of other functionals being trained on our test partition can be partially, though not completely, accounted for by comparing RMSEs instead of MAEs, since RMSE disfavors large error variances that might be indicative of over-training on part of a dataset. Indeed, Skala 1.1 and Skala 1.0 have higher ratios of NER-RMSE to NER-MAE than most other functionals tested (Extended Data Table~3). By NER-RMSE, CIDER26C-D4 is equivalent in accuracy to Skala 1.0 and has only $8\%$ higher error than Skala 1.1 (1.80 vs 1.66, Figure~\ref{fig:benchmarks}(d)), though again with the caveat that the overall NER is weighted toward noncovalent interactions. There remain significant accuracy differences for different properties, as discussed above.

\subsection*{Balancing accuracy for molecules, solids, and surfaces}

Encouraged by the performance of the CIDER26C and CIDER26C-D4 models trained only on molecular data, we turn to the surface-science targets of the CIDER26SS model trained on both molecular and surface-science data. CIDER26SS reproduces the EXX+RPA training data well (Figure~\ref{fig:benchmarks}(b) and Extended Data Table~4). In self-consistent calculations, the surface energies are reproduced to 0.09~eV (MAE) and the adsorption energies to 0.22~eV, reduced from 0.28 and 0.34~eV for PBE relative to the same reference. The adsorption error is comparable to EXX+RPA's own roughly 0.2~eV mean deviation from low-coverage adsorption experiments~\cite{schmidt2018benchmark}. The equation-of-state reactions are reproduced to 6~meV, a level of precision required to inherit the reference equilibrium lattice constants. Of all the reaction classes, the cohesive energies have the largest MAE at 0.28~eV. The EXX+RPA cohesive energies and atomization energies received the smallest weights in training, reflecting the fact that the EXX+RPA reference is less trustworthy for these properties than for the others (see Methods for details).

The degradation of accuracy for molecular systems due to including solid-state training data is modest. CIDER26SS gives a self-consistent GMTKN55 WTMAD-2 of 4.12~kcal~mol$^{-1}$ versus 3.22~kcal~mol$^{-1}$ for CIDER26C, and GSCDB137 NERs of 1.87 (validation) and 2.21 (test) versus 1.43 and 2.15 for CIDER26C (Extended Data Fig.~3). Even after this drop, CIDER26SS remains within the accuracy range of commonly used hybrid functionals on the molecular test set, sitting between B3LYP (2.08) and r$^2$SCAN0~\cite{bursch2022r2scan0} (2.25) in test NER. This is the expected cost of describing two regimes with one functional; the inconsistency between the coupled-cluster molecular reference energies and the EXX+RPA solid/surface reference energies may also contribute.

This accuracy trade is favorable because we maintain hybrid-DFT accuracy for molecules while gaining a marked improvement for bulk metals and surface chemistry, relative to the PBE baseline (see Figure~\ref{fig:benchmarks}(b)). The result of our joint training is a single functional that retains this molecular accuracy while simultaneously reproducing EXX+RPA surface and adsorption energetics near the fidelity of the reference method itself. To our knowledge, no existing exchange-correlation functional offers this unique combination of accuracies.

\subsection*{The CO/Pt puzzle}

\begin{figure}[!t]
    \centering
    \includegraphics[width=\linewidth]{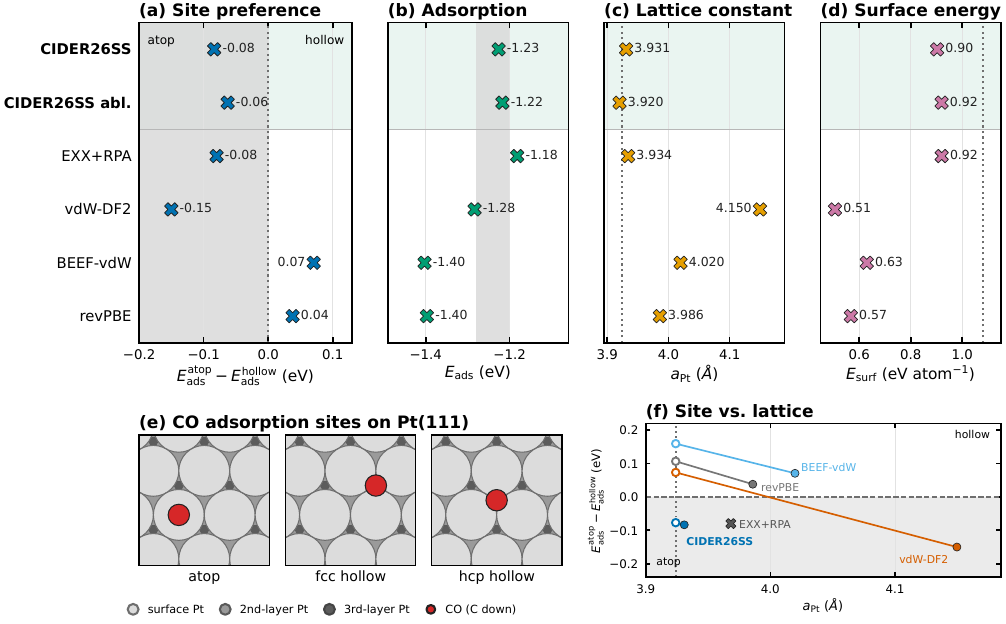}
    \caption{\textbf{\CIDERSS{} and its Pt-free ablation resolve the CO/Pt(111) puzzle, and accurately predict four observables.} \textbf{a}, Atop-site adsorption energy relative to the most stable hollow
    site at each method's optimized lattice; negative values favor atop adsorption.
    \textbf{b}, Preferred-site adsorption energy, with
    \(E_\text{ads} = E(\text{adsorbed slab}) - E(\text{clean slab}) - E(\text{CO})\), so more
    negative values are more strongly bound. \textbf{c}, Pt lattice constant. \textbf{d}, Pt(111)
    surface energy per surface atom. Values in \textbf{a--d} correspond to each method's optimized
    lattice, except that the EXX+RPA site-preference and adsorption results are at the PBE
    geometry, with the panel \textbf{c} lattice constant from a second-order fit to the platinum
    equation of state. \CIDERSS{} abl.\ is the otherwise identical model trained
    with all solid-state Pt reactions removed. \textbf{e}, Top views of the three high-symmetry
    adsorption sites on Pt(111): atop (above a surface atom), fcc hollow (above a third-layer
    atom), and hcp hollow (above a second-layer atom). Light to dark grey circles are first- to
    third-layer Pt atoms; red is CO (C down). \textbf{f}, Site-preference energy as a function of
    the Pt lattice constant at which the adsorption is evaluated: open circles at the
    experimental lattice (3.924~\AA{}) and filled symbols at each functional's own equilibrium
    lattice. Expanding from the experimental
    lattice constant to the too-large vdW-DF2 value of 4.150~\AA{} reverses the vdW-DF2 site ordering by 0.22~eV,
    whereas BEEF-vdW and revPBE shift toward atop without changing the site preference; \CIDERSS{} keeps the atop
    preference at both its own and the experimental lattice, and EXX+RPA (cross) gives atop at
    its PBE geometry. Grey shading in \textbf{a} marks
    atop preference and in \textbf{b} the experimental adsorption-energy interval, which
    originates from
    Refs.~\cite{kelemen1979binding,campbell1981molecular,poelsema1984thermal,schiesser2010thermodynamics}
    as tabulated with finite-temperature corrections in
    Refs.~\cite{wellendorff2015benchmark,schmidt2018benchmark}; dotted lines in \textbf{c,d,f}
    mark the experimental lattice constant~\cite{owen1933precision,swanson1953standard} and the
    liquid-metal-derived surface energy~\cite{tyson1977surface,vitos1998surface,stroppa2007co}.}
    \label{fig:co_pt}
\end{figure}

As a culminating test of CIDER26SS, we look at CO adsorption on Pt(111), which is a classic failure of semilocal density functionals that cannot be easily resolved by hybrid DFT due to its computational cost and poor description of metals. Semilocal DFT typically predicts adsorption at a threefold hollow site (fcc or hcp; Figure~\ref{fig:co_pt}(e)), whereas experiment and high-level periodic wave function methods find the atop site preferred~\cite{feibelman2001co,carbone2024co,hsing2019quantum}. We evaluated both CIDER26SS and an ablated variant trained identically but with every solid-state Pt reaction removed from the EXX+RPA data (Pt-containing molecular reactions from GSCDB137 were kept), to test whether the learned surface chemistry extends beyond the solids represented in training and to ensure that the success of CIDER26SS in resolving the CO/Pt puzzle does not simply occur due to the Pt reactions in the training set.

Figure~\ref{fig:co_pt} shows that CIDER26SS resolves the CO/Pt puzzle: CO correctly favors the atop site with an adsorption energy of $-1.23$~eV, with the hcp and fcc hollow sites higher in energy by 0.08 and 0.11~eV. This is in excellent agreement with experimental measurements, which give a CO/Pt(111) adsorption energy on the atop site of $-1.28$ to $-1.20$~eV~\cite{kelemen1979binding,campbell1981molecular,poelsema1984thermal,schiesser2010thermodynamics,wellendorff2015benchmark,schmidt2018benchmark}. In addition to obtaining the correct site preference, a reliable functional should simultaneously predict the correct lattice constant and surface energy of Pt. Indeed, the CIDER26SS Pt(111) surface energy (0.90~eV) and bulk Pt fcc lattice constant (3.931~\AA{}) also agree well with the experimental measurements of 1.08~eV (from measurements on liquid metal~\cite{tyson1977surface,vitos1998surface,stroppa2007co}) and 3.924~\AA{} (from X-ray diffraction~\cite{owen1933precision,swanson1953standard}). Notably, the correct binding site preference is not inherited from the semilocal baseline on which the model is built (Methods), which favors the hollow site, but learned from the EXX+RPA training data, and it is obtained at the $O(N^3)$ cost of a semilocal self-consistent calculation.

Other semilocal functionals, like revPBE~\cite{zhang_comment_1998} and BEEF-vdW~\cite{wellendorffDensityFunctionalsSurface2012}, incorrectly favor the fcc hollow site and predict the adsorption energy less accurately than CIDER26SS. vdW-DF2 does obtain the correct site preference, but its equilibrium lattice constant of 4.150~\AA{} is 5.8\% larger than the experimental value. At the experimental lattice constant, vdW-DF2 favors the fcc hollow site by 73 meV (Figure~\ref{fig:co_pt}(f)), so the previously noted correct site preference~\cite{lazic2010density} is an artifact of the known overestimation of lattice constants with vdW-DF2~\cite{klimes2011van}. CIDER26SS keeps the atop preference at both its own and the experimental lattice, which are quite similar in value (Figure~\ref{fig:co_pt}(c)), and it is the only functional we tested that simultaneously reproduces the lattice constant, surface energy, adsorption energy, and favored binding site.

The ablated model demonstrates that this accuracy transfers to data not seen in training. Trained without a single solid-state platinum reaction, it reproduces the atop site preference with an adsorption energy of $-1.22$~eV, with the fcc and hcp hollow sites higher by 0.06 and 0.10~eV (the two hollows swap order relative to the full model). The ablated model also predicts a lattice constant of 3.920~\AA{} and a surface energy of 0.92~eV per atom, within 0.02~eV of the full model's value. Obtaining the correct adsorption site and energy, on a metal whose solid state the model did not see in training or validation, implies that the combined training does not merely interpolate its adsorption data; it learns transferable trends for transition-metal surface chemistry. This demonstration shows that beyond the competitive aggregate benchmark scores reported in the previous section, the model can be applied to long-standing problems in heterogeneous catalysis. It does so without sacrificing the broad accuracy across chemistry and materials science expected of a surrogate for the universal exchange-correlation functional.

\subsection*{Other applications, transferability tests, and limitations}

\begin{figure}[!htbp]
    \centering
    \includegraphics[width=\linewidth]{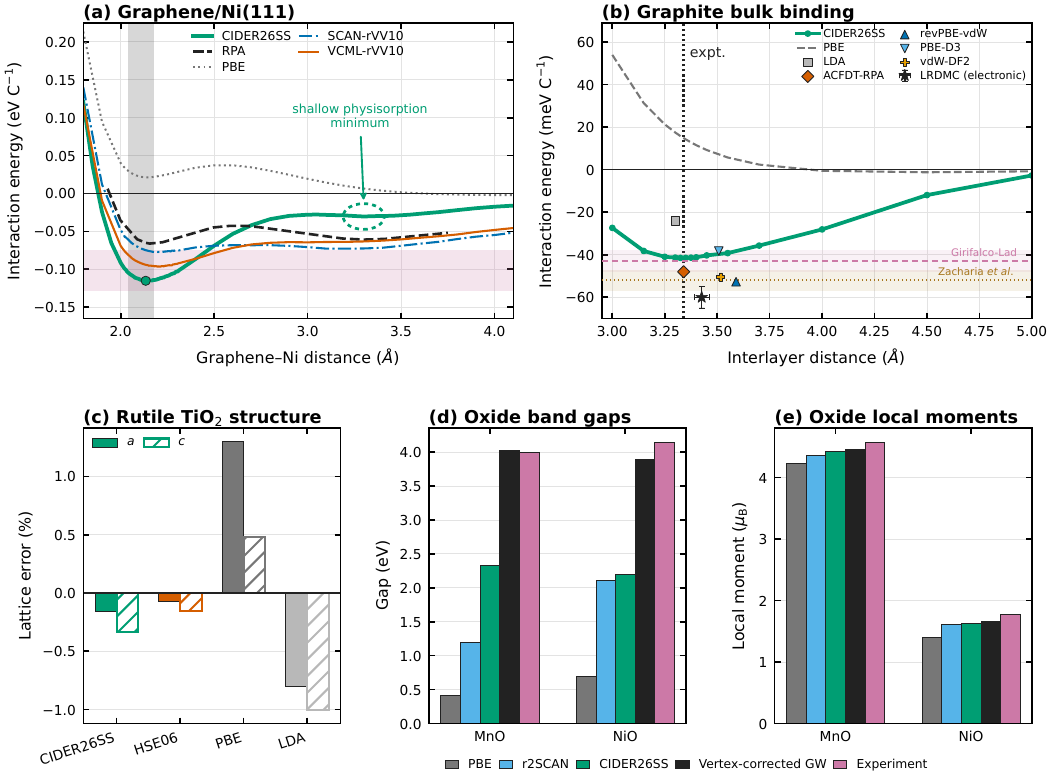}
    \caption{\textbf{\CIDERSS{} transfers to chemically distinct environments outside its
    training data.} \textbf{a}, Graphene/Ni(111) interaction curves; the vertical band is the
    direct LEED distance measurement~\cite{gamo1997graphene}, and the horizontal band is the
    model-dependent energy interval reconstructed from phase-equilibrium and exfoliation
    measurements (see text). RPA and the other density-functional curves are from Ref.~\cite{trepte2022vcml}.
    \textbf{b}, Graphite bulk interaction energy per carbon versus interlayer distance, with
    negative values denoting binding. Symbols mark literature bulk-binding minima at their
    reported equilibrium distances (LDA and electronic LRDMC from Ref.~\cite{spanu2009graphite},
    ACFDT-RPA from Ref.~\cite{lebegue2010graphite}, revPBE-vdW from
    Ref.~\cite{hazrati2014li}, and PBE-D3 and vdW-DF2 from Ref.~\cite{mchugh2020graphite}); the
    vertical dotted line marks the experimental interlayer spacing
    (3.34~\AA{})~\cite{lebegue2010graphite}. The plotted LRDMC value (\(60\pm5\)~meV per C) excludes
    zero-point and finite-temperature corrections, which reduce it to \(56\pm5\)~meV per
    C~\cite{spanu2009graphite}. Dashed and dotted lines mark the exfoliation-based experimental
    estimates~\cite{girifalco1956graphite,zacharia2004graphite}, which equal the bulk binding
    energy per layer by the identity of Ref.~\cite{jung2018exfoliation}; the Zacharia
    \textit{et al.} thermal-desorption value and the Girifalco--Lad estimate carry reported
    uncertainties of \(\pm5\)~meV per C~\cite{zacharia2004graphite,girifalco1956graphite}.
    \textbf{c}, Relative errors in the rutile \ce{TiO2} lattice constants compared to X-ray
    diffraction~\cite{burdett1987rutile,vasquez2018rutile}. \textbf{d,e}, Kohn--Sham gaps and
    projected local magnetic moments for antiferromagnetic MnO and NiO. The computed gaps
    include the zero-point renormalization of Ref.~\cite{abdallah2024qsgw} ($-0.30$~eV for MnO
    and $-0.25$~eV for NiO; the tabulated spin--orbit corrections are negligible for both
    oxides), matching the convention of the vertex-corrected \(GW\)
    values~\cite{abdallah2024qsgw}; the experimental
    gaps~\cite{vanelp1991mno,kurmaev2008oxides,sawatzky1984nio} and measured
    moments~\cite{cheetham1983mnni,roth1958monoxides,alperin1962nio,fender1968moments} are
    included.}
    \label{fig:transfer}
\end{figure}

CIDER26SS transfers well to chemical environments far from the training chemistry, which we demonstrate in this section with a series of applications ranging from transition metal chemistry to layered materials. However, as with any existing DFT functional, areas for potential improvement remain, and we also identify and analyze some shortcomings of CIDER26SS and identify potential solutions for future ML functionals.

Iron--sulfur clusters are prototypical strongly correlated systems with applications in biochemistry~\cite{zhai2023fes}. For the relative protonation energies of a dimeric iron--sulfur cluster (Extended Data Table~5), where small changes in proton position alter several strongly coupled metal centers, CIDER26SS gives a mean absolute error of 26.1~kJ~mol$^{-1}$ against the CCSD(T)+DMRG reference, far better than the semilocal PBE-D3 and r$^2$SCAN-D3 and comparable to hybrids TPSSh-D3 and B3LYP*-D3~\cite{zhai2023fes}. Skala~1.1 and a few hybrids like B3LYP-D3 have lower average error, though CIDER26SS more accurately predicts the protonation energy difference between the iron and iron bridge sites (28.1~kJ~mol$^{-1}$ vs the reference value of 30.7~kJ~mol$^{-1}$). CIDER26SS, Skala~1.1, and all hybrids except M06-D3 correctly capture the relative order of protonation energies.

We next turn to graphene and graphite, also a transferability test due to the lack of layered materials in training.
For graphene adsorbed on the Ni(111) surface (Figure~\ref{fig:transfer}(a)), CIDER26SS predicts an equilibrium graphene--Ni distance of 2.134~\AA{}, in good agreement with the experimental estimate of $2.11\pm0.07$~\AA{}~\cite{gamo1997graphene}, and an interaction energy of $-0.115$~eV per C atom, within the experimental interval of $-0.129$ to $-0.075$~eV per C reconstructed from phase-equilibrium and exfoliation measurements~\cite{shelton1974graphene,zacharia2004graphite,shepard2019graphene,trepte2022vcml}.\footnote{This experimental estimate is quite rough and relies on a single estimate of the relative stability of graphene/Ni(111) and graphene/graphite from an Arrhenius formula-based estimate~\cite{shelton1974graphene}, but it is nonetheless an approximate check of whether a functional provides a reasonable prediction. The experimental range comes from different estimates of the graphite exfoliation energy.} The curve also resolves a second, shallow physisorption minimum at 3.3~\AA{}, reproducing both the double-minimum structure that RPA predicts for graphene on Ni(111) and the position of the shallow physisorption minimum~\cite{mittendorfer2011graphene}, without incorporating explicit dispersion corrections.
For bulk graphite (Figure~\ref{fig:transfer}(b)), where PBE gives essentially no binding, the CIDER26SS total binding energy of $-42$~meV agrees well with RPA~\cite{lebegue2010graphite} and differs by only 18 meV from diffusion Monte Carlo~\cite{spanu2009graphite}. The location of the CIDER26SS minimum, 3.33~\AA{}, is in excellent agreement with both the experimental interlayer spacing and the ACFDT-RPA equilibrium distance (3.34~\AA{} for both)~\cite{lebegue2010graphite}, a closer match than any other functional shown in Figure~\ref{fig:transfer}(b), whereas the LRDMC minimum lies at a larger $3.43\pm0.04$~\AA{}~\cite{spanu2009graphite}. Because the bulk binding energy per layer rigorously equals the energy of exfoliating one surface sheet~\cite{jung2018exfoliation}, this value can be compared directly with the exfoliation-based experimental estimates of $43\pm5$~meV per C~\cite{girifalco1956graphite} and $52\pm 5$ meV per C~\cite{zacharia2004graphite} (Figure~\ref{fig:transfer}(b)). (Note that the theoretical values here are computed at 0 K without zero-point energy, while the experiments are at finite temperature. From PBE calculations~\cite{mounetFirstprinciplesDeterminationStructural2005,spanu2009graphite}, vibrational effects amount to roughly 4 meV at 300 K, which would change the total binding energy from $-42$ to $-38$~meV.)

As a case where CIDER could be improved in future versions, we also investigated \ce{O2} physisorption on graphene, which probes the long-range behavior of the dispersion interaction at a finer energy scale. CIDER26SS gives an equilibrium adsorption energy of $-0.033$~eV against the site-matched diffusion Monte Carlo value of $-0.126(4)$~eV~\cite{shin2019o2graphene}, while experiment places the adsorption energy near $-0.15$ to $-0.16$~eV~\cite{bagsican2017o2graphene}. Similar examples to this are the more rapid tail decay and the shallowness of the physisorption minimum relative to the RPA curve for graphene--Ni as shown in Figure~\ref{fig:transfer}(a). These long-range interaction underestimations could be explained by the lack of an asymptotically correct $r^{-6}$ dispersion term in CIDER26SS, which is missing because the nonlocal features decay exponentially in space~\cite{bystrom2024nonlocal}, and the D4 correction is currently available only in the molecular implementation for CIDER.

The accurate description of molecular oxygen is critical in a variety of applications, with one example being transition metal oxide formation energies. The \ce{O2} atomization energy enters the training through the W4-11 subset of GMTKN55, but these reaction energies do not fix the location or curvature of a potential minimum, so the equilibrium region of triplet \ce{O2} probes transferability between bond lengths (Extended Data Table~6). CIDER26SS predicts an equilibrium bond length of $R_e=1.2043$~\AA{} and vibrational frequency of $\omega_e=1606.2$~cm$^{-1}$, compared to spectroscopic measurements of 1.2075~\AA{} and 1580.2~cm$^{-1}$~\cite{krupenie1972o2}. The harmonic frequency has the closest agreement with experiment among the functionals in Extended Data Table~6. On the other hand, PBE0-D4, $\omega$B97M-V, $\omega$B97X-V, and both Skala versions overbind \ce{O2} with bond lengths of roughly $1.19$~\AA{} and frequencies of 1713--1726~cm$^{-1}$, more than 130~cm$^{-1}$ too stiff. The performance of Skala~1.1, at 1.1885~\AA{} and 1725.1~cm$^{-1}$, despite its accurate bond energy, is consistent with its relatively poor performance for vibrational frequencies (Extended Data Fig.~4). B3LYP-D4, B97M-V, and r$^2$SCAN-D4 are more accurate for bond length but remain 48--73~cm$^{-1}$ too stiff, while PBE-D4 errs in the opposite direction, with a longer bond and a softer frequency. Although the CIDER26SS well depth is 0.32~eV lower than the W4-17 reference~\cite{karton2017w417}, it corrects most of the \ce{O2} binding error of PBE-D4 (1.04 eV error) and improves on r$^2$SCAN-D4 (0.44 eV error), both commonly used in solid-state calculations. Because molecular oxygen is a highly spin-polarized system, further improvements could potentially be realized by modifying the correlation model to learn additional spin-dependence, rather than inheriting the spin-dependence of PBE correlation.

Transition metal oxides provide another set of systems which feature complex magnetic ordering and correlation effects, and for which semilocal DFT often performs poorly. With CIDER26SS, rutile \ce{TiO2} relaxes to lattice constants within 0.35\% of the experimental diffraction reference, comparable to the hybrid HSE06~\cite{heydHybridFunctionalsBased2003,heydErratumHybridFunctionals2006,krukauInfluenceExchangeScreening2006} and much closer than PBE or LDA~\cite{burdett1987rutile,vasquez2018rutile} (Figure~\ref{fig:transfer}(c)). For the correlated monoxides MnO and NiO, the Kohn--Sham gaps, a proxy for the fundamental gap and here corrected for the zero-point renormalization of Ref.~\cite{abdallah2024qsgw} (spin--orbit corrections are negligible for these oxides), open well beyond PBE (Figure~\ref{fig:transfer}(d)), and the local moments move close to experiment (Figure~\ref{fig:transfer}(e)), consistent with improved $d$-electron localization, although the gaps remain below the experimental and vertex-corrected $GW$ values~\cite{vanelp1991mno,kurmaev2008oxides,sawatzky1984nio,abdallah2024qsgw,cunningham2023qsgwhat,cheetham1983mnni,roth1958monoxides,alperin1962nio,fender1968moments}. None of these systems, nor any band-gap target, is represented in the training data. We also looked at the formation energies of these oxides. The raw CIDER26SS electronic formation energies, $-9.226$~eV per formula unit for rutile \ce{TiO2} and $-1.537$~eV for NiO, improve on the PBE values of $-8.963$ and $-1.111$~eV but remain well short of the experimental enthalpies of $-9.734$ and $-2.484$~eV~\cite{cox1989codata,boyle1954nio}. The demonstrated accuracy of r$^2$SCAN for solid formation energies~\cite{kingsbury2022r2scan} shows that a semilocal meta-GGA can describe this property substantially better than PBE. A plausible cause of the CIDER26SS error is that no insulator or semiconductor data is included in the training data. Adding such data could lead to significant improvements in future iterations of CIDER for oxides and other insulating materials.

\section*{Conclusion}
We have presented an end-to-end strategy for machine learning transferable, nonlocal, physically informed exchange-correlation functionals across heterogeneous datasets, and we have introduced the CIDER26SS functional to provide a balanced and accurate description of molecules, solids, and surfaces. Our models fit molecular and solid-state reference data within the same nonlocal representation, and the resulting functionals can be evaluated efficiently and self-consistently for both isolated and periodic chemical systems~\cite{bystrom2024nonlocal}. The resolution of the long-standing CO/Pt puzzle with CIDER26SS, even when Pt is excluded from training, shows that CIDER can address challenging surface-science problems without abandoning universality and transferability to other systems. In addition to verifying the accuracy of our approach across a broad range of datasets and applications, we have identified key directions for follow-up work on incorporating insulating solids and improving the long-range dispersion description, which could yield even more accurate and broadly applicable models.

While deep-learned representations are advancing XC modeling~\cite{dick2020machine,chen2021deepks,luise2025accurate}, our results show that hand-designed features and Gaussian process regression retain distinct advantages. The nonlocal descriptors are interpretable, admit exact constraint control, and depend on the electron density with no explicit atomic geometry information, preserving the universal form of the exact functional~\cite{hohenberg1964inhomogeneous} and enabling direct evaluation in both molecular and periodic DFT implementations. Gaussian process regression provides uncertainty and regularization tools that aid in data-efficiency, with CIDER26SS using only 2670 training data compared to roughly 400,000 for Skala 1.1~\cite{luise2025accurate}. This data-efficiency will make our approach more accessible to the academic community by not requiring massive compute resources. Its most expensive training steps, control-point selection and covariance-matrix construction, are readily parallelizable, offering further efficiency gains as more high-fidelity reference data become available. Together, these properties make the framework well suited to fitting heterogeneous reference data and developing transferable models where training data is limited. Several ingredients also offer room for refinement. Tuning the nonlocal feature length scales, carried over unchanged from Ref.~\cite{bystrom2024nonlocal} in this work, could improve the accuracy and richness of the nonlocal description, while density-matrix-based descriptors introduced for band-gap training~\cite{bystrom2024training} offer a natural extension of the feature set. CIDER's modular, open-source infrastructure allows these refinements to advance rapidly through community-driven efforts.

One of the questions that motivated this work was whether a single functional can describe molecules, metals, and metal--molecule interactions on an equal footing, a longstanding problem at the core of heterogeneous catalysis. We have shown that CIDER26SS marks a major milestone in the resolution of this problem, which opens the way to quantitatively reliable computational screening and design of catalysts and functional materials.

\clearpage

\section*{Methods}

\subsection*{Functional form and Gaussian process model}

Previous CIDER models applied machine learning only to the exchange
energy~\cite{bystrom2022cider,bystrom2024nonlocal,bystrom2024training}. Here the full XC energy density
is written as a machine-learned correction to an additive semilocal baseline,
\begin{equation}
\begin{aligned}
e_\text{xc}^\text{CIDER}(\mathbf{r})
&= \sum_\sigma e_\text{x}^\text{PBE}[n_\sigma](\mathbf{r}) + e_\text{c}^\text{PBE}[n_\uparrow,n_\downarrow](\mathbf{r}) \\
&\quad
 + \sum_\sigma e_\text{x}^\text{LDA}[n_\sigma](\mathbf{r})\,
   f_\text{x}^\text{CIDER}(\mathbf{X}^\text{x}[n_\sigma](\mathbf{r})) \\
 &\quad + e_\text{c}^\text{PBE}[n_\uparrow,n_\downarrow](\mathbf{r})\,f_\text{c}^\text{CIDER}
   \left(
   \tfrac{1}{2}\textstyle\sum_\sigma \mathbf{X}^\text{c}[n_\sigma](\mathbf{r})
   \right),
\end{aligned}
\end{equation}
where \(e_\text{x}^\text{LDA}\) and \(e_\text{x/c}^\text{PBE}\) are the LDA~\cite{Dirac1930NoteAtom}
and PBE~\cite{perdew1996generalized} energy densities. The enhancement factors \(f_\text{x}^\text{CIDER}\) and
\(f_\text{c}^\text{CIDER}\) are learned jointly by Gaussian process (GP) regression. The exchange
part is spin-separable; the correlation features are spin-averaged, with spin dependence entering
through the PBE correlation multiplier. The exchange feature vector \(\mathbf{X}^\text{x}\)
contains the reduced gradient $p=|\nabla n|^2/n^{8/3}$, the reduced kinetic energy descriptor $t=(\tau-\tau_0)/(\tau+\tau_0)$ (where $\tau_0=\frac{3}{10}(3\pi^2)^{2/3}n^{5/3}$ is the uniform electron gas kinetic energy density), and three nonlocal density features (NLDFs)
\begin{align}
    G_i(\mathbf{r}_1) &= \left(\frac{B_i+B_0}{2}\right)^{3/2}
    \int \mathrm{d}^3\mathbf{r}_2 \, \exp\left\{-\left[a(\mathbf{r}_2)+b_i(\mathbf{r}_1)\right]r_{12}^2\right\} n(\mathbf{r}_2),
\end{align}
with \(a[n]\) and \(b_i[n]\) semilocal functionals parameterized as in the nonlocal meta-GGA models of Ref.~\cite{bystrom2024nonlocal}. The kinetic descriptor $t$ is used in place of the SCAN
\(\alpha\)~\cite{sun2015strongly} for better numerical behavior, as in many
empirical functionals such as \(\omega\)B97M-V~\cite{mardirossian2016wB97MV}. Evaluating the NLDFs scales quasi-linearly to quadratically with
system size, so the features are computationally inexpensive relative to the \(O(N^3)\) Kohn--Sham
solver for large systems. The correlation vector \(\mathbf{X}^\text{c}\) adds the density \(n\) itself, because all
other features are invariant under uniform coordinate scaling~\cite{Levy1985,bystrom2024nonlocal}:
scale-invariant features enforce the exchange scaling constraint by construction, whereas
correlation may depend on the density scale. Also note that for the correlation feature vector, $n$, $|\nabla n|^2$, and $\tau$ are averaged over spin before using the spin-averaged quantities to construct $p$ and $t$. All features are regularized into finite intervals
before entering the GP kernel, and these regularized features are denoted by the lower-case $\mathbf{x}^\text{x}$ and $\mathbf{x}^\text{c}$.

The GP framework follows Refs.~\cite{bystrom2024nonlocal,bystrom2024training} with a squared
exponential covariance kernel per component
(\(\text{z}\in\{\text{x},\text{c}\}\)),
\begin{equation}
    k_\text{z}(\mathbf{x}, \mathbf{x}') = \Sigma_\text{z} \exp\left[-\tfrac{1}{2} \left|\mathbf{b}_\text{z} \odot (\mathbf{x} -\mathbf{x}')\right|^2 \right],
\end{equation}
whose hyperparameters are selected by the stability screening described below. Sets of control
points \(\tilde{\mathbf{x}}_a^\text{z}\), separate for exchange and correlation, are selected
with the sparse-GP procedure of Ref.~\cite{bystrom2024nonlocal}. In this procedure, grid points with spin
density above \(10^{-6}\)~bohr\(^{-3}\) are subsampled into a candidate pool, and the pivots retained by
a pivoted Cholesky factorization of the normalized kernel Gram matrix (tolerance \(10^{-5}\))
form the set of control points, giving 582--619 exchange and 1052--1097 correlation control points
depending on the functional. In the combined molecular--periodic self-consistent trainings the control-point sets are
carried and unioned across iterations (see below). The predictive mean of
each enhancement factor is a weighted sum of kernels evaluated against the control points,
\begin{equation}
    f_\text{z}^\text{CIDER}(\mathbf{X}_*^\text{z}) = \sum_a k_\text{z}(\mathbf{x}_*^\text{z}, \tilde{\mathbf{x}}_a^\text{z}) \, \alpha_a^\text{z},
    \label{eq:cider_pred_mean}
\end{equation}
with the trained weights \(\boldsymbol{\alpha}^\text{z}\) obtained from solving the GP regression as described in the next subsection. Each training label is the difference between the high-accuracy
reference reaction energy and the energy already supplied by the additive baseline and the
explicitly evaluated Kohn--Sham terms.

The hand-crafted features make key exact constraints straightforward to enforce. The exchange
model is spin-separable and obeys the uniform scaling rule, and all features can be evaluated
analytically for the uniform electron gas (UEG). We include UEG exchange and correlation energies
for a wide range of unpolarized densities (\(10^{-6}\) to \(10^{6}\)~bohr\(^{-3}\)) in the
training data for all three functionals introduced in this work, with noise hyperparameters small
enough that these points act as effectively exact constraints; the exchange values are the exact
UEG exchange energies~\cite{Dirac1930NoteAtom} and the correlation values are from the modified
PW92 parameterization~\cite{perdewAccurateSimpleAnalytic1992} as implemented in
libxc~\cite{Lehtola2018RecentTheory}.

\subsection*{Formulas for trained weight vectors}

Using the control point set from the previous section, the covariance of component ``z'' between two chemical systems $m$ and $n$ is
\begin{align}
    K_{mn}^\text{z} &= \tilde{\mathbf{k}}_m^\text{z} \left(\tilde{\mathbf{K}}^\text{z}\right)^{-1} \tilde{\mathbf{k}}^\text{z}_n
\end{align}
with
\begin{align}
    \tilde{K}_{ab}^\text{z} &= k_\text{z}(\tilde{\mathbf{x}}_a^\text{z}, \tilde{\mathbf{x}}_b^\text{z}) \\
    \left(\tilde{\mathbf{k}}_m^\text{z}\right)_a &= \sum_{g \in m} w_g e_g^\text{z,base} k(\tilde{\mathbf{x}}_g^\text{z}, \tilde{\mathbf{x}}_a^\text{z})
\end{align}
where $e_g^\text{x,base}$ is the LDA exchange energy density at grid point $g$, and $e_g^\text{c,base}$ is the PBE correlation energy density at grid point $g$. Note that for the exchange kernel, both the baseline energy and feature vectors are separated by spin. For the correlation kernel, the baseline energy is the full spin-polarized PBE correlation, and the feature vector is spin-averaged, as discussed above. From these components, we can construct the GP predictive mean for the enhancement factors $f_\text{x}^\text{CIDER}$ and $f_\text{c}^\text{CIDER}$:
\begin{align}
    f_\text{z}^\text{CIDER}(\mathbf{X}_*^\text{z}) &= \sum_a k_\text{z}(\mathbf{x}_*^\text{z}, \tilde{\mathbf{x}}_a^\text{z}) \alpha_a^\text{z} \\
    \boldsymbol{\alpha}^\text{z} &= \left(\tilde{\mathbf{K}}^\text{z}\right)^{-1} \tilde{\boldsymbol{\alpha}}^\text{z} \\
    \tilde{\boldsymbol{\alpha}}^\text{z} &= \sum_m \tilde{\mathbf{k}}_m^\text{z} \left\{\left[\mathbf{K}^\text{tot} + \boldsymbol{\Sigma}_\text{noise}\right]^{-1}\mathbf{y}\right\}_m \label{eq:alpha_tilde}
\end{align}
where $\mathbf{X}_*$ is the non-regularized feature vector, $\mathbf{x}_*$ is the corresponding regularized feature vector, and $\mathbf{K}^\text{tot} = \mathbf{K}^\text{x} + \mathbf{K}^\text{c}$. Because our training labels $\mathbf{y}$ correspond to the full XC energy, the covariance kernels $\mathbf{K}_\text{x}$ and $\mathbf{K}_\text{c}$ must be summed to obtain the total covariance kernel $\mathbf{K}^\text{tot}$ in Eq.~\ref{eq:alpha_tilde}.
Note that Eq. 43 of Ref.~\cite{bystrom2024nonlocal} and Eq. 29 of Ref.~\cite{bystrom2024training} both contain a mathematical typo: they are missing the multiplication by $\tilde{\mathbf{K}}^{-1}$.

\subsection*{EXX+RPA surface-science reference data}

To address the lack of high-fidelity solid-state training data for surface science, we constructed
a transition-metal reference set at the EXX+RPA level of theory, which gives a balanced and
accurate description of surface and adsorption energetics where semilocal functionals are
unreliable~\cite{schimka2010accurate,schmidt2018benchmark}. The set spans thirteen \(d\)-block
metals (Sc, Ti, V, Cr, Mn, Cu, Ru, Rh, Pd, Ag, Ir, Pt, and Au) organized as energy-difference
reactions in six classes: 576 primary reactions (415 adsorption, 115 surface, 26 phase-stability,
13 cohesive, and 7 atomization energies) together with 231 equation-of-state (EOS) reactions,
built from roughly 600 unique structures.

All reference energies were computed with VASP~6.4~\cite{kresse1996efficient,kresse1996efficiency}
using the PAW method~\cite{blochl1994projector,kresse1999ultrasoft} and GW-specific hard PAW data
sets~\cite{klimes2014predictive}. Geometries were relaxed with PBE, and the reference energies correspond to the one-shot EXX+RPA@PBE energy: full
unscreened exact exchange plus ACFDT-RPA correlation evaluated with the cubic-scaling
imaginary-time algorithm~\cite{kaltak2014cubic,kaltak2014low}, extrapolated
to the complete-basis limit~\cite{harl2008cohesive}, with the finite-temperature ACFDT
formalism~\cite{kaltak2020minimax} enabling RPA correlation for metals. Surfaces were modeled with four-layer slabs at
full (one monolayer) and low (1/4 monolayer) coverage; adsorption energies cover eight adsorbates
(\ce{CO}, \ce{NO}, \ce{N2}, \ce{OH}, \ce{CH}, \ce{O}, \ce{H}, and \ce{N}) on the fcc \{111\}
surface, with reactive fragments referenced through balanced formation reactions from stable
closed-shell molecules. The EOS reactions record EXX+RPA energy differences for isotropic strains
of \(\pm5\%\) about the PBE equilibrium cell; the phase-stability reactions order the bcc and hcp phases
of each metal against fcc. Structures were built with ASE~\cite{hjorthlarsen2017atomic}. Example input files for
the pipeline and the calculation settings will be provided on the Materials Cloud upon
publication.

\subsection*{Training, validation and test data}

The molecular training backbone is GMTKN55~\cite{goerigk2017look}, comprising 1505 relative
energies in 55 subdatasets. The GSCDB137 database~\cite{liang2025gscdb137} (8377 energy
differences in 137 subdatasets) is reserved for validation and testing, with two exceptions added
to training because GMTKN55 has no comparable data: the nine transition-metal subdatasets and the
two absolute-atomic-energy subdatasets (358 reactions in total).

Holding out GSCDB137 is meaningful only if no held-out reaction is contained in, or energetically
determined by, the training data. Overlap was removed at the level of geometries and
stoichiometries rather than dataset names. A direct-coincidence step (species equivalent when
charge, spin multiplicity, composition, and sorted interatomic-distance spectra agree within
\(0.01\)~\AA) eliminated 1427 reactions. A linear-span step then removed every reaction whose
stoichiometric vector lies in the exact rational linear span of the GMTKN55 reaction vectors
(using a tighter \(10^{-5}\)~\AA{} species tolerance), eliminating a further 844; this detects
implicit overlap, such as a reaction energy fixed by training atomization energies (an example: if the
atomization energies of \ce{H2}, \ce{O2}, and \ce{H2O} all appear in training,
the energy of $2\,\ce{H2}+\ce{O2}\rightarrow 2\,\ce{H2O}$ is energetically determined by them). In total 2271
of 8377 reactions were removed, leaving 6106 reactions with no direct or linear-combination
overlap with GMTKN55. The remaining reactions, less the 358 assigned to training, were divided into validation and test
splits by a deterministic multi-restart greedy optimization that keeps chemically linked reactions
in indivisible units and balances reaction counts, PBE0 MAE~\cite{perdew1996rationale,adamo1999toward}, reference
energy scales, and category fractions. A final manual audit of reactions whose overlap decision
is sensitive to the geometry tolerance removed a further 42 validation and 91 test reactions.
The result is 2571 validation and 3044 test reactions with closely matched size, difficulty, and
composition. As an additional audit of the partitioning, we assessed the uniqueness of molecules
in the GSCDB137 validation and test sets relative to the training set (GMTKN55 plus the
atomic-energy and transition-metal reactions). Excluding electric-field-containing reactions,
which GMTKN55 lacks entirely, 99.5\% of the validation reactions and 98.9\% of the test
reactions contain at least one state-specific system absent from training, and 98.7\% of the
test reactions contain at least one molecule absent from the validation set. The reaction and
system identities of these splits
will be released on the Materials Cloud upon publication.

\subsection*{Reaction weighting in the Gaussian process regression}

Each reaction enters the GP fit with a noise hyperparameter that sets its effective weight. For
the molecular data we set the noise so that the training loss mirrors the WTMAD-4 metric of
Bryenton and Johnson~\cite{bryenton2026wtmad4}, whose subdataset weights \(w_D\) are inversely
proportional to the typical error of representative functionals on subdataset \(D\). The noise for
reaction \(i\) in subdataset \(D\) is
\begin{equation}
    \sigma_i = \bar{\sigma}
    \left( \frac{\left\langle w/N \right\rangle}{w_D/N_D} \right)^{1/2},
    \label{eq:wtmad4_noise}
\end{equation}
where \(N_D\) is the subdataset size and \(\langle w/N\rangle\) the mean per-reaction weight, so
the GP minimizes a WTMAD-4-weighted squared error. Published weights are used for GMTKN55;
for the transition-metal and atomic-energy subsets we generate weights with the same construction,
\(w_D = (100/N_\text{bench})(3.5~\text{kcal~mol}^{-1}/\overline{\mathrm{MAD}}_D)\), where
\(N_\text{bench}\) is the number of subdatasets in the benchmark and
\(\overline{\mathrm{MAD}}_D\) averages over the sixteen hybrid and double-hybrid functionals whose
subset errors are provided with GSCDB137~\cite{liang2025gscdb137}. The overall scale
\(\bar{\sigma}\) was tightened as far as numerical stability allows, and the kernel
parameters \(\mathbf{b}_\text{z}\) and \(\Sigma_\text{z}\) were perturbed around their
values from Ref.~\cite{bystrom2024nonlocal}. Stability was assessed on a fixed probe set of 59
difficult-to-converge systems (stretched bonds, open-shell radicals and ions, transition-metal atoms and
monoxides), each attempted through a ladder of increasingly conservative SCF protocols (larger
DIIS subspaces, ADIIS, EDIIS, level shifting, and
damping)~\cite{pulay1984improved,pulay1993c2diis,hu2010adiis,kudin2002scfediis}. We adopt the
most flexible GP fit, that is the smallest \(\bar{\sigma}\) and largest
\(\mathbf{b}_\text{z}\) and \(\Sigma_\text{z}\), for which all 59 systems converge.

The construction of Eq.~\ref{eq:wtmad4_noise} does not carry over to the EXX+RPA set, because
its weights require a panel of established functionals and no comparable panel exists for
periodic transition-metal surface chemistry. We adopt the natural single-baseline analogue, in
which the periodic reactions are weighted by class (adsorption energies, surface energies, etc.) and the role of the panel error is played by
the residual of the PBE baseline itself. Each periodic reaction carries a residual label (the
difference between its EXX+RPA value and the PBE baseline reaction energy), so the six classes are
assigned the root-mean-square residual \(d_C\) of their members as a characteristic scale, and every
reaction in class \(C\) carries the noise
\begin{equation}
    \sigma_C = \max\left(\alpha\, d_C,\ \sigma_\text{min}\right).
    \label{eq:rpa_noise}
\end{equation}
Therefore, each class must be reproduced to the same fraction \(\alpha\) of the semilocal baseline's error
on that class. As in the molecular construction, classes with larger characteristic errors receive
looser tolerances, which automatically places the loosest tolerances on the cohesive and
atomization energies, the classes where EXX+RPA is least
reliable~\cite{furche2001molecular,harl2010assessing,ren2012random,schimka2013lattice,olsen2013beyond}
and also those with the largest PBE residuals. The floor \(\sigma_\text{min}=0.02\)~eV
ensures that the phase-stability energies, whose PBE residuals are below 0.1~eV, are not weighted
more tightly than the most strongly weighted molecular data, which would destabilize the trained
models in self-consistent calculations. The EOS reactions are assigned a fixed noise of 8.6~meV,
substantially tighter than any other periodic class, to pin the energy--volume curve near its
minimum, which determines the predicted equilibrium lattice constants and bulk moduli. In the combined trainings, the molecular reactions keep the noise of Eq.~\ref{eq:wtmad4_noise},
so \(\alpha\) is the single hyperparameter controlling the balance between the molecular and
periodic parts of the objective. Decreasing \(\alpha\) tightens the surface-science fit at the
cost of molecular accuracy and, at the smallest values considered, of numerical stability;
following the same philosophy as for \(\bar{\sigma}\), the production value \(\alpha=0.23\)
was selected from self-consistent-field numerical stability on a panel of molecular and
solid-state systems. The resulting mean noise of the periodic non-EOS reactions, 0.131~eV, is
about twice the 0.067~eV mean of the GMTKN55 assignments, with the tightest weights on the
surface and phase-stability classes, where the reference is most reliable and the required
precision is highest.

\subsection*{Performance scoring for GMTKN55 and GSCDB137}
Training performance on GMTKN55 is reported with its standard weighted error metric WTMAD-2~\cite{goerigk2017look},
\begin{equation}
    \text{WTMAD-2} = \frac{1}{\sum_{i}^{55} N_i} \sum_{i}^{55} N_i \,
    \frac{56.84~\text{kcal~mol}^{-1}}{|\overline{\Delta E}|_i} \, \text{MAD}_i,
    \label{eq:wtmad2}
\end{equation}
where \(\text{MAD}_i\) is the mean absolute deviation on subset \(i\), \(N_i\) is its number of
reactions, \(|\overline{\Delta E}|_i\) is its mean absolute reference reaction energy, and
56.84~kcal~mol\(^{-1}\) is the average of \(|\overline{\Delta E}|_i\) over the 55 subsets. This ensures that
subsets with smaller energy scales receive proportionally larger weights. Validation and test
performance is reported with the normalized error ratio (NER) of
GSCDB137~\cite{liang2025gscdb137},
\begin{equation}
    \mathrm{NER}_D = \frac{E_D}{s_D^\mathrm{hyb}},
    \qquad
    \mathrm{NER} = \frac{1}{N_\text{bench}}\sum_D \mathrm{NER}_D ,
    \label{eq:ner}
\end{equation}
where \(E_D\) is the benchmark's per-subdataset error metric, the mean absolute error in
kcal~mol\(^{-1}\) for most subdatasets, with relative or unit-converted variants for the dipole,
polarizability, electric-field, and frequency sets~\cite{liang2025gscdb137}. The
\(s_D^\mathrm{hyb}\) denominator is the standard error published with the benchmark, the average
of the second-through-fourth lowest errors from its panel of fourteen dispersion-corrected hybrid
functionals computed with the same metric (double hybrids are not included). NER is dimensionless, with lower values being better, and a value
near one corresponds to a strong hybrid baseline. Unless noted otherwise, NER denotes this
MAE-based quantity. We also report an RMSE-based variant, NER-RMSE, obtained by replacing
\(E_D\) with the root-mean-square error of each subdataset and \(s_D^\mathrm{hyb}\) with the
corresponding RMSE-based standard error of Ref.~\cite{liang2025gscdb137}. Since both variants
are normalized to the same hybrid panel, the ratio of NER-RMSE to NER-MAE measures the prominence of large individual errors relative to the model's MAE, and values above unity indicate a heavier error tail than that characteristic of the best hybrid functionals in the panel. MoM-MAE, the unweighted mean of per-subdataset MAEs,
keeps comparisons in kcal~mol\(^{-1}\), and MoM-RMSE is defined in the same way from the
per-subdataset RMSEs. All GSCDB137 results quoted from
Ref.~\cite{liang2025gscdb137} use that work's dispersion treatment, in which functionals without
built-in dispersion carry a D4 correction, or a D3
variant~\cite{grimme2010d3,grimme2011d3bj} where no D4 parametrization exists. Throughout the
GSCDB137 comparisons, functional names written without a correction label still denote these
dispersion-corrected variants. Functionals with the VV10 dispersion correction~\cite{vydrov2010vv10} are marked
with -V.

\subsection*{Dispersion corrections for CIDER}

Much of the noncovalent binding in molecular benchmarks arises at near-equilibrium geometries and
can be learned by a flexible nonlocal functional directly; only the strict \(-C_6/R^6\) asymptote
lies outside the reach of the short-ranged CIDER convolution kernels. To assess the contribution
of explicit long-range dispersion, the D4 correction~\cite{caldeweyher2019d4} (with the PBE damping
parameters) is computed for every training structure and subtracted
from the training label, so the GP learns only the remainder; the same correction is added back in
every subsequent calculation. The molecular functionals are trained both with (\CIDERCD{}) and
without (\CIDERC{}) this correction. The additive correction is currently implemented in the
molecular code only; the periodic \CIDERSS{} calculations contain no dispersion term.

\subsection*{Self-consistent training}

For a full XC functional trained on reaction energies, the ground-state density is part of the
prediction: in a self-consistent calculation the orbitals relax away from the densities used in
training, and density-driven errors are known to be significant for the anions, radicals,
stretched bonds, and transition states that molecular benchmarks
emphasize~\cite{kim2013understanding,sim2022improving}. Prior learned-functional work has
addressed self-consistency through derivative-free optimization with an SCF solve in the
objective~\cite{nagai2020completing,nagai2022machine}, iterative refitting on model-generated
densities~\cite{dick2020machine,chen2021deepks,luise2025accurate}, and end-to-end differentiation
through the Kohn--Sham equations~\cite{li2021kohn,kasim2021learning,dick2021highly,kalita2022well}.
We use a fixed-point iteration on the training densities, similar to the approach of Dick and
Fernandez-Serra~\cite{dick2020machine}, implemented in the CiderPress
code~\cite{ciderpress}. Iteration-0 seed models are fit on
PBE densities for GMTKN55 systems and PBE0 densities for GSCDB137
systems~\cite{perdew1996generalized,perdew1996rationale,adamo1999toward}. At each subsequent iteration \(k\) the
descriptors are regenerated on the current densities \(\{n^{(k)}\}\) and each label is reassembled
as
\begin{equation}
    y_i^{(k)} = \Delta E_i^{\text{ref}}
    - \Delta E_i^{\text{base}}\big[\{n^{(k)}\}\big],
    \label{eq:sc_label}
\end{equation}
where \(\Delta E_i^{\text{ref}}\) never changes and \(\Delta E_i^{\text{base}}\) collects the
additive baseline (including any dispersion correction) and explicitly evaluated Kohn--Sham
contributions. The GP is refit with all noise assignments, the UEG constraint, and kernel
hyperparameters held fixed, self-consistent calculations with the refit model define
\(\{n^{(k+1)}\}\), and the iteration is judged converged when the fixed-density training error and
the full self-consistent error agree per subdataset and in aggregate. For the combined
molecular--periodic trainings, successive refits carry a soft Gaussian prior centered on the
previous iteration's prediction (width 0.2 on the scale of the learned correction), imposed at the
control points and at probe points across the metallic region of feature space, since
unconstrained self-consistent training was found to lead to numerical instabilities in the
trained model. Control-point sets are carried between iterations, each refit inheriting the union of
its predecessor's points and those selected from the regenerated data. This procedure does not
guarantee accurate densities, merely self-consistent ones; the dipole-moment benchmarks
(Extended Data Table~2) provide the corresponding accuracy check.

Molecular calculations used PySCF~\cite{sun2020pyscf,sunPythonSimulationsChemistry2026}, while periodic calculations used GPAW~\cite{enkovaara2010gpaw,mortensenGPAWOpenPython2024a}.
Nonconverging systems were retried through the conservative SCF ladder described under Reaction weighting. For particularly difficult-to-converge systems, we used an in-house
implementation of geometric direct minimization (GDM)~\cite{vanvoorhis2002gdm,dunietz2002gdm} within
PySCF.

\subsection*{Computational details}

For the EXX+RPA reference data, the hard PAW data sets were chosen for an accurate
description of the unoccupied states entering the response function (for example
\texttt{Pt\_sv\_GW}, with \texttt{C/N/O/H\_GW} for the adsorbates). The RPA correlation
energy was evaluated on a 16-point imaginary-frequency grid at a finite electronic
temperature of 0.01~eV (about 120~K), with the basis-limit
extrapolation taken with respect to the response-function cutoff (two thirds of the orbital
cutoff by default) and with a 500~eV plane-wave
cutoff for the metals and slabs (650~eV for Cu-containing systems). Bulk metals were relaxed
with full cell and ionic relaxation on $\Gamma$-centered $16\times16\times16$ $k$-point
meshes at a 700~eV cutoff, which set the PBE lattice constants used to generate every periodic
structure. The slabs had 8~\AA{} of vacuum on each side and were relaxed with
PBE at fixed cell with the bottom layers held at bulk positions, with $\Gamma$-centered
$16\times16\times1$ ($8\times8\times1$) $k$-point sampling at full (quarter) coverage. All
PBE relaxations used Methfessel--Paxton smearing of width 0.1~eV, with forces converged to
0.02~eV~\AA$^{-1}$. Gas-phase references
(isolated atoms and the adsorbate-forming closed-shell molecules) were computed spin-polarized
at the $\Gamma$ point, the atoms in slightly distorted 15~\AA{} cells and the molecules in
10--12~\AA{} boxes, with cutoffs and smearing matched to the periodic calculations.

For the self-consistent training and evaluation pipelines, the molecular calculations used the
def2-QZVPPD basis~\cite{weigend2005balanced,rappoport2010property} with matching effective core
potentials and level-1 integration grids for the GMTKN55 systems (level-3 grids for the
electron-affinity and ionization-potential reactions),
and the per-subset basis sets and integration grids prescribed by the GSCDB137
database for its systems~\cite{liang2025gscdb137} (level-3 grids, or finer where prescribed),
removing nearly linearly dependent basis functions. Since CIDER evaluates no exact exchange, density fitting enters only through the
Coulomb problem, treated with the def2-universal-jfit auxiliary basis. Density fitting was disabled for systems
with fewer than about 800 atomic orbitals, for finite-field response systems, and for the
electron-affinity and ionization-potential subsets, whose charge-state energy differences are
sensitive to fitting errors that do not cancel, always keeping the density-fitting choice
consistent among the reactants of a reaction. Open-shell and charged molecular species were treated
spin-unrestricted with the database charges and multiplicities.
The periodic calculations used plane waves at a
700~eV cutoff with standard PAW data sets. Bulk and equation-of-state calculations used
$16\times16\times16$ Monkhorst--Pack meshes with 0.01~eV Fermi--Dirac smearing, and the
four-layer training slabs used $2\times2$ surface cells with $8\times8\times1$ meshes at 1/4
monolayer and $1\times1$ cells with $16\times16\times1$ meshes at full coverage, with the
adsorbates on one side of the slab, 0.05~eV
smearing and about 16~\AA{} of vacuum between periodic slab images. Extended systems were
treated non-spin-polarized,
consistent with the EXX+RPA reference protocol, while isolated atoms and molecules were
computed spin-polarized at a single shifted $k$ point in slightly distorted 10--15~\AA{}
boxes, with Hund's-rule initial moments for the atoms.

For the transferability applications, CO adsorption used one CO on a $2\times2$ four-layer Pt(111) slab with
the bottom two layers fixed and the top two layers and CO relaxed at fixed lattice constant
to forces below 0.04~eV~\AA$^{-1}$
(700~eV, $8\times8\times1$, 0.02~eV smearing), otherwise mirroring the training-slab settings.
Pt lattice constants were obtained by hydrostatic relaxation of the primitive fcc cell (700~eV,
$8\times8\times8$, 0.02~eV smearing), the surface energy is referenced to bulk Pt per surface
atom on the two slab faces, and the comparison functionals were evaluated with the identical
slab protocol at their own, the CIDER26SS, and the experimental lattice constants. The
graphene/Ni(111) interaction curves are rigid top-fcc scans of a two-atom graphene sheet on a
five-layer $1\times1$ Ni(111) cell at the CIDER26SS-relaxed Ni lattice (3.530~\AA{}),
spin-polarized with about 28~\AA{} of vacuum, at a 1000~eV cutoff with
$28\times28\times1$ meshes and 0.02~eV smearing, otherwise following the calculation methodology of
Ref.~\cite{trepte2022vcml}. The graphite bulk-binding construction follows Spanu \textit{et
al.}~\cite{spanu2009graphite}, with AB stacking, the experimental in-plane bonds fixed at
1.420~\AA{}, and
$20\times20\times8$ $k$-point sampling (700~eV, 0.02~eV smearing), with the isolated-graphene
reference in a cell with matching in-plane parameters and a 30~\AA{} out-of-plane length. The \ce{O2}/graphene scans
follow the methodology of
Shin \textit{et al.}~\cite{shin2019o2graphene}, with rigid height scans of horizontal and
vertical \ce{O2} above a $3\times3$ graphene cell with fixed bond lengths and a fixed total
moment of 2~$\mu_\mathrm{B}$, sampled $6\times6\times1$ at 700~eV and with the reported minimum
recomputed at 1400~eV. The MnO and NiO results use fully relaxed
antiferromagnetic (AFM-II) rhombohedral cells at fixed zero total moment (700~eV,
$10\times10\times10$ relaxations to forces below 0.005~eV~\AA$^{-1}$ and stresses below
0.001~eV~\AA$^{-3}$, and
$12\times12\times12$ final electronic structure calculations with improved-tetrahedron
integration), with PBE and r$^2$SCAN evaluated on the same CIDER26SS-relaxed structures. All
structures entering the oxide formation energies (rutile \ce{TiO2}, AFM-II NiO, hcp Ti, fcc
Ni, and triplet \ce{O2} in a 20~\AA{} box) were relaxed with CIDER26SS at 700~eV to the same
force and stress targets, with every reaction term recomputed at 800~eV at fixed geometry and
no Hubbard $U$, oxygen correction, dispersion, or zero-point term applied. The iron--sulfur protonation energies were computed all-electron with
spin-free exact two-component scalar relativity on the fixed
broken-symmetry geometries of Ref.~\cite{zhai2023fes} with the cc-pVQZ-DK
basis~\cite{dejong2001parallel,balabanov2005dk}, as antiferromagnetically coupled open-shell
singlets with exact two-electron integrals rather than density fitting. The triplet \ce{O2} equilibrium properties come
from spin-unrestricted fixed-distance scans (1.170--1.240~\AA{} in 0.005~\AA{} steps) in the
aug-cc-pV7Z basis with matching triplet atomic references, with $R_e$ and $\omega_e$ obtained
from local quadratic fits.

\subsection*{Software}

The functionals are distributed with the open-source CiderPress
package~\cite{ciderpress,ciderpress_docs} through its PySCF and GPAW interfaces. For this work, the
GPAW nonlocal-feature generator was rewritten with an optimized C-level implementation based on
libvdwxc~\cite{Larsen_2017} and the convolution strategy of Rom{\'a}n-P{\'e}rez and
Soler~\cite{romanperez2009efficient}. In addition, the PAW feature-projector sets were revised element-by-element for
numerical stability~\cite{bystrom2024nonlocal}, and the sparse GP of Eq.~\ref{eq:cider_pred_mean}
was evaluated directly during SCF without a spline or neural-network mapping step. Baseline correlation
functionals are generated through libxc~\cite{Lehtola2018RecentTheory}, which is also used to
evaluate other functionals in this work. Parts of the code development and scripting were assisted by Claude Code and Codex, with all code reviewed and validated by the authors.

\section*{Data availability}
The machine-readable data underlying all figures and tables are provided with this manuscript.
The GMTKN55 and GSCDB137 benchmark data is available from their original
publications~\cite{goerigk2017look,liang2025gscdb137}. We plan to make the EXX+RPA surface-science reference set publicly available in a
forthcoming publication.

\section*{Code availability}
The CIDER functionals introduced in this work are distributed with the open-source CiderPress
package (\url{https://cider-dft.github.io/CiderPress/})~\cite{ciderpress,ciderpress_docs} and can
be selected by their packaged model names through its PySCF and GPAW interfaces.

\section*{Acknowledgements}
The authors thank Diptarka Hait for kindly providing a PySCF implementation and documentation of the GDM algorithm, which informed the validation and refinement of the GDM solver used in this work for difficult-to-converge systems. The Flatiron Institute is a division of the Simons Foundation. 

This work was supported by the National Defense Science and Engineering  Graduate (NDSEG) Fellowship under Contract No.~FA9550-21-F-0003; Tata Steel Ltd; the Camille and Henry Dreyfus Foundation under Award No.~ML-22-075; the U.S. Department of the Navy, Office of Naval Research under Award No.~N00014-20-1-2418; the U.S. Department of Energy, Office of Science, Office of Basic Energy Sciences under Award No.~DE-SC0022199; and the National Science Foundation Institute for Data Driven Dynamical Design (ID4) under Award No.~OAC-2118201.

The computations in this paper were run in part on the FASRC Cannon cluster supported by the FAS Division of Science Research Computing Group at Harvard University. This research used resources of the National Energy Research Scientific Computing Center (NERSC), a U.S.\ Department of Energy Office of Science User Facility, using NERSC award BES-ERCAP0024206. An award of computer time was provided by the Innovative and Novel Computational Impact on Theory and Experiment (INCITE) program. This research also used resources of the Oak Ridge Leadership Computing Facility, which is a U.S.\ Department of Energy Office of Science User Facility supported under Contract No.~DE-AC05-00OR22725.


\section*{Competing interests}
The authors declare no competing interests.

\bibliographystyle{unsrtnat}
\bibliography{references}

@article{hohenberg1964inhomogeneous,
  author = {Hohenberg, P. and Kohn, W.},
  title = {Inhomogeneous Electron Gas},
  journal = {Phys. Rev.},
  volume = {136},
  number = {3B},
  pages = {B864--B871},
  year = {1964},
  doi = {10.1103/PhysRev.136.B864}
}

@article{kohn1965self,
  author = {Kohn, W. and Sham, L. J.},
  title = {Self-Consistent Equations Including Exchange and Correlation Effects},
  journal = {Phys. Rev.},
  volume = {140},
  number = {4A},
  pages = {A1133--A1138},
  year = {1965},
  doi = {10.1103/PhysRev.140.A1133}
}

@article{becke1988exchange,
  author = {Becke, Axel D.},
  title = {Density-functional exchange-energy approximation with correct asymptotic behavior},
  journal = {Phys. Rev. A},
  volume = {38},
  number = {6},
  pages = {3098--3100},
  year = {1988},
  doi = {10.1103/PhysRevA.38.3098}
}

@article{lee1988development,
  author = {Lee, Chengteh and Yang, Weitao and Parr, Robert G.},
  title = {Development of the {Colle--Salvetti} correlation-energy formula into a functional of the electron density},
  journal = {Phys. Rev. B},
  volume = {37},
  number = {2},
  pages = {785--789},
  year = {1988},
  doi = {10.1103/PhysRevB.37.785}
}

@article{perdew1996generalized,
  author = {Perdew, John P. and Burke, Kieron and Ernzerhof, Matthias},
  title = {Generalized Gradient Approximation Made Simple},
  journal = {Phys. Rev. Lett.},
  volume = {77},
  number = {18},
  pages = {3865--3868},
  year = {1996},
  doi = {10.1103/PhysRevLett.77.3865}
}

@article{taoperdew2003tpss,
  author = {Tao, Jianmin and Perdew, John P. and Staroverov, Viktor N. and Scuseria, Gustavo E.},
  title = {Climbing the Density Functional Ladder: Nonempirical Meta-Generalized Gradient Approximation Designed for Molecules and Solids},
  journal = {Phys. Rev. Lett.},
  volume = {91},
  number = {14},
  pages = {146401},
  year = {2003},
  doi = {10.1103/PhysRevLett.91.146401}
}

@article{becke1993density,
  author = {Becke, Axel D.},
  title = {Density-functional thermochemistry. {III}. The role of exact exchange},
  journal = {J. Chem. Phys.},
  volume = {98},
  number = {7},
  pages = {5648--5652},
  year = {1993},
  doi = {10.1063/1.464913}
}

@article{becke1997systematic,
  author = {Becke, Axel D.},
  title = {Density-functional thermochemistry. {V}. Systematic optimization of exchange-correlation functionals},
  journal = {J. Chem. Phys.},
  volume = {107},
  number = {20},
  pages = {8554--8560},
  year = {1997},
  doi = {10.1063/1.475007}
}

@article{grimme2006semiempirical,
  author = {Grimme, Stefan},
  title = {Semiempirical hybrid density functional with perturbative second-order correlation},
  journal = {J. Chem. Phys.},
  volume = {124},
  number = {3},
  pages = {034108},
  year = {2006},
  doi = {10.1063/1.2148954}
}

@article{stephens1994abinitio,
  author = {Stephens, P. J. and Devlin, F. J. and Chabalowski, C. F. and Frisch, M. J.},
  title = {Ab Initio Calculation of Vibrational Absorption and Circular Dichroism Spectra Using Density Functional Force Fields},
  journal = {J. Phys. Chem.},
  volume = {98},
  number = {45},
  pages = {11623--11627},
  year = {1994},
  doi = {10.1021/j100096a001}
}

@article{sun2015strongly,
  author = {Sun, Jianwei and Ruzsinszky, Adrienn and Perdew, John P.},
  title = {Strongly Constrained and Appropriately Normed Semilocal Density Functional},
  journal = {Phys. Rev. Lett.},
  volume = {115},
  number = {3},
  pages = {036402},
  year = {2015},
  doi = {10.1103/PhysRevLett.115.036402}
}

@article{furness2020accurate,
  author = {Furness, James W. and Kaplan, Aaron D. and Ning, Jinliang and Perdew, John P. and Sun, Jianwei},
  title = {Accurate and Numerically Efficient {r2SCAN} Meta-Generalized Gradient Approximation},
  journal = {J. Phys. Chem. Lett.},
  volume = {11},
  number = {19},
  pages = {8208--8215},
  year = {2020},
  doi = {10.1021/acs.jpclett.0c02405}
}

@article{mardirossian2014wb97xv,
  author = {Mardirossian, Narbe and Head-Gordon, Martin},
  title = {{$\omega$B97X-V}: A 10-parameter, range-separated hybrid, generalized gradient approximation density functional with nonlocal correlation, designed by a survival-of-the-fittest strategy},
  journal = {Phys. Chem. Chem. Phys.},
  volume = {16},
  number = {21},
  pages = {9904--9924},
  year = {2014},
  doi = {10.1039/C3CP54374A}
}

@article{mardirossian2016wB97MV,
  author = {Mardirossian, Narbe and Head-Gordon, Martin},
  title = {{$\omega$B97M-V}: A combinatorially optimized, range-separated hybrid, meta-{GGA} density functional with {VV10} nonlocal correlation},
  journal = {J. Chem. Phys.},
  volume = {144},
  number = {21},
  pages = {214110},
  year = {2016},
  doi = {10.1063/1.4952647}
}

@article{goerigk2017look,
  author = {Goerigk, Lars and Hansen, Andreas and Bauer, Christoph and Ehrlich, Stephan and Najibi, Asim and Grimme, Stefan},
  title = {A look at the density functional theory zoo with the advanced {GMTKN55} database for general main group thermochemistry, kinetics and noncovalent interactions},
  journal = {Phys. Chem. Chem. Phys.},
  volume = {19},
  number = {48},
  pages = {32184--32215},
  year = {2017},
  doi = {10.1039/C7CP04913G}
}

@article{liang2025gscdb137,
  author = {Liang, Jiashu and Head-Gordon, Martin},
  title = {Gold-Standard Chemical Database 137 ({GSCDB137}): A Diverse Set of Accurate Energy Differences for Assessing and Developing Density Functionals},
  journal = {J. Chem. Theory Comput.},
  volume = {21},
  number = {24},
  pages = {12601--12621},
  year = {2025},
  doi = {10.1021/acs.jctc.5c01380},
  eprint = {2508.13468},
  archivePrefix = {arXiv}
}

@misc{liang2026coach,
  author = {Liang, Jiashu and Head-Gordon, Martin},
  title = {Reaching for the performance limit of hybrid density functional theory for molecular chemistry},
  year = {2026},
  eprint = {2603.23466},
  archivePrefix = {arXiv},
  doi = {10.48550/arXiv.2603.23466}
}

@article{schimka2010accurate,
  author = {Schimka, Laurids and Harl, Judith and Stroppa, Alessandro and Gr{\"u}neis, Andreas and Marsman, Martijn and Mittendorfer, Florian and Kresse, Georg},
  title = {Accurate surface and adsorption energies from many-body perturbation theory},
  journal = {Nat. Mater.},
  volume = {9},
  number = {9},
  pages = {741--744},
  year = {2010},
  doi = {10.1038/nmat2806}
}

@article{schmidt2018benchmark,
  author = {Schmidt, Per Simmendefeldt and Thygesen, Kristian Sommer},
  title = {Benchmark Database of Transition Metal Surface and Adsorption Energies from Many-Body Perturbation Theory},
  journal = {J. Phys. Chem. C},
  volume = {122},
  number = {8},
  pages = {4381--4390},
  year = {2018},
  doi = {10.1021/acs.jpcc.7b12258}
}

@article{wellendorff2015benchmark,
  author = {Wellendorff, Jess and Silbaugh, Trent L. and Garcia-Pintos, Delfina and N{\o}rskov, Jens K. and Bligaard, Thomas and Studt, Felix and Campbell, Charles T.},
  title = {A benchmark database for adsorption bond energies to transition metal surfaces and comparison to selected {DFT} functionals},
  journal = {Surf. Sci.},
  volume = {640},
  pages = {36--44},
  year = {2015},
  doi = {10.1016/j.susc.2015.03.023}
}

@article{kelemen1979binding,
  author = {Kelemen, S. R. and Fischer, T. E. and Schwarz, J. A.},
  title = {The binding energy of {CO} on clean and sulfur covered platinum surfaces},
  journal = {Surf. Sci.},
  volume = {81},
  number = {2},
  pages = {440--450},
  year = {1979},
  doi = {10.1016/0039-6028(79)90111-0}
}

@article{campbell1981molecular,
  author = {Campbell, C. T. and Ertl, G. and Kuipers, H. and Segner, J.},
  title = {A molecular beam investigation of the interactions of {CO} with a {Pt(111)} surface},
  journal = {Surf. Sci.},
  volume = {107},
  number = {1},
  pages = {207--219},
  year = {1981},
  doi = {10.1016/0039-6028(81)90621-X}
}

@article{poelsema1984thermal,
  author = {Poelsema, Bene and Palmer, Robert L. and Comsa, George},
  title = {A thermal {He} scattering study of {CO} adsorption on {Pt(111)}},
  journal = {Surf. Sci.},
  volume = {136},
  number = {1},
  pages = {1--14},
  year = {1984},
  doi = {10.1016/0039-6028(84)90651-4}
}

@article{schiesser2010thermodynamics,
  author = {Schie{\ss}er, Alexander and H{\"o}rtz, Peter and Sch{\"a}fer, Rolf},
  title = {Thermodynamics and kinetics of {CO} and benzene adsorption on {Pt(111)} studied with pulsed molecular beams and microcalorimetry},
  journal = {Surf. Sci.},
  volume = {604},
  number = {23--24},
  pages = {2098--2105},
  year = {2010},
  doi = {10.1016/j.susc.2010.09.001}
}

@article{tyson1977surface,
  author = {Tyson, W. R. and Miller, W. A.},
  title = {Surface free energies of solid metals: Estimation from liquid surface tension measurements},
  journal = {Surf. Sci.},
  volume = {62},
  number = {1},
  pages = {267--276},
  year = {1977},
  doi = {10.1016/0039-6028(77)90442-3}
}

@article{vitos1998surface,
  author = {Vitos, L. and Ruban, A. V. and Skriver, H. L. and Koll{\'a}r, J.},
  title = {The surface energy of metals},
  journal = {Surf. Sci.},
  volume = {411},
  number = {1--2},
  pages = {186--202},
  year = {1998},
  doi = {10.1016/S0039-6028(98)00363-X}
}

@article{stroppa2007co,
  author = {Stroppa, Alessandro and Termentzidis, Konstantinos and Paier, Joachim and Kresse, Georg and Hafner, J{\"u}rgen},
  title = {{CO} adsorption on metal surfaces: A hybrid functional study with plane-wave basis set},
  journal = {Phys. Rev. B},
  volume = {76},
  number = {19},
  pages = {195440},
  year = {2007},
  doi = {10.1103/PhysRevB.76.195440}
}

@techreport{swanson1953standard,
  author = {Swanson, Howard E. and Tatge, Eleanor},
  title = {Standard {X-ray} Diffraction Powder Patterns},
  institution = {National Bureau of Standards},
  type = {NBS Circular},
  number = {539},
  volume = {1},
  pages = {31--32},
  url = {https://www.govinfo.gov/content/pkg/GOVPUB-C13-f2a5ff42fffeb0ad99cc234023c5e011/pdf/GOVPUB-C13-f2a5ff42fffeb0ad99cc234023c5e011.pdf},
  year = {1953}
}

@article{owen1933precision,
  author = {Owen, E. A. and Yates, E. L.},
  title = {Precision measurements of crystal parameters},
  journal = {Philos. Mag.},
  volume = {15},
  number = {98},
  pages = {472--488},
  year = {1933},
  doi = {10.1080/14786443309462199}
}

@article{carbone2024co,
  author = {Carbone, Johanna P. and Irmler, Andreas and Gallo, Alejandro and Sch{\"a}fer, Tobias and {Van Benschoten}, William Z. and Shepherd, James J. and Gr{\"u}neis, Andreas},
  title = {{CO} adsorption on {Pt}(111) studied by periodic coupled cluster theory},
  journal = {Faraday Discuss.},
  volume = {254},
  pages = {586--597},
  year = {2024},
  doi = {10.1039/D4FD00085D}
}

@article{ceperley1980ground,
  author = {Ceperley, D. M. and Alder, B. J.},
  title = {Ground State of the Electron Gas by a Stochastic Method},
  journal = {Phys. Rev. Lett.},
  volume = {45},
  number = {7},
  pages = {566--569},
  year = {1980},
  doi = {10.1103/PhysRevLett.45.566}
}

@article{kent2020qmcpack,
  author = {Kent, P. R. C. and Annaberdiyev, Abdulgani and Benali, Anouar and Bennett, M. Chandler and {Landinez Borda}, Edgar Josu{\'e} and Doak, Peter and Hao, Hongxia and Jordan, Kenneth D. and Krogel, Jaron T. and Kyl{\"a}np{\"a}{\"a}, Ilkka and Lee, Joonho and Luo, Ye and Malone, Fionn D. and Melton, Cody A. and Mitas, Lubos and Morales, Miguel A. and Neuscamman, Eric and Reboredo, Fernando A. and Rubenstein, Brenda and Saritas, Kayahan and Upadhyay, Shiv and Wang, Guangming and Zhang, Shuai and Zhao, Luning},
  title = {{QMCPACK}: Advances in the development, efficiency, and application of auxiliary field and real-space variational and diffusion quantum Monte Carlo},
  journal = {J. Chem. Phys.},
  volume = {152},
  number = {17},
  pages = {174105},
  year = {2020},
  doi = {10.1063/5.0004860}
}

@article{hsing2019quantum,
  author = {Hsing, Cheng-Rong and Chang, Chun-Ming and Cheng, Ching and Wei, Ching-Ming},
  title = {Quantum Monte Carlo Studies of {CO} Adsorption on Transition Metal Surfaces},
  journal = {J. Phys. Chem. C},
  volume = {123},
  number = {25},
  pages = {15659--15664},
  year = {2019},
  doi = {10.1021/acs.jpcc.9b03780}
}

@article{snyder2012finding,
  author = {Snyder, John C. and Rupp, Matthias and Hansen, Katja and M{\"u}ller, Klaus-Robert and Burke, Kieron},
  title = {Finding Density Functionals with Machine Learning},
  journal = {Phys. Rev. Lett.},
  volume = {108},
  number = {25},
  pages = {253002},
  year = {2012},
  doi = {10.1103/PhysRevLett.108.253002}
}

@article{snyder2013orbital,
  author = {Snyder, John C. and Rupp, Matthias and Hansen, Katja and Blooston, Leo and M{\"u}ller, Klaus-Robert and Burke, Kieron},
  title = {Orbital-free bond breaking via machine learning},
  journal = {J. Chem. Phys.},
  volume = {139},
  number = {22},
  pages = {224104},
  year = {2013},
  doi = {10.1063/1.4834075}
}

@article{nagai2020completing,
  author = {Nagai, Ryo and Akashi, Ryosuke and Sugino, Osamu},
  title = {Completing density functional theory by machine learning hidden messages from molecules},
  journal = {npj Comput. Mater.},
  volume = {6},
  pages = {43},
  year = {2020},
  doi = {10.1038/s41524-020-0310-0}
}

@article{dick2020machine,
  author = {Dick, Sebastian and Fernandez-Serra, Marivi},
  title = {Machine learning accurate exchange and correlation functionals of the electronic density},
  journal = {Nat. Commun.},
  volume = {11},
  pages = {3509},
  year = {2020},
  doi = {10.1038/s41467-020-17265-7}
}

@article{dick2021highly,
  author = {Dick, Sebastian and Fernandez-Serra, Marivi},
  title = {Highly accurate and constrained density functional obtained with differentiable programming},
  journal = {Phys. Rev. B},
  volume = {104},
  number = {16},
  pages = {L161109},
  year = {2021},
  doi = {10.1103/PhysRevB.104.L161109}
}

@article{li2021kohn,
  author = {Li, Li and Hoyer, Stephan and Pederson, Ryan and Sun, Ruoxi and Cubuk, Ekin D. and Riley, Patrick and Burke, Kieron},
  title = {{Kohn-Sham} Equations as Regularizer: Building Prior Knowledge into Machine-Learned Physics},
  journal = {Phys. Rev. Lett.},
  volume = {126},
  number = {3},
  pages = {036401},
  year = {2021},
  doi = {10.1103/PhysRevLett.126.036401}
}

@article{chen2021deepks,
  author = {Chen, Yixiao and Zhang, Linfeng and Wang, Han and E, Weinan},
  title = {{DeePKS}: A Comprehensive Data-Driven Approach toward Chemically Accurate Density Functional Theory},
  journal = {J. Chem. Theory Comput.},
  volume = {17},
  number = {1},
  pages = {170--181},
  year = {2021},
  doi = {10.1021/acs.jctc.0c00872}
}

@article{kirkpatrick2021pushing,
  author = {Kirkpatrick, James and McMorrow, Brendan and Turban, David H. P. and Gaunt, Alexander L. and Spencer, James S. and Matthews, Alexander G. D. G. and Obika, Annette and Thiry, Louis and Fortunato, Meire and Pfau, David and Castellanos, Lara Rom{\'a}n and Petersen, Stig and Nelson, Alexander W. R. and Kohli, Pushmeet and Mori-S{\'a}nchez, Paula and Hassabis, Demis and Cohen, Aron J.},
  title = {Pushing the frontiers of density functionals by solving the fractional electron problem},
  journal = {Science},
  volume = {374},
  number = {6573},
  pages = {1385--1389},
  year = {2021},
  doi = {10.1126/science.abj6511}
}

@misc{luise2025accurate,
  author = {Luise, Giulia and Huang, Chin-Wei and Vogels, Thijs and Kooi, Derk P. and Ehlert, Sebastian and Lanius, Stephanie and Giesbertz, Klaas J. H. and Karton, Amir and Gunceler, Deniz and Battaglia, Stefano and Simm, Gregor N. C. and Szab{\'o}, P. Bern{\'a}t and Stanley, Megan and Bruinsma, Wessel P. and Huang, Lin and Wei, Xinran and Garrido Torres, Jos{\'e} and Katbashev, Abylay and Chavez Zavaleta, Rodrigo and M{\'a}t{\'e}, B{\'a}lint and Kaba, S{\'e}kou-Oumar and Sordillo, Roberto and Chen, Yingrong and Williams-Young, David B. and Bishop, Christopher M. and Hermann, Jan and van den Berg, Rianne and Gori-Giorgi, Paola},
  title = {Accurate and scalable exchange-correlation with deep learning},
  year = {2025},
  eprint = {2506.14665},
  archivePrefix = {arXiv},
  doi = {10.48550/arXiv.2506.14665},
  note = {arXiv:2506.14665v6, revised April 21, 2026}
}

@misc{luise2025accuratev3,
  author = {Luise, Giulia and Huang, Chin-Wei and Vogels, Thijs and Kooi, Derk P. and Ehlert, Sebastian and Lanius, Stephanie and Giesbertz, Klaas J. H. and Karton, Amir and Gunceler, Deniz and Stanley, Megan and Bruinsma, Wessel P. and Huang, Lin and Wei, Xinran and Garrido Torres, Jos{\'e} and Katbashev, Abylay and Chavez Zavaleta, Rodrigo and M{\'a}t{\'e}, B{\'a}lint and Kaba, S{\'e}kou-Oumar and Sordillo, Roberto and Chen, Yingrong and Williams-Young, David B. and Bishop, Christopher M. and Hermann, Jan and van den Berg, Rianne and Gori-Giorgi, Paola},
  title = {Accurate and scalable exchange-correlation with deep learning},
  year = {2025},
  eprint = {2506.14665v3},
  archivePrefix = {arXiv},
  url = {https://arxiv.org/abs/2506.14665v3},
  note = {arXiv:2506.14665v3, revised June 23, 2025}
}

@misc{skala_hf_model_card,
  author = {{Microsoft Research AI for Science}},
  title = {{Skala 1.1} model card},
  howpublished = {{Hugging Face} model card},
  url = {https://huggingface.co/microsoft/skala-1.1},
  year = {2026},
  note = {Accessed July 10, 2026}
}

@article{riemelmoser2023machine,
  author = {Riemelmoser, Stefan and Verdi, Carla and Kaltak, Merzuk and Kresse, Georg},
  title = {Machine Learning Density Functionals from the Random-Phase Approximation},
  journal = {J. Chem. Theory Comput.},
  volume = {19},
  number = {20},
  pages = {7287--7299},
  year = {2023},
  doi = {10.1021/acs.jctc.3c00848}
}

@article{bystrom2022cider,
  author = {Bystrom, Kyle and Kozinsky, Boris},
  title = {{CIDER}: An Expressive, Nonlocal Feature Set for Machine Learning Density Functionals with Exact Constraints},
  journal = {J. Chem. Theory Comput.},
  volume = {18},
  number = {4},
  pages = {2180--2192},
  year = {2022},
  doi = {10.1021/acs.jctc.1c00904}
}

@article{bystrom2024nonlocal,
  author = {Bystrom, Kyle and Kozinsky, Boris},
  title = {Nonlocal machine-learned exchange functional for molecules and solids},
  journal = {Phys. Rev. B},
  volume = {110},
  number = {7},
  pages = {075130},
  year = {2024},
  doi = {10.1103/PhysRevB.110.075130}
}

@article{bystrom2024training,
  author = {Bystrom, Kyle and Falletta, Stefano and Kozinsky, Boris},
  title = {Training Machine-Learned Density Functionals on Band Gaps},
  journal = {J. Chem. Theory Comput.},
  volume = {20},
  number = {17},
  pages = {7516--7532},
  year = {2024},
  doi = {10.1021/acs.jctc.4c00999}
}

@article{feibelman2001co,
  author = {Feibelman, Peter J. and Hammer, Bj{\o}rk and N{\o}rskov, Jens K. and Wagner, Florian and Scheffler, Matthias and Stumpf, Roland and Watwe, Ravindra and Dumesic, James},
  title = {The {CO/Pt(111)} Puzzle},
  journal = {J. Phys. Chem. B},
  volume = {105},
  number = {18},
  pages = {4018--4025},
  year = {2001},
  doi = {10.1021/jp002302t}
}

@article{lazic2010density,
  author = {Lazi{\'c}, Predrag and Alaei, M. and Atodiresei, N. and Caciuc, V. and Brako, R. and Bl{\"u}gel, S.},
  title = {Density functional theory with nonlocal correlation: A key to the solution of the {CO} adsorption puzzle},
  journal = {Phys. Rev. B},
  volume = {81},
  number = {4},
  pages = {045401},
  year = {2010},
  doi = {10.1103/PhysRevB.81.045401}
}

@article{tozer1996exchange,
  author = {Tozer, David J. and Ingamells, V. E. and Handy, Nicholas C.},
  title = {Exchange-correlation potentials},
  journal = {J. Chem. Phys.},
  volume = {105},
  number = {20},
  pages = {9200--9213},
  year = {1996},
  doi = {10.1063/1.472753}
}

@article{snyder2015nonlinear,
  author = {Snyder, John C. and Rupp, Matthias and M{\"u}ller, Klaus-Robert and Burke, Kieron},
  title = {Nonlinear gradient denoising: Finding accurate extrema from inaccurate functional derivatives},
  journal = {Int. J. Quantum Chem.},
  volume = {115},
  number = {16},
  pages = {1102--1114},
  year = {2015},
  doi = {10.1002/qua.24937}
}

@article{li2016understanding,
  author = {Li, Li and Snyder, John C. and Pelaschier, Isabelle M. and Huang, Jessica and Niranjan, Uma-Naresh and Duncan, Paul and Rupp, Matthias and M{\"u}ller, Klaus-Robert and Burke, Kieron},
  title = {Understanding machine-learned density functionals},
  journal = {Int. J. Quantum Chem.},
  volume = {116},
  number = {11},
  pages = {819--833},
  year = {2016},
  doi = {10.1002/qua.25040}
}

@article{li2016pure,
  author = {Li, Li and Baker, Thomas E. and White, Steven R. and Burke, Kieron},
  title = {Pure density functional for strong correlation and the thermodynamic limit from machine learning},
  journal = {Phys. Rev. B},
  volume = {94},
  number = {24},
  pages = {245129},
  year = {2016},
  doi = {10.1103/PhysRevB.94.245129}
}

@article{hollingsworth2018can,
  author = {Hollingsworth, Jacob and Li, Li and Baker, Thomas E. and Burke, Kieron},
  title = {Can exact conditions improve machine-learned density functionals?},
  journal = {J. Chem. Phys.},
  volume = {148},
  number = {24},
  pages = {241743},
  year = {2018},
  doi = {10.1063/1.5025668}
}

@article{lei2019design,
  author = {Lei, Xiangyun and Medford, Andrew J.},
  title = {Design and analysis of machine learning exchange-correlation functionals via rotationally invariant convolutional descriptors},
  journal = {Phys. Rev. Materials},
  volume = {3},
  number = {6},
  pages = {063801},
  year = {2019},
  doi = {10.1103/PhysRevMaterials.3.063801}
}

@article{zhou2019toward,
  author = {Zhou, Yi and Wu, Jiang and Chen, Shuguang and Chen, GuanHua},
  title = {Toward the Exact Exchange-Correlation Potential: A Three-Dimensional Convolutional Neural Network Construct},
  journal = {J. Phys. Chem. Lett.},
  volume = {10},
  number = {22},
  pages = {7264--7269},
  year = {2019},
  doi = {10.1021/acs.jpclett.9b02838}
}

@article{kasim2021learning,
  author = {Kasim, Muhammad F. and Vinko, Sam M.},
  title = {Learning the Exchange-Correlation Functional from Nature with Fully Differentiable Density Functional Theory},
  journal = {Phys. Rev. Lett.},
  volume = {127},
  number = {12},
  pages = {126403},
  year = {2021},
  doi = {10.1103/PhysRevLett.127.126403}
}

@article{margraf2021pure,
  author = {Margraf, Johannes T. and Reuter, Karsten},
  title = {Pure non-local machine-learned density functional theory for electron correlation},
  journal = {Nat. Commun.},
  volume = {12},
  pages = {344},
  year = {2021},
  doi = {10.1038/s41467-020-20471-y}
}

@article{pokharel2022exact,
  author = {Pokharel, Kanun and Furness, James W. and Yao, Yi and Blum, Volker and Irons, Tom J. P. and Teale, Andrew M. and Sun, Jianwei},
  title = {Exact constraints and appropriate norms in machine-learned exchange-correlation functionals},
  journal = {J. Chem. Phys.},
  volume = {157},
  number = {17},
  pages = {174106},
  year = {2022},
  doi = {10.1063/5.0111183}
}

@article{nagai2022machine,
  author = {Nagai, Ryo and Akashi, Ryosuke and Sugino, Osamu},
  title = {Machine-learning-based exchange-correlation functional with physical asymptotic constraints},
  journal = {Phys. Rev. Research},
  volume = {4},
  number = {1},
  pages = {013106},
  year = {2022},
  doi = {10.1103/PhysRevResearch.4.013106}
}

@article{cuierrier2021constructing,
  author = {Cuierrier, Etienne and Roy, Pierre-Olivier and Ernzerhof, Matthias},
  title = {Constructing and representing exchange-correlation holes through artificial neural networks},
  journal = {J. Chem. Phys.},
  volume = {155},
  number = {17},
  pages = {174121},
  year = {2021},
  doi = {10.1063/5.0062940}
}

@article{ma2022evolving,
  author = {Ma, He and Narayanaswamy, Arunachalam and Riley, Patrick and Li, Li},
  title = {Evolving symbolic density functionals},
  journal = {Sci. Adv.},
  volume = {8},
  number = {36},
  pages = {eabq0279},
  year = {2022},
  doi = {10.1126/sciadv.abq0279}
}

@article{klimes2011van,
  author = {Klime{\v{s}}, Ji{\v{r}}{\'i} and Bowler, David R. and Michaelides, Angelos},
  title = {Van der Waals density functionals applied to solids},
  journal = {Phys. Rev. B},
  volume = {83},
  number = {19},
  pages = {195131},
  year = {2011},
  doi = {10.1103/PhysRevB.83.195131}
}

@article{liu2017improving,
  author = {Liu, Qin and Wang, Jingchun and Du, PengLi and Hu, Lihong and Zheng, Xiao and Chen, GuanHua},
  title = {Improving the Performance of Long-Range-Corrected Exchange-Correlation Functional with an Embedded Neural Network},
  journal = {J. Phys. Chem. A},
  volume = {121},
  number = {38},
  pages = {7273--7281},
  year = {2017},
  doi = {10.1021/acs.jpca.7b07045}
}

@article{ryabov2020neural,
  author = {Ryabov, Alexander and Akhatov, Iskander and Zhilyaev, Petr},
  title = {Neural network interpolation of exchange-correlation functional},
  journal = {Sci. Rep.},
  volume = {10},
  pages = {8000},
  year = {2020},
  doi = {10.1038/s41598-020-64619-8}
}

@article{vargas2020bayesian,
  author = {Vargas-Hern{\'a}ndez, Rodrigo A.},
  title = {Bayesian Optimization for Calibrating and Selecting Hybrid-Density Functional Models},
  journal = {J. Phys. Chem. A},
  volume = {124},
  number = {20},
  pages = {4053--4061},
  year = {2020},
  doi = {10.1021/acs.jpca.0c01375}
}

@article{kalita2022well,
  author = {Kalita, Bhupalee and Pederson, Ryan and Chen, Jielun and Li, Li and Burke, Kieron},
  title = {How well does {Kohn--Sham} regularizer work for weakly correlated systems?},
  journal = {J. Phys. Chem. Lett.},
  volume = {13},
  number = {11},
  pages = {2540--2547},
  year = {2022},
  doi = {10.1021/acs.jpclett.2c00371}
}

@article{palos2022density,
  author = {Palos, Etienne and Lambros, Eleftherios and Dasgupta, Saswata and Paesani, Francesco},
  title = {Density functional theory of water with the machine-learned {DM21} functional},
  journal = {J. Chem. Phys.},
  volume = {156},
  number = {16},
  pages = {161103},
  year = {2022},
  doi = {10.1063/5.0090862}
}

@article{zhao2024dm21,
  author = {Zhao, Heng and Gould, Tim and Vuckovic, Stefan},
  title = {{Deep Mind 21} functional does not extrapolate to transition metal chemistry},
  journal = {Phys. Chem. Chem. Phys.},
  volume = {26},
  pages = {12289--12298},
  year = {2024},
  doi = {10.1039/D4CP00878B}
}

@inproceedings{gao2024learning,
  author = {Gao, Nicholas and Eberhard, Eike and G{\"u}nnemann, Stephan},
  title = {Learning Equivariant Non-Local Electron Density Functionals},
  booktitle = {International Conference on Learning Representations ({ICLR})},
  year = {2025},
  note = {Spotlight},
  url = {https://openreview.net/forum?id=FhBT596F1X},
  eprint = {2410.07972},
  archivePrefix = {arXiv}
}

@article{kulaev2025dm21,
  author = {Kulaev, Kirill and Ryabov, Alexander and Medvedev, Michael G. and Burnaev, Evgeny and Vanovskiy, Vladimir},
  title = {On the practical applicability of {DM21} neural-network {DFT} functional for chemical calculations: Focus on geometry optimization},
  journal = {J. Chem. Phys.},
  volume = {163},
  number = {7},
  pages = {074105},
  year = {2025},
  doi = {10.1063/5.0266500}
}

@article{kauppDataDrivenLearningOptimal2025,
  title = {Data-{{Driven Learning}} of {{Optimal Position-Dependent Exact-Exchange Energy Density Mixing}} for {{Improved Density Functionals}}},
  author = {Kaupp, Martin and Kov{\'a}cs, N{\'o}ra and Wody{\'n}ski, Artur},
  year = {2026},
  month = mar,
  journal = {The Journal of Physical Chemistry A},
  volume = {130},
  number = {10},
  pages = {1978--1987},
  issn = {1089-5639},
  doi = {10.1021/acs.jpca.5c06845},
  urldate = {2026-08-17}
}

@article{kovacs2026doubly,
  author = {Kov{\'a}cs, N{\'o}ra and {\'S}miga, Szymon and Kaupp, Martin and Wody{\'n}ski, Artur},
  title = {Toward Doubly Local Double Hybrid Functionals Using Neural-Network Local Mixing Functions},
  journal = {J. Chem. Theory Comput.},
  volume = {22},
  number = {7},
  pages = {3268--3281},
  year = {2026},
  doi = {10.1021/acs.jctc.5c01952}
}

@article{romanperez2009efficient,
  author = {{Rom{\'a}n-P{\'e}rez}, Guillermo and Soler, Jos{\'e} M.},
  title = {Efficient Implementation of a van der Waals Density Functional: Application to Double-Wall Carbon Nanotubes},
  journal = {Phys. Rev. Lett.},
  volume = {103},
  number = {9},
  pages = {096102},
  year = {2009},
  doi = {10.1103/PhysRevLett.103.096102}
}

@article{sun2020pyscf,
  author = {Sun, Qiming and Zhang, Xing and Banerjee, Samragni and Bao, Peng and Barbry, Marc and Blunt, Nick S. and Bogdanov, Nikolay A. and Booth, George H. and Chen, Jia and Cui, Zhi-Hao and Eriksen, Janus J. and Gao, Yang and Guo, Sheng and Hermann, Jan and Hermes, Matthew R. and Koh, Kevin and Koval, Peter and Lehtola, Susi and Li, Zhendong and Liu, Junzi and Mardirossian, Narbe and McClain, James D. and Motta, Mario and Mussard, Bastien and Pham, Hung Q. and Pulkin, Artem and Purwanto, Wirawan and Robinson, Paul J. and Ronca, Enrico and Sayfutyarova, Elvira R. and Scheurer, Maximilian and Schurkus, Henry F. and Smith, James E. T. and Sun, Chong and Sun, Shi-Ning and Upadhyay, Shiv and Wagner, Lucas K. and Wang, Xiao and White, Alec and Whitfield, James Daniel and Williamson, Mark J. and Wouters, Sebastian and Yang, Jun and Yu, Jason M. and Zhu, Tianyu and Berkelbach, Timothy C. and Sharma, Sandeep and Sokolov, Alexander Yu. and Chan, Garnet Kin-Lic},
  title = {Recent developments in the {PySCF} program package},
  journal = {J. Chem. Phys.},
  volume = {153},
  number = {2},
  pages = {024109},
  year = {2020},
  doi = {10.1063/5.0006074}
}

@article{enkovaara2010gpaw,
  author = {Enkovaara, J. and Rostgaard, C. and Mortensen, J. J. and Chen, J. and Dulak, M. and Ferrighi, L. and Gavnholt, J. and Glinsvad, C. and Haikola, V. and Hansen, H. A. and Kristoffersen, H. H. and Kuisma, M. and Larsen, A. H. and Lehtovaara, L. and Ljungberg, M. and Lopez-Acevedo, O. and Moses, P. G. and Ojanen, J. and Olsen, T. and Petzold, V. and Romero, N. A. and Stausholm-M{\o}ller, J. and Strange, M. and Tritsaris, G. A. and Vanin, M. and Walter, M. and Hammer, B. and H{\"a}kkinen, H. and Madsen, G. K. H. and Nieminen, R. M. and N{\o}rskov, J. K. and Puska, M. and Rantala, T. T. and Schi{\o}tz, J. and Thygesen, K. S. and Jacobsen, K. W.},
  title = {Electronic structure calculations with {GPAW}: a real-space implementation of the projector augmented-wave method},
  journal = {J. Phys.: Condens. Matter},
  volume = {22},
  number = {25},
  pages = {253202},
  year = {2010},
  doi = {10.1088/0953-8984/22/25/253202}
}

@article{perdewAccurateSimpleAnalytic1992,
  title = {Accurate and Simple Analytic Representation of the Electron-Gas Correlation Energy},
  author = {Perdew, John P. and Wang, Yue},
  year = 1992,
  month = jun,
  journal = {Phys. Rev. B},
  volume = {45},
  number = {23},
  pages = {13244--13249},
  publisher = {American Physical Society},
  doi = {10.1103/PhysRevB.45.13244},
  urldate = {2025-08-27}
}

@misc{ciderpress,
  author = {{MIR Group}},
  title = {{CiderPress}: A high-performance software package for training and evaluating machine-learned {XC} functionals using the {CIDER} framework},
  howpublished = {\url{https://github.com/cider-dft/CiderPress}},
  url = {https://github.com/cider-dft/CiderPress},
  year = {2026},
  note = {GitHub repository, accessed July 29, 2026}
}

@article{kresse1993ab,
  author = {Kresse, G. and Hafner, J.},
  title = {Ab initio molecular dynamics for liquid metals},
  journal = {Phys. Rev. B},
  volume = {47},
  number = {1},
  pages = {558--561},
  year = {1993},
  doi = {10.1103/PhysRevB.47.558}
}

@article{kresse1996efficiency,
  author = {Kresse, G. and Furthm{\"u}ller, J.},
  title = {Efficiency of ab-initio total energy calculations for metals and semiconductors using a plane-wave basis set},
  journal = {Comput. Mater. Sci.},
  volume = {6},
  number = {1},
  pages = {15--50},
  year = {1996},
  doi = {10.1016/0927-0256(96)00008-0}
}

@article{kresse1996efficient,
  author = {Kresse, G. and Furthm{\"u}ller, J.},
  title = {Efficient iterative schemes for ab initio total-energy calculations using a plane-wave basis set},
  journal = {Phys. Rev. B},
  volume = {54},
  number = {16},
  pages = {11169--11186},
  year = {1996},
  doi = {10.1103/PhysRevB.54.11169}
}

@article{blochl1994projector,
  author = {Bl{\"o}chl, P. E.},
  title = {Projector augmented-wave method},
  journal = {Phys. Rev. B},
  volume = {50},
  number = {24},
  pages = {17953--17979},
  year = {1994},
  doi = {10.1103/PhysRevB.50.17953}
}

@article{kresse1999ultrasoft,
  author = {Kresse, G. and Joubert, D.},
  title = {From ultrasoft pseudopotentials to the projector augmented-wave method},
  journal = {Phys. Rev. B},
  volume = {59},
  number = {3},
  pages = {1758--1775},
  year = {1999},
  doi = {10.1103/PhysRevB.59.1758}
}

@article{harl2008cohesive,
  author = {Harl, Judith and Kresse, Georg},
  title = {Cohesive energy curves for noble gas solids calculated by adiabatic connection fluctuation-dissipation theory},
  journal = {Phys. Rev. B},
  volume = {77},
  number = {4},
  pages = {045136},
  year = {2008},
  doi = {10.1103/PhysRevB.77.045136}
}

@article{kaltak2014low,
  author = {Kaltak, Merzuk and Klime{\v{s}}, Ji{\v{r}}{\'i} and Kresse, Georg},
  title = {Low scaling algorithms for the random phase approximation: Imaginary time and Laplace transformations},
  journal = {J. Chem. Theory Comput.},
  volume = {10},
  number = {6},
  pages = {2498--2507},
  year = {2014},
  doi = {10.1021/ct5001268}
}

@article{kaltak2014cubic,
  author = {Kaltak, Merzuk and Klime{\v{s}}, Ji{\v{r}}{\'i} and Kresse, Georg},
  title = {Cubic scaling algorithm for the random phase approximation: Self-interstitials and vacancies in {Si}},
  journal = {Phys. Rev. B},
  volume = {90},
  number = {5},
  pages = {054115},
  year = {2014},
  doi = {10.1103/PhysRevB.90.054115}
}

@article{klimes2014predictive,
  author = {Klime{\v{s}}, Ji{\v{r}}{\'i} and Kaltak, Merzuk and Kresse, Georg},
  title = {Predictive {GW} calculations using plane waves and pseudopotentials},
  journal = {Phys. Rev. B},
  volume = {90},
  number = {7},
  pages = {075125},
  year = {2014},
  doi = {10.1103/PhysRevB.90.075125}
}

@article{hjorthlarsen2017atomic,
  author = {Hjorth Larsen, Ask and Mortensen, Jens J{\o}rgen and Blomqvist, Jakob and Castelli, Ivano E. and Christensen, Rune and Du{\l}ak, Marcin and Friis, Jesper and Groves, Michael N. and Hammer, Bj{\o}rk and Hargus, Cory and Hermes, Eric D. and Jennings, Paul C. and Bjerre Jensen, Peter and Kermode, James and Kitchin, John R. and Kolsbjerg, Esben L. and Kubal, Joseph and Kaasbjerg, Kristen and Lysgaard, Steen and Maronsson, J{\'o}n Bergmann and Maxson, Tristan and Olsen, Thomas and Pastewka, Lars and Peterson, Andrew and Rostgaard, Carsten and Schi{\o}tz, Jakob and Sch{\"u}tt, Ole and Strange, Mikkel and Thygesen, Kristian S. and Vegge, Tejs and Vilhelmsen, Lasse and Walter, Michael and Zeng, Zhenhua and Jacobsen, Karsten W.},
  title = {The atomic simulation environment---a Python library for working with atoms},
  journal = {J. Phys.: Condens. Matter},
  volume = {29},
  number = {27},
  pages = {273002},
  year = {2017},
  doi = {10.1088/1361-648X/aa680e}
}

@article{adamo1999toward,
  author = {Adamo, Carlo and Barone, Vincenzo},
  title = {Toward reliable density functional methods without adjustable parameters: The {PBE0} model},
  journal = {J. Chem. Phys.},
  volume = {110},
  number = {13},
  pages = {6158--6170},
  year = {1999},
  doi = {10.1063/1.478522}
}

@article{perdew1996rationale,
  author  = {Perdew, John P. and Ernzerhof, Matthias and Burke, Kieron},
  title   = {Rationale for mixing exact exchange with density functional approximations},
  journal = {J. Chem. Phys.},
  volume  = {105},
  number  = {22},
  pages   = {9982--9985},
  year    = {1996},
  doi     = {10.1063/1.472933}
}

@article{Lehtola2018RecentTheory,
    title = {{Recent developments in LIBXC — A comprehensive library of functionals for density functional theory}},
    year = {2018},
    journal = {SoftwareX},
    author = {Lehtola, Susi and Steigemann, Conrad and Oliveira, Micael J.T. and Marques, Miguel A.L.},
    volume = {7},
    pages = {1--5},
    doi = {10.1016/j.softx.2017.11.002},
    issn = {23527110}
}

@article{Larsen_2017,
  title = {Libvdwxc: A Library for Exchange--Correlation Functionals in the {{vdW-DF}} Family},
  author = {Larsen, Ask Hjorth and Kuisma, Mikael and L{\"o}fgren, Joakim and Pouillon, Yann and Erhart, Paul and Hyldgaard, Per},
  year = 2017,
  month = jun,
  journal = {Modelling and Simulation in Materials Science and Engineering},
  volume = {25},
  number = {6},
  pages = {065004},
  publisher = {IOP Publishing},
  doi = {10.1088/1361-651X/aa7320}
}

@article{bryenton2026wtmad4,
  author = {Bryenton, Kyle R. and Johnson, Erin R.},
  title = {{WTMAD-4}: a fair weighting scheme for {GMTKN55}},
  journal = {Phys. Chem. Chem. Phys.}, volume = {28}, number = {2},
  pages = {1463--1469}, year = {2026}, doi = {10.1039/D5CP03741G}
}

@article{caldeweyher2019d4,
  author = {Caldeweyher, Eike and Ehlert, Sebastian and Hansen, Andreas and Neugebauer, Hagen and Spicher, Sebastian and Bannwarth, Christoph and Grimme, Stefan},
  title = {A generally applicable atomic-charge dependent {London} dispersion correction},
  journal = {J. Chem. Phys.}, volume = {150}, number = {15},
  pages = {154122}, year = {2019}, doi = {10.1063/1.5090222}
}

@article{vydrov2010vv10,
  author = {Vydrov, Oleg A. and Van Voorhis, Troy},
  title = {Nonlocal van der {Waals} density functional: The simpler the better},
  journal = {J. Chem. Phys.}, volume = {133}, number = {24},
  pages = {244103}, year = {2010}, doi = {10.1063/1.3521275}
}

@article{furche2001molecular,
    author = {Furche, Filipp},
    title = {Molecular tests of the random phase approximation to the exchange-correlation energy functional},
    journal = {Phys. Rev. B},
    volume = {64}, number = {19}, pages = {195120}, year = {2001},
    doi = {10.1103/PhysRevB.64.195120}
}

@article{harl2010assessing,
    author = {Harl, Judith and Schimka, Laurids and Kresse, Georg},
    title = {Assessing the quality of the random phase approximation for lattice constants and atomization energies of
    solids},
    journal = {Phys. Rev. B},
    volume = {81}, number = {11}, pages = {115126}, year = {2010},
    doi = {10.1103/PhysRevB.81.115126}
}

@article{ren2012random,
    author = {Ren, Xinguo and Rinke, Patrick and Joas, Christian and Scheffler, Matthias},
    title = {Random-phase approximation and its applications in computational chemistry and materials science},
    journal = {J. Mater. Sci.},
    volume = {47}, number = {21}, pages = {7447--7471}, year = {2012},
    doi = {10.1007/s10853-012-6570-4}
}

@article{schimka2013lattice,
  author = {Schimka, Laurids and Gaudoin, Ren{\'e} and Klime{\v s}, Ji{\v r}{\'i} and Marsman, Martijn and Kresse, Georg},
  title = {Lattice constants and cohesive energies of alkali, alkaline-earth, and transition metals: Random phase approximation and density functional theory results},
  journal = {Phys. Rev. B},
  volume = {87},
  number = {21},
  pages = {214102},
  year = {2013},
  doi = {10.1103/PhysRevB.87.214102}
}

@article{olsen2013beyond,
  author = {Olsen, Thomas and Thygesen, Kristian S.},
  title = {Beyond the random phase approximation: Improved description of short-range correlation by a renormalized adiabatic local density approximation},
  journal = {Phys. Rev. B},
  volume = {88},
  number = {11},
  pages = {115131},
  year = {2013},
  doi = {10.1103/PhysRevB.88.115131}
}

@article{kim2013understanding,
  author = {Kim, Min-Cheol and Sim, Eunji and Burke, Kieron},
  title = {Understanding and Reducing Errors in Density Functional Calculations},
  journal = {Phys. Rev. Lett.},
  volume = {111},
  number = {7},
  pages = {073003},
  year = {2013},
  doi = {10.1103/PhysRevLett.111.073003}
}

@article{sim2022improving,
  author = {Sim, Eunji and Song, Suhwan and Vuckovic, Stefan and Burke, Kieron},
  title = {Improving Results by Improving Densities: Density-Corrected Density Functional Theory},
  journal = {J. Am. Chem. Soc.},
  volume = {144},
  number = {15},
  pages = {6625--6639},
  year = {2022},
  doi = {10.1021/jacs.1c11506}
}

@article{Levy1985,
author = {Levy, Mel and Perdew, John P.},
doi = {10.1103/PhysRevA.32.2010},
journal = {Phys. Rev. A},
number = {4},
pages = {2010--2021},
title = {{Hellmann-Feynman, virial, and scaling requisites for the exact universal density functionals. Shape of the correlation potential and diamagnetic susceptibility for atoms}},
volume = {32},
year = {1985}
}

@article{pulay1984improved,
  author = {Pulay, Peter},
  title = {Improved SCF convergence acceleration},
  journal = {J. Comput. Chem.},
  volume = {3},
  number = {4},
  pages = {556--560},
  year = {1982},
  doi = {10.1002/jcc.540030413}
}

@article{kudin2002scfediis,
  author = {Kudin, Konstantin N. and Scuseria, Gustavo E. and Cancès, Eric},
  title = {A black-box self-consistent field convergence algorithm: One step closer},
  journal = {J. Chem. Phys.},
  volume = {116},
  number = {19},
  pages = {8255--8261},
  year = {2002},
  doi = {10.1063/1.1470195}
}

@article{hu2010adiis,
  author = {Hu, Xiangqian and Yang, Weitao},
  title = {Accelerating self-consistent field convergence with the augmented {Roothaan--Hall} energy function},
  journal = {J. Chem. Phys.},
  volume = {132},
  number = {5},
  pages = {054109},
  year = {2010},
  doi = {10.1063/1.3304922}
}

@article{pulay1993c2diis,
  author = {Sellers, Harrell},
  title = {The {$C^2$}-{DIIS} convergence acceleration algorithm},
  journal = {Int. J. Quantum Chem.},
  volume = {45},
  number = {1},
  pages = {31--41},
  year = {1993},
  doi = {10.1002/qua.560450106}
}

@article{grimme2010d3,
  author = {Grimme, Stefan and Antony, Jens and Ehrlich, Stephan and Krieg, Helge},
  title = {A consistent and accurate ab initio parametrization of density functional dispersion correction ({DFT-D}) for the 94 elements {H-Pu}},
  journal = {J. Chem. Phys.},
  volume = {132},
  number = {15},
  pages = {154104},
  year = {2010},
  doi = {10.1063/1.3382344}
}

@article{grimme2011d3bj,
  author = {Grimme, Stefan and Ehrlich, Stephan and Goerigk, Lars},
  title = {Effect of the damping function in dispersion corrected density functional theory},
  journal = {J. Comput. Chem.},
  volume = {32},
  number = {7},
  pages = {1456--1465},
  year = {2011},
  doi = {10.1002/jcc.21759}
}

@article{Dirac1930NoteAtom,
    title = {{Note on Exchange Phenomena in the Thomas Atom}},
    year = {1930},
    journal = {Mathematical Proceedings of the Cambridge Philosophical Society},
    author = {Dirac, P. A. M.},
    number = {3},
    month = {7},
    pages = {376--385},
    volume = {26},
    publisher = {Cambridge University Press},
    url = {https://www.cambridge.org/core/journals/mathematical-proceedings-of-the-cambridge-philosophical-society/article/note-on-exchange-phenomena-in-the-thomas-atom/6C5FF7297CD96F49A8B8E9E3EA50E412},
    doi = {10.1017/S0305004100016108},
    issn = {0305-0041}
}

@misc{ciderpress_docs,
    author  = {{CiderPress Developers}},
    title   = {{CiderPress Documentation}},
    year    = {2026},
    url     = {https://cider-dft.github.io/CiderPress/},
    urldate = {2026-08-09}
}

@article{zhai2023fes,
  author = {Zhai, Huanchen and Lee, Seunghoon and Cui, Zhi-Hao and Cao, Lili and Ryde, Ulf and Chan, Garnet Kin-Lic},
  title = {Multireference Protonation Energetics of a Dimeric Model of Nitrogenase Iron--Sulfur Clusters},
  journal = {J. Phys. Chem. A},
  volume = {127},
  number = {47},
  pages = {9974--9984},
  year = {2023},
  doi = {10.1021/acs.jpca.3c06142}
}

@article{bytautas2010o2,
  author = {Bytautas, Laimutis and Matsunaga, Nikita and Ruedenberg, Klaus},
  title = {Accurate Ab Initio Potential Energy Curve of {O2}. {II}. Core-Valence Correlations, Relativistic Contributions, and Vibration-Rotation Spectrum},
  journal = {J. Chem. Phys.},
  volume = {132},
  number = {7},
  pages = {074307},
  year = {2010},
  doi = {10.1063/1.3298376}
}

@article{trepte2022vcml,
  author = {Trepte, Kai and Voss, Johannes},
  title = {Data-Driven and Constrained Optimization of Semilocal Exchange and Nonlocal Correlation Functionals for Materials and Surface Chemistry},
  journal = {J. Comput. Chem.},
  volume = {43},
  number = {16},
  pages = {1104--1112},
  year = {2022},
  doi = {10.1002/jcc.26872}
}

@article{gamo1997graphene,
  author = {Gamo, Y. and Nagashima, A. and Wakabayashi, M. and Terai, M. and Oshima, C.},
  title = {Atomic Structure of Monolayer Graphite Formed on {Ni(111)}},
  journal = {Surf. Sci.},
  volume = {374},
  number = {1--3},
  pages = {61--64},
  year = {1997},
  doi = {10.1016/S0039-6028(96)00785-6}
}

@article{shelton1974graphene,
  author = {Shelton, J. C. and Patil, H. R. and Blakely, J. M.},
  title = {Equilibrium Segregation of Carbon to a Nickel (111) Surface: A Surface Phase Transition},
  journal = {Surf. Sci.},
  volume = {43},
  number = {2},
  pages = {493--520},
  year = {1974},
  doi = {10.1016/0039-6028(74)90272-6}
}

@article{spanu2009graphite,
  author = {Spanu, Leonardo and Sorella, Sandro and Galli, Giulia},
  title = {Nature and Strength of Interlayer Binding in Graphite},
  journal = {Phys. Rev. Lett.},
  volume = {103},
  number = {19},
  pages = {196401},
  year = {2009},
  doi = {10.1103/PhysRevLett.103.196401}
}

@article{jung2018exfoliation,
  author  = {Jung, Jong Hyun and Park, Cheol-Hwan and Ihm, Jisoon},
  title   = {A Rigorous Method of Calculating Exfoliation Energies from First Principles},
  journal = {Nano Letters},
  volume  = {18},
  number  = {5},
  pages   = {2759--2765},
  year    = {2018},
  doi     = {10.1021/acs.nanolett.7b04201}
}

@article{zacharia2004graphite,
  author = {Zacharia, Renju and Ulbricht, Hendrik and Hertel, Tobias},
  title = {Interlayer Cohesive Energy of Graphite from Thermal Desorption of Polyaromatic Hydrocarbons},
  journal = {Phys. Rev. B},
  volume = {69},
  number = {15},
  pages = {155406},
  year = {2004},
  doi = {10.1103/PhysRevB.69.155406}
}

@article{abdallah2024qsgw,
  author = {Abdallah, Mohamed S. and Pasquarello, Alfredo},
  title = {Quasiparticle Self-Consistent {$GW$} with Effective Vertex Corrections in the Polarizability and the Self-Energy Applied to {MnO}, {FeO}, {CoO}, and {NiO}},
  journal = {Phys. Rev. B},
  volume = {110},
  number = {15},
  pages = {155105},
  year = {2024},
  doi = {10.1103/PhysRevB.110.155105}
}

@article{cunningham2023qsgwhat,
  author = {Cunningham, Brian and Gr{\"u}ning, Myrta and Pashov, Dimitar and van Schilfgaarde, Mark},
  title = {{QS$G\widehat{W}$}: Quasiparticle Self-Consistent {$GW$} with Ladder Diagrams in {$W$}},
  journal = {Phys. Rev. B},
  volume = {108},
  number = {16},
  pages = {165104},
  year = {2023},
  doi = {10.1103/PhysRevB.108.165104}
}

@article{bagsican2017o2graphene,
  author = {Bagsican, Filchito Renee and Winchester, Andrew and Ghosh, Sujoy and Zhang, Xiang and Ma, Lulu and Wang, Minjie and Murakami, Hironaru and Talapatra, Saikat and Vajtai, Robert and Ajayan, Pulickel M. and Kono, Junichiro and Tonouchi, Masayoshi and Kawayama, Iwao},
  title = {Adsorption Energy of Oxygen Molecules on Graphene and Two-Dimensional Tungsten Disulfide},
  journal = {Sci. Rep.},
  volume = {7},
  number = {1},
  pages = {1774},
  year = {2017},
  doi = {10.1038/s41598-017-01883-1}
}

@article{shin2019o2graphene,
  author = {Shin, Hyeondeok and Luo, Ye and Benali, Anouar and Kwon, Yongkyung},
  title = {Diffusion Monte Carlo Study of {O2} Adsorption on Single-Layer Graphene},
  journal = {Phys. Rev. B},
  volume = {100},
  number = {7},
  pages = {075430},
  year = {2019},
  doi = {10.1103/PhysRevB.100.075430}
}

@article{krupenie1972o2,
  author = {Krupenie, Paul H.},
  title = {The Spectrum of Molecular Oxygen},
  journal = {J. Phys. Chem. Ref. Data},
  volume = {1},
  number = {2},
  pages = {423--534},
  year = {1972},
  doi = {10.1063/1.3253101}
}

@article{karton2017w417,
  author = {Karton, Amir and Sylvetsky, Nitai and Martin, Jan M. L.},
  title = {{W4-17}: A Diverse and High-Confidence Dataset of Atomization Energies for Benchmarking High-Level Electronic Structure Methods},
  journal = {J. Comput. Chem.},
  volume = {38},
  number = {24},
  pages = {2063--2075},
  year = {2017},
  doi = {10.1002/jcc.24854}
}

@article{burdett1987rutile,
  author = {Burdett, Jeremy K. and Hughbanks, Timothy and Miller, Gordon J. and Richardson, Jr., James W. and Smith, Joseph V.},
  title = {Structural-electronic relationships in inorganic solids: powder neutron diffraction studies of the rutile and anatase polymorphs of titanium dioxide at 15 and 295 {K}},
  journal = {J. Am. Chem. Soc.},
  volume = {109},
  number = {12},
  pages = {3639--3646},
  year = {1987},
  doi = {10.1021/ja00246a021}
}

@article{vasquez2018rutile,
  author = {V{\'a}squez, G. Cristian and Maestre, David and Cremades, Ana and Ram{\'i}rez-Castellanos, Julio and Magnano, Elena and Nappini, Silvia and Karazhanov, Smagul Zh.},
  title = {Understanding the Effects of {Cr} Doping in Rutile {TiO2} by {DFT} Calculations and X-Ray Spectroscopy},
  journal = {Sci. Rep.},
  volume = {8},
  pages = {8740},
  year = {2018},
  doi = {10.1038/s41598-018-26728-3}
}

@article{vanelp1991mno,
  author = {van Elp, J. and Potze, R. H. and Eskes, H. and Berger, R. and Sawatzky, G. A.},
  title = {Electronic Structure of {MnO}},
  journal = {Phys. Rev. B},
  volume = {44},
  number = {4},
  pages = {1530--1537},
  year = {1991},
  doi = {10.1103/PhysRevB.44.1530}
}

@article{kurmaev2008oxides,
  author = {Kurmaev, E. Z. and Wilks, R. G. and Moewes, A. and Finkelstein, L. D. and Shamin, S. N. and Kune{\v{s}}, J.},
  title = {Oxygen X-Ray Emission and Absorption Spectra as a Probe of the Electronic Structure of Strongly Correlated Oxides},
  journal = {Phys. Rev. B},
  volume = {77},
  number = {16},
  pages = {165127},
  year = {2008},
  doi = {10.1103/PhysRevB.77.165127}
}

@article{cheetham1983mnni,
  author = {Cheetham, A. K. and Hope, D. A. O.},
  title = {Magnetic Ordering and Exchange Effects in the Antiferromagnetic Solid Solutions {$\mathrm{Mn}_x\mathrm{Ni}_{1-x}\mathrm{O}$}},
  journal = {Phys. Rev. B},
  volume = {27},
  number = {11},
  pages = {6964--6967},
  year = {1983},
  doi = {10.1103/PhysRevB.27.6964}
}

@article{sawatzky1984nio,
  author = {Sawatzky, G. A. and Allen, J. W.},
  title = {Magnitude and Origin of the Band Gap in {NiO}},
  journal = {Phys. Rev. Lett.},
  volume = {53},
  number = {24},
  pages = {2339--2342},
  year = {1984},
  doi = {10.1103/PhysRevLett.53.2339}
}

@article{roth1958monoxides,
  author = {Roth, W. L.},
  title = {Magnetic Structures of {MnO}, {FeO}, {CoO}, and {NiO}},
  journal = {Phys. Rev.},
  volume = {110},
  number = {6},
  pages = {1333--1341},
  year = {1958},
  doi = {10.1103/PhysRev.110.1333}
}

@article{alperin1962nio,
  author = {Alperin, H. A.},
  title = {The Magnetic Form Factor of Nickel Oxide},
  journal = {J. Phys. Soc. Jpn.},
  volume = {17},
  pages = {12--15},
  year = {1962},
  note = {Supplement B-III}
}

@article{boyle1954nio,
  author = {Boyle, B. J. and King, E. G. and Conway, K. C.},
  title = {Heats of Formation of Nickel and Cobalt Oxides ({NiO} and {CoO}) of Combustion Calorimetry},
  journal = {J. Am. Chem. Soc.},
  volume = {76},
  number = {14},
  pages = {3835--3837},
  year = {1954},
  doi = {10.1021/ja01643a072}
}

@book{cox1989codata,
  editor = {Cox, J. D. and Wagman, D. D. and Medvedev, V. A.},
  title = {{CODATA} Key Values for Thermodynamics},
  publisher = {Hemisphere Publishing Corporation},
  address = {New York},
  year = {1989}
}

@article{girifalco1956graphite,
  author = {Girifalco, L. A. and Lad, R. A.},
  title = {Energy of Cohesion, Compressibility, and the Potential Energy Functions of the Graphite System},
  journal = {J. Chem. Phys.},
  volume = {25},
  number = {4},
  pages = {693--697},
  year = {1956},
  doi = {10.1063/1.1743030}
}

@article{hazrati2014li,
    author = {Hazrati, E. and de Wijs, G. A. and Brocks, G.},
    title = {Li Intercalation in Graphite: A van der Waals Density-Functional Study},
    journal = {Phys. Rev. B},
    volume = {90},
    number = {15},
    pages = {155448},
    year = {2014},
    doi = {10.1103/PhysRevB.90.155448},
    url = {https://doi.org/10.1103/PhysRevB.90.155448}
}

@article{lebegue2010graphite,
  author = {Leb{\`e}gue, S. and Harl, J. and Gould, Tim and {\'A}ngy{\'a}n, J. G. and Kresse, G. and Dobson, J. F.},
  title = {Cohesive Properties and Asymptotics of the Dispersion Interaction in Graphite by the Random Phase Approximation},
  journal = {Phys. Rev. Lett.},
  volume = {105},
  number = {19},
  pages = {196401},
  year = {2010},
  doi = {10.1103/PhysRevLett.105.196401}
}

@article{mchugh2020graphite,
  author = {McHugh, James G. and Jolley, Kenny and Mouratidis, Pavlos},
  title = {Ab-initio Calculations of Fission Product Diffusion on Graphene},
  journal = {J. Nucl. Mater.},
  volume = {533},
  pages = {152123},
  year = {2020},
  doi = {10.1016/j.jnucmat.2020.152123}
}

@article{vanvoorhis2002gdm,
  author = {Van Voorhis, Troy and Head-Gordon, Martin},
  title = {A Geometric Approach to Direct Minimization},
  journal = {Mol. Phys.},
  volume = {100},
  number = {11},
  pages = {1713--1721},
  year = {2002},
  doi = {10.1080/00268970110103642}
}

@article{dunietz2002gdm,
  author = {Dunietz, Barry D. and Van Voorhis, Troy and Head-Gordon, Martin},
  title = {Geometric Direct Minimization of {Hartree--Fock} Calculations Involving Open-Shell Wavefunctions with Spin-Restricted Orbitals},
  journal = {J. Theor. Comput. Chem.},
  volume = {1},
  number = {2},
  pages = {255--261},
  year = {2002},
  doi = {10.1142/S0219633602000233}
}

@article{hait2018dipoles,
  author = {Hait, Diptarka and Head-Gordon, Martin},
  title = {How Accurate Is Density Functional Theory at Predicting Dipole Moments? An Assessment Using a New Database of 200 Benchmark Values},
  journal = {J. Chem. Theory Comput.},
  volume = {14},
  number = {4},
  pages = {1969--1981},
  year = {2018},
  doi = {10.1021/acs.jctc.7b01252}
}

@article{kaltak2020minimax,
  author  = {Kaltak, Merzuk and Kresse, Georg},
  title   = {Minimax isometry method: A compressive sensing approach for {Matsubara} summation in many-body perturbation theory},
  journal = {Phys. Rev. B},
  volume  = {101},
  pages   = {205145},
  year    = {2020},
  doi     = {10.1103/PhysRevB.101.205145}
}

@article{fender1968moments,
  author  = {Fender, B. E. F. and Jacobson, A. J. and Wedgwood, F. A.},
  title   = {Covalency Parameters in {MnO}, $\alpha$-{MnS}, and {NiO}},
  journal = {J. Chem. Phys.},
  volume  = {48},
  number  = {3},
  pages   = {990--994},
  year    = {1968},
  doi     = {10.1063/1.1668855}
}

@article{shepard2019graphene,
  author  = {Shepard, Stuart and Smeu, Manuel},
  title   = {First principles study of graphene on metals with the {SCAN} and {SCAN+rVV10} functionals},
  journal = {J. Chem. Phys.},
  volume  = {150},
  number  = {15},
  pages   = {154702},
  year    = {2019},
  doi     = {10.1063/1.5046855}
}

@article{mittendorfer2011graphene,
  author  = {Mittendorfer, Florian and Garhofer, Andreas and Redinger, Josef and Klime{\v{s}}, Ji{\v{r}}{\'i} and Harl, Judith and Kresse, Georg},
  title   = {Graphene on {Ni}(111): Strong interaction and weak adsorption},
  journal = {Phys. Rev. B},
  volume  = {84},
  number  = {20},
  pages   = {201401},
  year    = {2011},
  doi     = {10.1103/PhysRevB.84.201401}
}

@article{paier2006screened,
  author  = {Paier, Joachim and Marsman, Martijn and Hummer, Kerstin and Kresse, Georg and Gerber, Iann C. and {\'A}ngy{\'a}n, J{\'a}nos G.},
  title   = {Screened hybrid density functionals applied to solids},
  journal = {J. Chem. Phys.},
  volume  = {124},
  number  = {15},
  pages   = {154709},
  year    = {2006},
  doi     = {10.1063/1.2187006}
}

@article{paier2006screenederratum,
  author  = {Paier, Joachim and Marsman, Martijn and Hummer, Kerstin and Kresse, Georg and Gerber, Iann C. and {\'A}ngy{\'a}n, J{\'a}nos G.},
  title   = {Erratum: ``Screened hybrid density functionals applied to solids'' [{J. Chem. Phys.} {124}, 154709 (2006)]},
  journal = {J. Chem. Phys.},
  volume  = {125},
  number  = {24},
  pages   = {249901},
  year    = {2006},
  doi     = {10.1063/1.2403866}
}

@article{paier2007b3lyp,
  author  = {Paier, Joachim and Marsman, Martijn and Kresse, Georg},
  title   = {Why does the {B3LYP} hybrid functional fail for metals?},
  journal = {J. Chem. Phys.},
  volume  = {127},
  number  = {2},
  pages   = {024103},
  year    = {2007},
  doi     = {10.1063/1.2747249}
}

@article{maxson2026dftu,
  author  = {Maxson, Tristan and Szilv{\'a}si, Tibor},
  title   = {Self-consistent {DFT+U} for {CO} adsorption on transition metal surfaces: A practical correction and benchmark},
  journal = {J. Catal.},
  volume  = {458},
  pages   = {116875},
  month   = jun,
  year    = {2026},
  doi     = {10.1016/j.jcat.2026.116875}
}

@article{li2021thermal,
  author  = {Li, Wan-Lu and Lininger, Christianna N. and Chen, Kaixuan and Vaissier Welborn, Valerie and Rossomme, Elliot and Bell, Alexis T. and Head-Gordon, Martin and Head-Gordon, Teresa},
  title   = {Critical Role of Thermal Fluctuations for {CO} Binding on Electrocatalytic Metal Surfaces},
  journal = {JACS Au},
  volume  = {1},
  number  = {10},
  pages   = {1708--1718},
  year    = {2021},
  doi     = {10.1021/jacsau.1c00300}
}

@article{norskov2009towards,
  author  = {N{\o}rskov, J. K. and Bligaard, T. and Rossmeisl, J. and Christensen, C. H.},
  title   = {Towards the computational design of solid catalysts},
  journal = {Nat. Chem.},
  volume  = {1},
  number  = {1},
  pages   = {37--46},
  year    = {2009},
  doi     = {10.1038/nchem.121}
}

@article{medford2014assessing,
  author  = {Medford, Andrew J. and Wellendorff, Jess and Vojvodic, Aleksandra and Studt, Felix and Abild-Pedersen, Frank and Jacobsen, Karsten W. and Bligaard, Thomas and N{\o}rskov, Jens K.},
  title   = {Assessing the reliability of calculated catalytic ammonia synthesis rates},
  journal = {Science},
  volume  = {345},
  number  = {6193},
  pages   = {197--200},
  year    = {2014},
  doi     = {10.1126/science.1253486}
}

@article{guoFirstPrinciplesDeterminationCO2018,
  title = {First-{{Principles Determination}} of {{CO Adsorption}} and {{Desorption}} on {{Pt}}(111) in the {{Free Energy Landscape}}},
  author = {Guo, Chenxi and Wang, Ziyun and Wang, Dong and Wang, Hai-Feng and Hu, P.},
  year = 2018,
  month = sep,
  journal = {The Journal of Physical Chemistry C},
  volume = {122},
  number = {37},
  pages = {21478--21483},
  issn = {1932-7447, 1932-7455},
  doi = {10.1021/acs.jpcc.8b06782},
  urldate = {2026-08-16},
  langid = {english}
}

@article{kresseSignificanceSingleelectronEnergies2003,
  title = {Significance of Single-Electron Energies for the Description of {{CO}} on {{Pt}}(111)},
  author = {Kresse, G. and Gil, A. and Sautet, P.},
  year = 2003,
  month = aug,
  journal = {Physical Review B},
  volume = {68},
  number = {7},
  pages = {073401},
  publisher = {American Physical Society},
  doi = {10.1103/PhysRevB.68.073401},
  urldate = {2026-08-16}
}

@article{weiResolvingCOPuzzle2025,
  title = {Resolving the ``{{CO}} Puzzle'': {{Disentangling}} Electronic Structure and Dynamic Effects via an Operando Dynamics Framework},
  shorttitle = {Resolving the ``{{CO}} Puzzle''},
  author = {Wei, Zhiyuan and He, Jin-En and Li, Siwu and Chen, Zhe-Ning and Zhang, Lu and Chen, Jun},
  year = 2025,
  month = dec,
  journal = {The Journal of Chemical Physics},
  volume = {163},
  number = {23},
  pages = {234705},
  issn = {0021-9606},
  doi = {10.1063/5.0299897},
  urldate = {2026-08-16}
}

@article{Kaplan2023TheTheory,
    title = {{The Predictive Power of Exact Constraints and Appropriate Norms in Density Functional Theory}},
    year = {2023},
    journal = {Annual Review of Physical Chemistry},
    author = {Kaplan, Aaron D. and Levy, Mel and Perdew, John P.},
    number = {1},
    month = {4},
    pages = {193--218},
    volume = {74},
    url = {https://www.annualreviews.org/doi/10.1146/annurev-physchem-062422-013259},
    doi = {10.1146/annurev-physchem-062422-013259},
    issn = {0066-426X}
}

@article{dionVanWaalsDensity2004,
  title = {Van Der {{Waals Density Functional}} for {{General Geometries}}},
  author = {Dion, M. and Rydberg, H. and Schr{\"o}der, E. and Langreth, D. C. and Lundqvist, B. I.},
  year = 2004,
  month = jun,
  journal = {Physical Review Letters},
  volume = {92},
  number = {24},
  pages = {246401},
  publisher = {American Physical Society},
  doi = {10.1103/PhysRevLett.92.246401},
  urldate = {2026-08-17}
}

@article{wellendorffDensityFunctionalsSurface2012,
  title = {Density Functionals for Surface Science: {{Exchange-correlation}} Model Development with {{Bayesian}} Error Estimation},
  shorttitle = {Density Functionals for Surface Science},
  author = {Wellendorff, Jess and Lundgaard, Keld T. and M{\o}gelh{\o}j, Andreas and Petzold, Vivien and Landis, David D. and N{\o}rskov, Jens K. and Bligaard, Thomas and Jacobsen, Karsten W.},
  year = 2012,
  month = jun,
  journal = {Physical Review B},
  volume = {85},
  number = {23},
  pages = {235149},
  issn = {1098-0121, 1550-235X},
  doi = {10.1103/PhysRevB.85.235149},
  urldate = {2026-08-16},
  copyright = {http://link.aps.org/licenses/aps-default-license},
  langid = {english}
}

@article{mortensenGPAWOpenPython2024a,
  title = {{{GPAW}}: {{An}} Open {{Python}} Package for Electronic Structure Calculations},
  shorttitle = {{{GPAW}}},
  author = {Mortensen, Jens J{\o}rgen and Larsen, Ask Hjorth and Kuisma, Mikael and Ivanov, Aleksei V. and Taghizadeh, Alireza and Peterson, Andrew and Haldar, Anubhab and Dohn, Asmus Ougaard and Sch{\"a}fer, Christian and J{\'o}nsson, Elvar {\"O}rn and Hermes, Eric D. and Nilsson, Fredrik Andreas and Kastlunger, Georg and Levi, Gianluca and J{\'o}nsson, Hannes and H{\"a}kkinen, Hannu and Fojt, Jakub and Kangsabanik, Jiban and S{\o}dequist, Joachim and Lehtom{\"a}ki, Jouko and Heske, Julian and Enkovaara, Jussi and Winther, Kirsten Tr{\o}strup and Dulak, Marcin and Melander, Marko M. and Ovesen, Martin and Louhivuori, Martti and Walter, Michael and Gjerding, Morten and {Lopez-Acevedo}, Olga and Erhart, Paul and Warmbier, Robert and W{\"u}rdemann, Rolf and Kaappa, Sami and Latini, Simone and Boland, Tara Maria and Bligaard, Thomas and Skovhus, Thorbj{\o}rn and Susi, Toma and Maxson, Tristan and Rossi, Tuomas and Chen, Xi and Schmerwitz, Yorick Leonard A. and Schi{\o}tz, Jakob and Olsen, Thomas and Jacobsen, Karsten Wedel and Thygesen, Kristian Sommer},
  year = 2024,
  month = mar,
  journal = {The Journal of Chemical Physics},
  volume = {160},
  number = {9},
  pages = {092503},
  issn = {0021-9606, 1089-7690},
  doi = {10.1063/5.0182685},
  urldate = {2026-08-17},
  langid = {english}
}

@misc{sunPythonSimulationsChemistry2026,
  title = {The {{Python Simulations}} of {{Chemistry Framework}}: 10 Years of an Open-Source Quantum Chemistry Project},
  shorttitle = {The {{Python Simulations}} of {{Chemistry Framework}}},
  author = {Sun, Qiming and Hermes, Matthew R. and Wu, Xiaojie and Zhai, Huanchen and Zhang, Xing and Ahmed, Abdelrahman M. and Aucar, Juan Jos{\'e} and Backhouse, Oliver J. and Banerjee, Samragni and Bao, Peng and Bogdanov, Nikolay A. and Bystrom, Kyle and Chapoton, Fr{\'e}d{\'e}ric and Chen, Ning-Yuan and Chernyshov, Ivan Yu and Clifford, Helen S. and {Cohen-Janes}, Sander and Cui, Zhi-Hao and Damour, Yann D. and Dattani, Nike and Dittmer, Linus Bjarne and Ehlert, Sebastian and Eriksen, Janus Juul and Evangelista, Francesco A. and Ewing, Simon A. and Farahvash, Ardavan and Focke, Kevin and Gao, Yang and Gasperich, Kevin E. and Gillispie, Nathan and Greiner, Jonas and Hennefarth, Matthew R. and Hermann, Jan and Hillenbrand, Christopher and Huhtasalo, Joonatan and Ibrahim, Basil and Jangid, Bhavnesh and Javaremi, Alireza Nejati and Jenkins, Andrew J. and Jin, Yu and King, Daniel S. and Kooi, Derk Pieter and Kurian, Jo S. and Larsson, Henrik R. and Lau, Bryan Tak Gwong and Lee, Seunghoon and Lehtola, Susi and Li, Chenghan and Li, Hao and Li, Jiachen and Li, Rui and Li, Shuhang and Lykhin, Aleksandr O. and Mahajan, Ankit and Mauger, Nastasia and del {Mazo-Sevillano}, Pablo and Moussa, Jonathan and Nakano, Kousuke and Neufeld, Verena A. and Peng, Linqing and Pham, Hung Q. and Pinski, Peter and Pokhilko, Pavel and Pu, Zhichen and Qian, Yubing and Quiton, Stephen Jon and Schulze, Wanja T. and Scott, Thais R. and Seal, Aniruddha and Serna, James D. and Smith, James E. T. and Smyser, Kori E. and Stahl, Terrence and Sun, Chong and Sung, Kevin J. and Trushin, Egor and Upadhyay, Shiv and Vo, Ethan A. and Vogels, Thijs and Wang, Shirong and Wang, Tai and Wang, Xiao and Wang, Xubo and Wang, Yuanheng and Williamson, Mark and Yang, Junjie and Ye, Hong-Zhou and Yeh, Chia-Nan and Yu, Haiyang and Yu, Jincheng and Yu, Victor Wen-zhe and Zhang, Chaoqun and Zhang, Dayou and Zhang, Yichi and Zhao, Zijun and Zhou, Zehao and Zhu, Andrew J. and Zhu, Tianyu and Berkelbach, Timothy C. and Gagliardi, Laura and Sharma, Sandeep and Sokolov, Alexander Yu. and Chan, Garnet Kin-Lic},
  year = 2026,
  month = apr,
  number = {arXiv:2603.14155},
  eprint = {2603.14155},
  primaryclass = {physics.chem-ph},
  publisher = {arXiv},
  doi = {10.48550/arXiv.2603.14155},
  urldate = {2026-08-17},
  archiveprefix = {arXiv}
}

@article{skubicReviewMultiscaleModelling2024,
  title = {A Review of Multiscale Modelling Approaches for Understanding Catalytic Ammonia Synthesis and Decomposition},
  author = {Skubic, Luka and Gyergyek, Sa{\v s}o and Hu{\v s}, Matej and Likozar, Bla{\v z}},
  year = 2024,
  month = jan,
  journal = {Journal of Catalysis},
  volume = {429},
  pages = {115217},
  issn = {0021-9517},
  doi = {10.1016/j.jcat.2023.115217},
  urldate = {2026-08-17}
}

@article{thorarinsdottirSelfhealingOxygenEvolution2022,
  title = {Self-Healing Oxygen Evolution Catalysts},
  author = {Thorarinsdottir, Agnes E. and Veroneau, Samuel S. and Nocera, Daniel G.},
  year = 2022,
  month = mar,
  journal = {Nature Communications},
  volume = {13},
  number = {1},
  pages = {1243},
  publisher = {Nature Publishing Group},
  issn = {2041-1723},
  doi = {10.1038/s41467-022-28723-9},
  urldate = {2026-08-17},
  copyright = {2022 The Author(s)},
  langid = {english}
}

@misc{dingCoupledReactionDiffusion2025,
  title = {Coupled Reaction and Diffusion Governing Interface Evolution in Solid-State Batteries},
  author = {Ding, Jingxuan and Zichi, Laura and Carli, Matteo and Wang, Menghang and Musaelian, Albert and Xie, Yu and Kozinsky, Boris},
  year = 2025,
  month = jun,
  number = {arXiv:2506.10944},
  eprint = {2506.10944},
  primaryclass = {cond-mat.mtrl-sci},
  publisher = {arXiv},
  doi = {10.48550/arXiv.2506.10944},
  urldate = {2026-08-17},
  archiveprefix = {arXiv}
}

@article{coulterEngineeringIdealHelical2024,
  title = {Engineering Ideal Helical Topological Networks in Stanene via {{Zn}} Decoration},
  author = {Coulter, Jennifer and Hirsbrunner, Mark R. and Dubinkin, Oleg and Hughes, Taylor L. and Kozinsky, Boris},
  year = 2024,
  month = aug,
  journal = {Communications Physics},
  volume = {7},
  number = {1},
  pages = {284},
  publisher = {Nature Publishing Group},
  issn = {2399-3650},
  doi = {10.1038/s42005-024-01764-w},
  urldate = {2026-08-17},
  copyright = {2024 The Author(s)},
  langid = {english}
}

@article{dahalGrapheneNickelInterfaces2014,
  title = {Graphene--Nickel Interfaces: A Review},
  shorttitle = {Graphene--Nickel Interfaces},
  author = {Dahal, Arjun and Batzill, Matthias},
  year = 2014,
  month = mar,
  journal = {Nanoscale},
  volume = {6},
  number = {5},
  pages = {2548--2562},
  issn = {2040-3364},
  doi = {10.1039/c3nr05279f},
  urldate = {2026-08-17}
}

@article{linAdaptivelyCompressedExchange2016,
  title = {Adaptively {{Compressed Exchange Operator}}},
  author = {Lin, Lin},
  year = 2016,
  month = may,
  journal = {Journal of Chemical Theory and Computation},
  volume = {12},
  number = {5},
  pages = {2242--2249},
  issn = {1549-9618, 1549-9626},
  doi = {10.1021/acs.jctc.6b00092},
  urldate = {2026-08-17},
  langid = {english}
}

@article{neeseEfficientApproximateParallel2009a,
  title = {Efficient, Approximate and Parallel {{Hartree}}--{{Fock}} and Hybrid {{DFT}} Calculations. {{A}} `Chain-of-Spheres' Algorithm for the {{Hartree}}--{{Fock}} Exchange},
  author = {Neese, Frank and Wennmohs, Frank and Hansen, Andreas and Becker, Ute},
  year = 2009,
  month = feb,
  journal = {Chemical Physics},
  volume = {356},
  number = {1-3},
  pages = {98--109},
  issn = {03010104},
  doi = {10.1016/j.chemphys.2008.10.036},
  urldate = {2026-08-17},
  copyright = {https://www.elsevier.com/tdm/userlicense/1.0/},
  langid = {english}
}

@article{perdewJacobsLadderDensity2001,
  title = {Jacob's Ladder of Density Functional Approximations for the Exchange-Correlation Energy},
  author = {Perdew, John P. and Schmidt, Karla},
  year = 2001,
  month = jul,
  journal = {AIP Conference Proceedings},
  volume = {577},
  number = {1},
  pages = {1--20},
  issn = {0094-243X},
  doi = {10.1063/1.1390175},
  urldate = {2026-08-17}
}

@article{bartlettCoupledclusterTheoryQuantum2007,
  title = {Coupled-Cluster Theory in Quantum Chemistry},
  author = {Bartlett, Rodney J. and Musia{\l}, Monika},
  year = 2007,
  month = feb,
  journal = {Reviews of Modern Physics},
  volume = {79},
  number = {1},
  pages = {291--352},
  publisher = {American Physical Society},
  doi = {10.1103/RevModPhys.79.291},
  urldate = {2026-08-17}
}

@article{mardirossian_thirty_2017-1,
  title = {Thirty Years of Density Functional Theory in Computational Chemistry: An Overview and Extensive Assessment of 200 Density Functionals},
  shorttitle = {Thirty Years of Density Functional Theory in Computational Chemistry},
  author = {Mardirossian, Narbe and {Head-Gordon}, Martin},
  year = 2017,
  month = oct,
  journal = {Molecular Physics},
  volume = {115},
  number = {19},
  pages = {2315--2372},
  publisher = {Taylor \& Francis},
  issn = {0026-8976},
  doi = {10.1080/00268976.2017.1333644},
  urldate = {2026-07-27}
}

@article{Wellendorff2014MBEEF:Functional,
    title = {{mBEEF}: An accurate semi-local Bayesian error estimation density functional},
    year = {2014},
    journal = {Journal of Chemical Physics},
    author = {Wellendorff, Jess and Lundgaard, Keld T. and Jacobsen, Karsten W. and Bligaard, Thomas},
    number = {14},
    month = {4},
    pages = {144107},
    volume = {140},
    publisher = {AIP Publishing},
    doi = {10.1063/1.4870397},
    issn = {00219606}
}

@book{Rasmussen2005GaussianLearning,
    title = {{Gaussian Processes for Machine Learning}},
    year = {2006},
    booktitle = {Gaussian Processes for Machine Learning},
    author = {Rasmussen, Carl Edward and Williams, Christopher K. I.},
    publisher = {The MIT Press},
    url = {https://direct.mit.edu/books/book/2320/gaussian-processes-for-machine-learning},
    isbn = {9780262256834},
    doi = {10.7551/mitpress/3206.001.0001}
}

@article{kingsbury2022r2scan,
  author  = {Kingsbury, Ryan and Gupta, Ayush S. and Bartel, Christopher J. and Munro, Jason M. and Dwaraknath, Shyam and Horton, Matthew and Persson, Kristin A.},
  title   = {Performance comparison of {r$^2$SCAN} and {SCAN} {metaGGA} density functionals for solid materials via an automated, high-throughput computational workflow},
  journal = {Phys. Rev. Mater.},
  volume  = {6},
  number  = {1},
  pages   = {013801},
  year    = {2022},
  doi     = {10.1103/PhysRevMaterials.6.013801}
}

@article{mardirossian2015b97mv,
  author  = {Mardirossian, Narbe and Head-Gordon, Martin},
  title   = {Mapping the genome of meta-generalized gradient approximation density functionals: The search for {B97M-V}},
  journal = {J. Chem. Phys.},
  volume  = {142},
  number  = {7},
  pages   = {074111},
  year    = {2015},
  doi     = {10.1063/1.4907719}
}

@article{yu_mn15_2016,
  author  = {Yu, Haoyu S. and He, Xiao and Li, Shaohong L. and Truhlar, Donald G.},
  title   = {{MN15}: A {Kohn--Sham} Global-Hybrid Exchange--Correlation Density Functional with Broad Accuracy for Multi-Reference and Single-Reference Systems and Noncovalent Interactions},
  journal = {Chem. Sci.},
  volume  = {7},
  number  = {8},
  pages   = {5032--5051},
  year    = {2016},
  doi     = {10.1039/C6SC00705H}
}

@article{yanai2004new,
  author  = {Yanai, Takeshi and Tew, David P. and Handy, Nicholas C.},
  title   = {A New Hybrid Exchange--Correlation Functional Using the Coulomb-Attenuating Method ({CAM-B3LYP})},
  journal = {Chem. Phys. Lett.},
  volume  = {393},
  number  = {1--3},
  pages   = {51--57},
  year    = {2004},
  doi     = {10.1016/j.cplett.2004.06.011}
}

@article{zhang_comment_1998,
  author  = {Zhang, Yingkai and Yang, Weitao},
  title   = {Comment on ``Generalized Gradient Approximation Made Simple''},
  journal = {Phys. Rev. Lett.},
  volume  = {80},
  number  = {4},
  pages   = {890},
  year    = {1998},
  doi     = {10.1103/PhysRevLett.80.890}
}

@article{heydErratumHybridFunctionals2006,
  title = {Erratum: ``{{Hybrid}} Functionals Based on a Screened {{Coulomb}} Potential'' [{{J}}. {{Chem}}. {{Phys}}. 118, 8207 (2003)]},
  shorttitle = {Erratum},
  author = {Heyd, Jochen and Scuseria, Gustavo E. and Ernzerhof, Matthias},
  year = 2006,
  month = jun,
  journal = {The Journal of Chemical Physics},
  volume = {124},
  number = {21},
  pages = {219906},
  issn = {0021-9606},
  doi = {10.1063/1.2204597},
  urldate = {2026-08-19}
}

@article{heydHybridFunctionalsBased2003,
  title = {Hybrid Functionals Based on a Screened {{Coulomb}} Potential},
  author = {Heyd, Jochen and Scuseria, Gustavo E. and Ernzerhof, Matthias},
  year = 2003,
  month = may,
  journal = {The Journal of Chemical Physics},
  volume = {118},
  number = {18},
  pages = {8207--8215},
  issn = {0021-9606},
  doi = {10.1063/1.1564060},
  urldate = {2026-08-19}
}

@article{krukauInfluenceExchangeScreening2006,
  title = {Influence of the Exchange Screening Parameter on the Performance of Screened Hybrid Functionals},
  author = {Krukau, Aliaksandr V. and Vydrov, Oleg A. and Izmaylov, Artur F. and Scuseria, Gustavo E.},
  year = 2006,
  month = dec,
  journal = {The Journal of Chemical Physics},
  volume = {125},
  number = {22},
  pages = {224106},
  issn = {0021-9606},
  doi = {10.1063/1.2404663},
  urldate = {2026-08-19}
}

@article{mounetFirstprinciplesDeterminationStructural2005,
  title = {First-Principles Determination of the Structural, Vibrational and Thermodynamic Properties of Diamond, Graphite, and Derivatives},
  author = {Mounet, Nicolas and Marzari, Nicola},
  year = 2005,
  month = may,
  journal = {Physical Review B},
  volume = {71},
  number = {20},
  pages = {205214},
  issn = {1098-0121, 1550-235X},
  doi = {10.1103/PhysRevB.71.205214},
  urldate = {2026-08-19},
  copyright = {http://link.aps.org/licenses/aps-default-license},
  langid = {english}
}

@article{bursch2022r2scan0,
  author  = {Bursch, Markus and Neugebauer, Hagen and Ehlert, Sebastian and Grimme, Stefan},
  title   = {Dispersion Corrected r$^2${SCAN} Based Global Hybrid Functionals: r$^2${SCAN}h, r$^2${SCAN}0, and r$^2${SCAN}50},
  journal = {J. Chem. Phys.},
  volume  = {156},
  number  = {13},
  pages   = {134105},
  year    = {2022},
  doi     = {10.1063/5.0086040}
}

@article{weigend2005balanced,
  author = {Weigend, Florian and Ahlrichs, Reinhart},
  title = {Balanced basis sets of split valence, triple zeta valence and quadruple zeta valence quality for {H} to {Rn}: Design and assessment of accuracy},
  journal = {Phys. Chem. Chem. Phys.},
  volume = {7},
  number = {18},
  pages = {3297--3305},
  year = {2005},
  doi = {10.1039/b508541a}
}

@article{rappoport2010property,
  author = {Rappoport, Dmitrij and Furche, Filipp},
  title = {Property-optimized {Gaussian} basis sets for molecular response calculations},
  journal = {J. Chem. Phys.},
  volume = {133},
  number = {13},
  pages = {134105},
  year = {2010},
  doi = {10.1063/1.3484283}
}

@article{dejong2001parallel,
  author = {de Jong, W. A. and Harrison, R. J. and Dixon, D. A.},
  title = {Parallel {Douglas--Kroll} energy and gradients in {NWChem}: Estimating scalar relativistic effects using {Douglas--Kroll} contracted basis sets},
  journal = {J. Chem. Phys.},
  volume = {114},
  number = {1},
  pages = {48--53},
  year = {2001},
  doi = {10.1063/1.1329891}
}

@article{balabanov2005dk,
  author = {Balabanov, Nikolai B. and Peterson, Kirk A.},
  title = {Systematically convergent basis sets for transition metals. {I}. All-electron correlation consistent basis sets for the 3d elements {Sc--Zn}},
  journal = {J. Chem. Phys.},
  volume = {123},
  number = {6},
  pages = {064107},
  year = {2005},
  doi = {10.1063/1.1998907}
}

\clearpage
\section*{Extended Data}

\begin{center}
    \includegraphics[width=\linewidth]{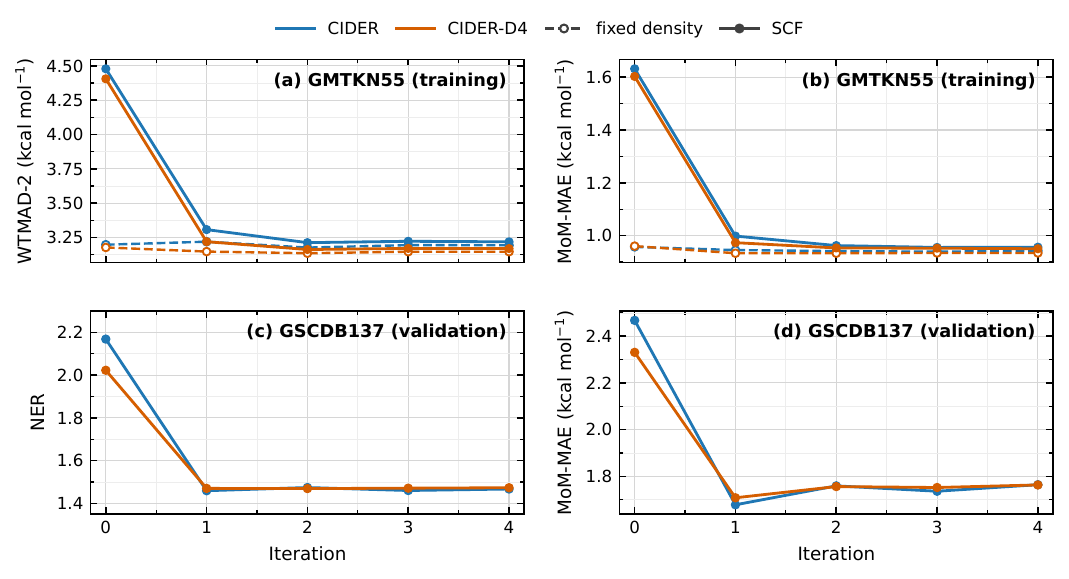}
\end{center}
\noindent\textbf{Extended Data Fig.~1 $|$ Self-consistent molecular training.}
\textbf{a,b}, GMTKN55 training WTMAD-2 and MoM-MAE through iteration 4; GMTKN55 is training data
for the CIDER models. Dashed lines with open markers show each iteration's model evaluated on the
fixed densities it was fit on; solid lines with filled markers show full self-consistent
calculations. \textbf{c,d}, Self-consistent NER and MoM-MAE on the held-out GSCDB137
validation split (2571 reactions). CIDER and CIDER-D4 denote the dispersion-free and
D4-corrected iteration families that produced CIDER26C and CIDER26C-D4; the loop mainly corrects
the large seed-model density shift within one iteration. The selected production models are
separate refits after the final atomic and transition-metal density refresh and are not shown
on these curves. Lower is better in all panels.

\clearpage
\begin{center}
    \includegraphics[width=\linewidth]{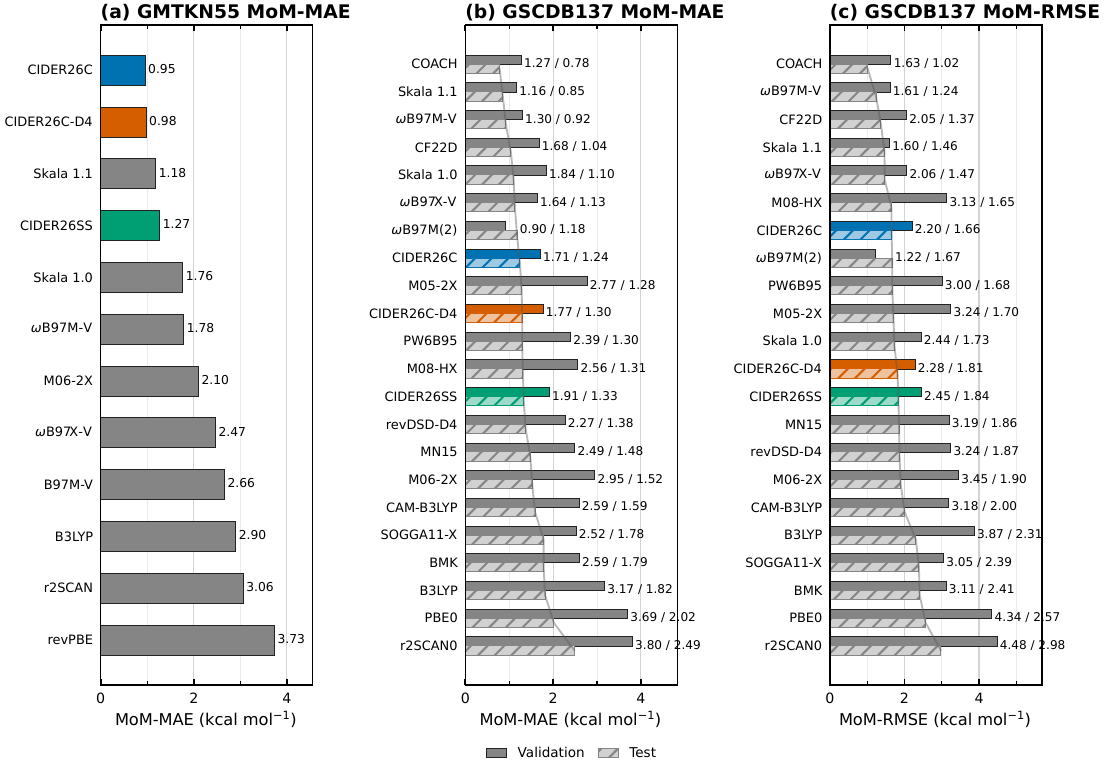}
\end{center}
\noindent\textbf{Extended Data Fig.~2 $|$ Mean-of-means benchmark errors.}
\textbf{a}, GMTKN55 MoM-MAE, the unweighted mean of the 55 subset MADs, in kcal~mol\(^{-1}\);
GMTKN55 is training data for the CIDER models, and the values for Skala and all other non-CIDER
functionals are computed from the data of Refs.~\cite{luise2025accurate,luise2025accuratev3}.
\textbf{b,c}, GSCDB137 MoM-MAE and MoM-RMSE on the held-out validation (2571 reactions) and test
(3044 reactions) splits; solid bars are validation and hatched bars test, with labels giving
validation/test values. Rows are ordered by test error, and the grey line traces the test values
as a guide to the eye. Skala~1.1 covers 3041 of the test reactions; the other displayed CIDER
and Skala models cover both complete splits. Hybrid-functional and COACH values are computed
from the data of Refs.~\cite{liang2025gscdb137,liang2026coach}, and the Skala values are this
work's self-consistent evaluations. Lower is better in all panels.

\clearpage
\begin{center}
    \includegraphics[width=\linewidth]{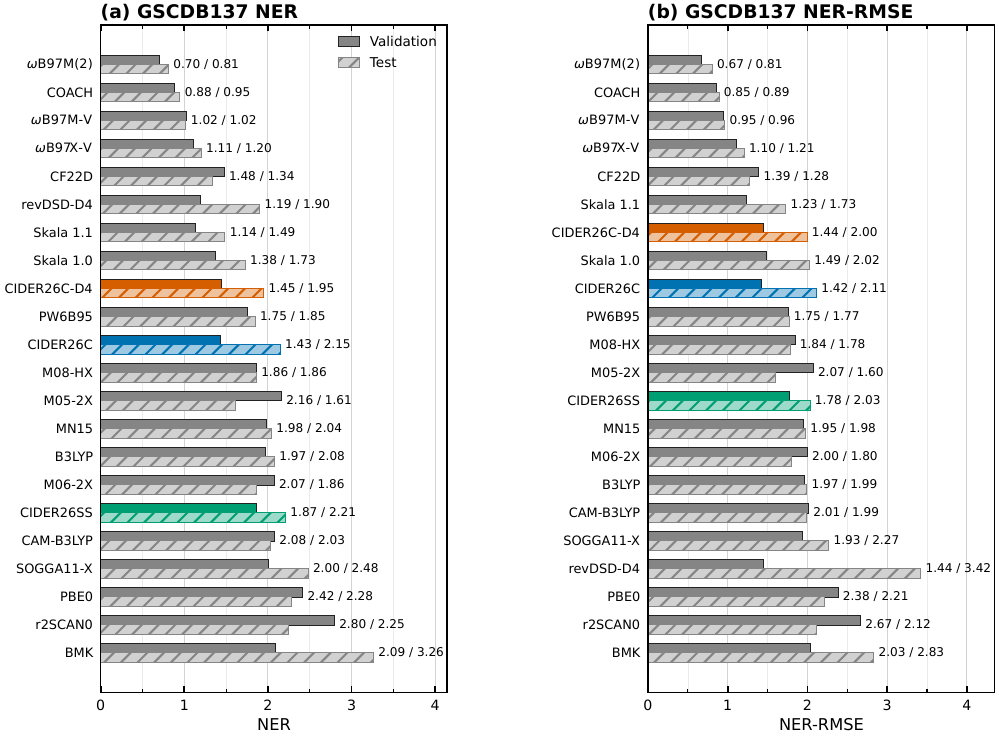}
\end{center}
\noindent\textbf{Extended Data Fig.~3 $|$ Validation and test benchmark errors.}
\textbf{a,b}, GSCDB137 NER and NER-RMSE on the held-out validation (2571 reactions, 38
subdatasets) and test (3044 reactions, 51 subdatasets) splits, with 21 subdatasets
contributing reactions to both; solid bars are validation and
hatched bars test, with labels giving validation/test values. Rows follow the combined
validation+test ordering of the corresponding panels of Figure~\ref{fig:benchmarks}. Skala~1.1
covers 3041 of the test reactions; the other displayed CIDER and Skala models cover both
complete splits. Hybrid-functional and COACH values are computed from the data of
Refs.~\cite{liang2025gscdb137,liang2026coach}, and the Skala values are this work's
self-consistent evaluations. Lower is better in both panels.

\clearpage
\begin{center}
    \includegraphics[width=\linewidth]{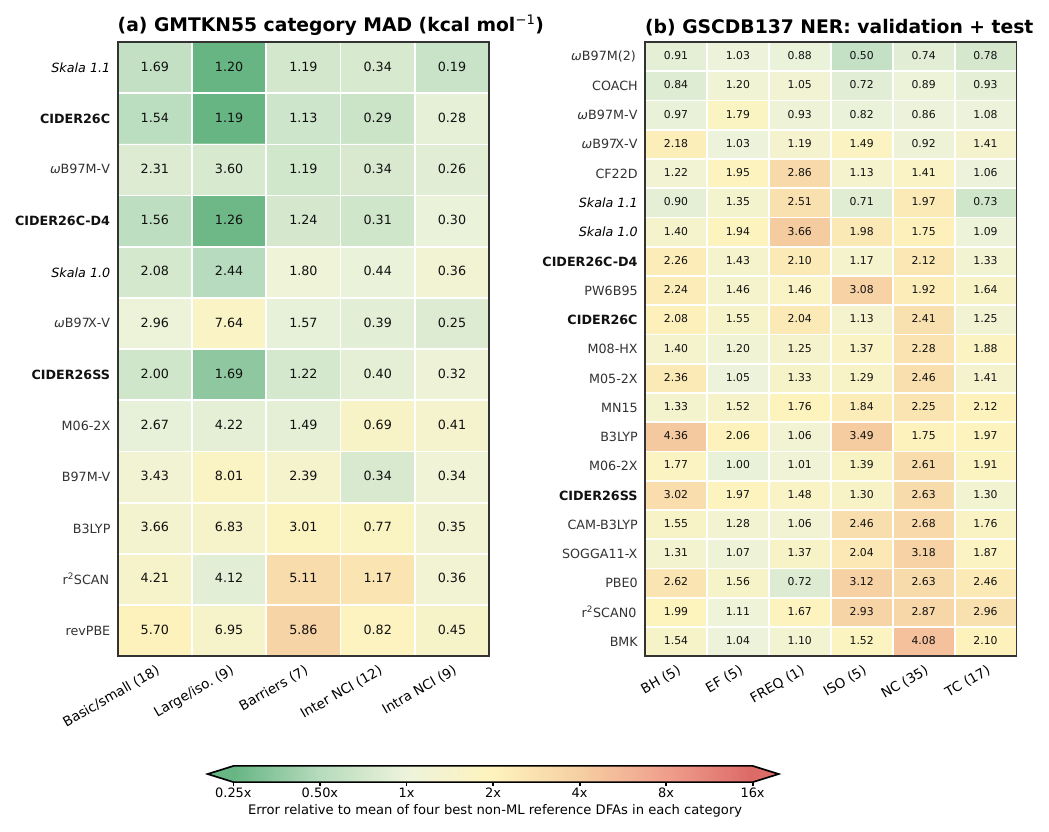}
\end{center}
\noindent\textbf{Extended Data Fig.~4 $|$ Category-resolved molecular benchmarks.}
\textbf{a}, GMTKN55 category mean absolute deviations (kcal~mol\(^{-1}\)) over the five GMTKN55
top-level categories (basic and small-system properties; large-system and isomerization
reactions; barrier heights; intermolecular noncovalent interactions; and intramolecular
noncovalent interactions); GMTKN55 is training data for the CIDER models. Values for Skala and
all other non-CIDER functionals are from Refs.~\cite{luise2025accurate,luise2025accuratev3}. \textbf{b}, GSCDB137 per-category NER on the combined
validation+test panel, with rows ordered by combined overall NER (BH, barrier heights; EF,
electric-field response; FREQ, harmonic frequencies; ISO, isomerization; NC, noncovalent
interactions, with the inter- and intramolecular subdatasets combined; TC, thermochemistry);
hybrid-functional and COACH values are computed from the
data of Refs.~\cite{liang2025gscdb137,liang2026coach}, the Skala values are this work's
self-consistent evaluations, and revDSD-PBEP86-D4 is omitted from this panel. In both panels,
the number in parentheses after each category label gives the number of subdatasets in that
category. Cell colors give
the error relative to the mean of the four best non-ML reference functionals in each category.

\clearpage
\noindent\textbf{Extended Data Table~1 $|$ Selected molecular model performance.}
Performance on the GMTKN55 training set and the GSCDB137 validation and test sets. Model
selection used the validation set only. All entries except NER are self-consistent
kcal~mol$^{-1}$ errors. AE and TM denote the atomic-energy and transition-metal GSCDB137 subsets, which are part of
the training set; category rows report GMTKN55 MAD only.

\begin{center}
\begin{tabular}{lcc}
\toprule
Metric & CIDER26C & CIDER26C-D4 \\
\midrule
GMTKN55 WTMAD-2 & 3.215 & 3.350 \\
GMTKN55 MoM-MAE & 0.953 & 0.975 \\
AE MoM-MAE & 2.850 & 2.820 \\
TM MoM-MAE & 2.529 & 3.454 \\
GSCDB137 NER (validation) & 1.427 & 1.446 \\
GSCDB137 NER (test) & 2.147 & 1.950 \\
GSCDB137 MoM-MAE (validation) & 1.705 & 1.768 \\
GSCDB137 MoM-MAE (test) & 1.243 & 1.295 \\
\midrule
\multicolumn{3}{l}{GMTKN55 category MAD} \\
Basic/small & 1.54 & 1.56 \\
Large/iso. & 1.19 & 1.26 \\
Barriers & 1.13 & 1.24 \\
Inter NCI & 0.29 & 0.31 \\
Intra NCI & 0.28 & 0.30 \\
\bottomrule
\end{tabular}
\end{center}

\clearpage
\noindent\textbf{Extended Data Table~2 $|$ Dipole-density quality on the GSCDB137 Dip146 subset.}
Values are Dip146 NER on the full 190-reaction panel (189 reactions for Skala~1.1), computed
with the regularized dipole metric and the official GSCDB137 standard-error normalization of
Ref.~\cite{liang2025gscdb137}. All non-CIDER, non-Skala values are obtained from Ref.~\cite{liang2025gscdb137}, except COACH,
which comes from Ref.~\cite{liang2026coach}.

\begin{center}
\begin{tabular}{lc}
\toprule
Functional & Dip146 NER \\
\midrule
$\omega$B97M(2) & 0.75 \\
revDSD-PBEP86-D4 & 0.89 \\
COACH & 0.91 \\
SOGGA11-X & 0.93 \\
Skala~1.1 & 0.93 \\
PW6B95 & 0.95 \\
$\omega$B97X-V & 1.00 \\
CAM-B3LYP & 1.05 \\
PBE0 & 1.05 \\
M06-2X & 1.05 \\
$\omega$B97M-V & 1.05 \\
M08-HX & 1.08 \\
CF22D & 1.12 \\
B3LYP & 1.20 \\
r$^2$SCAN0 & 1.21 \\
BMK & 1.24 \\
CIDER26C & 1.30 \\
M05-2X & 1.30 \\
CIDER26C-D4 & 1.30 \\
CIDER26SS & 1.32 \\
MN15 & 1.33 \\
Skala~1.0 & 1.47 \\
\midrule
\multicolumn{2}{c}{Semilocal functionals} \\
\midrule
r$^2$SCAN & 1.79 \\
B97M-V & 2.07 \\
revPBE & 2.44 \\
\bottomrule
\end{tabular}
\end{center}

\clearpage
\noindent\textbf{Extended Data Table~3 $|$ Combined-panel NER-MAE, NER-RMSE, and the ratio of the latter to the former.}
Values on the combined GSCDB137 validation+test panel; the ratio measures error spread relative
to the hybrid reference scale (Methods). Non-CIDER values are computed from the data of
Refs.~\cite{liang2025gscdb137,liang2026coach}; the Skala values are this work's evaluations.
DH marks the two double hybrids, which are excluded from the standard-error normalization
(Methods).

\begin{center}
\begin{tabular}{lccc}
\toprule
Functional & NER-MAE & NER-RMSE & Ratio \\
\midrule
$\omega$B97M(2) (DH) & 0.767 & 0.767 & 1.00 \\
COACH & 0.907 & 0.870 & 0.96 \\
$\omega$B97M-V & 0.986 & 0.945 & 0.96 \\
$\omega$B97X-V & 1.187 & 1.186 & 1.00 \\
CF22D & 1.350 & 1.295 & 0.96 \\
revDSD-PBEP86-D4 (DH) & 1.427 & 2.405 & 1.69 \\
Skala~1.1 & 1.453 & 1.657 & 1.14 \\
Skala~1.0 & 1.619 & 1.813 & 1.12 \\
CIDER26C-D4 & 1.809 & 1.800 & 1.00 \\
PW6B95 & 1.917 & 1.880 & 0.98 \\
CIDER26C & 1.931 & 1.865 & 0.97 \\
M08-HX & 1.953 & 1.891 & 0.97 \\
M05-2X & 1.984 & 1.921 & 0.97 \\
MN15 & 2.057 & 2.014 & 0.98 \\
B3LYP & 2.137 & 2.081 & 0.97 \\
M06-2X & 2.144 & 2.047 & 0.95 \\
CIDER26SS & 2.165 & 2.012 & 0.93 \\
CAM-B3LYP & 2.221 & 2.184 & 0.98 \\
SOGGA11-X & 2.450 & 2.269 & 0.93 \\
PBE0 & 2.513 & 2.474 & 0.98 \\
r$^2$SCAN0 & 2.685 & 2.542 & 0.95 \\
BMK & 2.943 & 2.617 & 0.89 \\
\bottomrule
\end{tabular}
\end{center}

\clearpage
\noindent\textbf{Extended Data Table~4 $|$ Errors against the EXX+RPA surface-science training
targets.} Mean absolute errors by reaction class, in eV. For CIDER26SS, the fixed-density
column evaluates the model on the training densities it was fit on, while the self-consistent
column uses full SCF calculations at the model's own densities.

\begin{center}
\begin{tabular}{lccc}
\toprule
 & \multicolumn{2}{c}{CIDER26SS} & PBE \\
Reaction class & Fixed & SCF & \\
\midrule
Adsorption & 0.196 & 0.216 & 0.338 \\
Surface & 0.089 & 0.093 & 0.277 \\
Cohesive & 0.299 & 0.283 & 0.550 \\
Phase stability & 0.049 & 0.050 & 0.048 \\
Atomization & 0.248 & 0.216 & 0.763 \\
Equation of state & 0.006 & 0.006 & 0.036 \\
\bottomrule
\end{tabular}
\end{center}

\clearpage
\noindent\textbf{Extended Data Table~5 $|$ Relative protonation energies for the dimeric
iron--sulfur model of Ref.~\cite{zhai2023fes}.} HC, HS, HFe, and HFe2 denote protonation at
the carbon, sulfur, single-iron, and iron--iron bridging sites of the cluster. Values are in
kJ~mol$^{-1}$ with the HC structure as zero; the final column is the mean absolute error over
HS, HFe, and HFe2. The DFT values and the CCSD(T)+DMRG reference are from that work; CIDER,
Skala, $\omega$B97M-V, and $\omega$B97X-V values were evaluated here on the same structures.

\begin{center}
\begin{tabular}{lrrrr}
\toprule
Method & HS & HFe & HFe2 & MAE \\
\midrule
CCSD(T)+DMRG & 135.5 & 230.6 & 199.9 & --- \\
PBE-D3 & 93.1 & 145.6 & 87.8 & 79.8 \\
r$^2$SCAN-D3 & 114.5 & 163.5 & 128.3 & 53.2 \\
TPSSh-D3 & 115.6 & 200.9 & 156.0 & 31.2 \\
B3LYP*-D3 & 113.9 & 203.4 & 176.0 & 24.2 \\
M06-D3 & 135.5 & 190.6 & 194.0 & 15.3 \\
PBE0-D3 & 132.0 & 239.5 & 225.0 & 12.5 \\
B3LYP-D3 & 122.1 & 223.8 & 209.1 & 9.8 \\
\textbf{CIDER26SS} & \textbf{130.5} & \textbf{192.7} & \textbf{164.5} & \textbf{26.1} \\
$\omega$B97M-V & 135.1 & 281.2 & 272.4 & 41.2 \\
$\omega$B97X-V & 135.6 & 279.7 & 267.2 & 38.8 \\
Skala~1.0 & 127.9 & 150.4 & 127.1 & 53.5 \\
Skala~1.1 & 120.1 & 257.3 & 196.6 & 15.1 \\
\bottomrule
\end{tabular}
\end{center}

\clearpage

\noindent\textbf{Extended Data Table~6 $|$ Equilibrium properties of triplet \ce{O2}.}
Bond lengths ($R_e$) are in \AA{}, harmonic frequencies ($\omega_e$) in cm$^{-1}$, and
electronic well depths ($D_e$) in eV. The spectroscopic, CEEIS-FCI/CBS, and W4-17 references
are from Refs.~\cite{krupenie1972o2,bytautas2010o2,karton2017w417}.

\begin{center}
\begin{tabular}{lccc}
\toprule
Method & $R_e$ (\AA{}) & $\omega_e$ (cm$^{-1}$) & $D_e$ (eV) \\
\midrule
\textbf{CIDER26SS} & \textbf{1.2043} & \textbf{1606.2} & \textbf{4.893} \\
PBE-D4 & 1.2176 & 1543.9 & 6.254 \\
B3LYP-D4 & 1.2033 & 1628.2 & 5.392 \\
PBE0-D4 & 1.1918 & 1726.0 & 5.421 \\
$\omega$B97M-V & 1.1964 & 1712.8 & 5.355 \\
$\omega$B97X-V & 1.1953 & 1724.1 & 5.505 \\
B97M-V & 1.1991 & 1653.6 & 5.313 \\
r$^2$SCAN-D4 & 1.2052 & 1629.5 & 5.650 \\
Skala~1.0 & 1.1936 & 1725.3 & 5.314 \\
Skala~1.1 & 1.1885 & 1725.1 & 5.185 \\
\midrule
CEEIS-FCI/CBS & 1.2078 & --- & 5.210 \\
W4-17 & --- & --- & 5.213 \\
Expt. & 1.2075 & 1580.2 & --- \\
\bottomrule
\end{tabular}
\end{center}

\end{document}